\documentclass[%
reprint,
superscriptaddress,
nobibnotes,
amsmath,amssymb,
aps,pre
floatfix,longbibliography
]{revtex4-1}

\usepackage{graphicx}% Include figure files
\usepackage{dcolumn}% Align table columns on decimal point
\usepackage{bm}% bold math
\usepackage[hidelinks]{hyperref}

\usepackage{tipa}
\usepackage{epsfig}
\usepackage{amsmath}   
\usepackage{amsfonts}
\usepackage{amssymb}
\usepackage{graphicx}  
\usepackage{verbatim} 
\usepackage{color}   
\usepackage{subfigure}  
\usepackage{hyperref}  
\usepackage[T1]{fontenc}
\usepackage{xcolor}
\usepackage{bbm}

\usepackage[bbgreekl]{mathbbol}
\DeclareMathSymbol\bDelta  \mathord{bbold}{"01}
\newcommand{\bbDelta}{\text{\large$\bDelta$}}

\DeclareMathOperator{\Tr}{Tr}

\newcommand{\Mean}[1]{\underset{#1}{\text{Mean}}}
\newcommand{\Sum}[2]{\sum\limits_{#1}^{#2}}

\newcommand{\avRS}[2]{ \bm{\{ } #1 \bm{\} }_{\text{\tiny #2}} }
\newcommand{\mQ}[2]{ {\mathbbm{Q}^{#1}_{\text{\tiny #2}} } }

\newcommand{\qEA}{q_{\text{\tiny EA}}}
\newcommand{\TMCT}{T_{\text{\tiny MCT}}}

\newcommand{\MCT}{\text{\tiny MCT}}

\newcommand{\M}{\text{\tiny M}}
\newcommand{\FP}{\text{\tiny FP}}

\begin{document}

%\tableofcontents
%\clearpage

\title{Equilibrium Fluctuations in Mean-field Disordered Models}

\author{Giampaolo Folena}
\affiliation{Laboratoire de Physique de l'Ecole Normale Sup\'erieure, ENS, Universit\'e PSL, CNRS, Sorbonne Universit\'e, Universit\'e de Paris, F-75005 Paris, France
}
\affiliation{James Franck Institute and Department of Physics, University of Chicago, Chicago, IL 60637, U.S.A.}
\author{Giulio Biroli}
\affiliation{Laboratoire de Physique de l'Ecole Normale Sup\'erieure, ENS, Universit\'e PSL, CNRS, Sorbonne Universit\'e, Universit\'e de Paris, F-75005 Paris, France
}
\author{Patrick Charbonneau}
\affiliation{Department of Chemistry, Duke University, Durham, North Carolina 27708, USA}
\affiliation{Department of Physics, Duke University, Durham, North Carolina 27708, USA}

\author{Yi Hu}
\affiliation{Department of Chemistry, Duke University, Durham, North Carolina 27708, USA}
\author{Francesco Zamponi}
\affiliation{Laboratoire de Physique de l'Ecole Normale Sup\'erieure, ENS, Universit\'e PSL, CNRS, Sorbonne Universit\'e, Universit\'e de Paris, F-75005 Paris, France
}

\begin{abstract}
	 Mean-field models of glasses that present a random first order transition exhibit highly non-trivial  fluctuations. Building on previous studies that focused on the critical scaling regime, we here obtain a fully quantitative framework for all equilibrium conditions. By means of the replica method we evaluate Gaussian fluctuations of the overlaps around the thermodynamic limit, decomposing them in thermal fluctuations inside each state and heterogeneous fluctuations between different states. 
	We first test and compare our analytical results with numerical simulation results for the $p$-spin spherical model and the random orthogonal model, and then analyze the random Lorentz gas. In all cases, a strong quantitative agreement is obtained. Our analysis thus provides a robust scheme for identifying the key finite-size (or finite-dimensional) corrections to the mean-field treatment of these paradigmatic glass models.
\end{abstract}
	
\maketitle

\section{Introduction}
\label{intro}
\medskip

Equilibrium fluctuations of macroscopic observables are a key concern of statistical mechanics. The chaotic microscopic dynamics, once averaged over a large number of degrees of freedom, is macroscopically described by averaged observables, such as energy, pressure and density. Away from phase transitions, these quantities fluctuate relatively little %show only very small dynamical fluctuations 
around their average value.  At phase transitions, however, competition between different phases leads to very different microscopic states being sampled, and thus to large fluctuations. %of macroscopic observables, driven by the coexistence of 
Certain phase transition lines also terminate at critical points, at which the free energy cost of changing phase vanishes. Fluctuations are then even more significant.

More formally, away from phase transitions a given macroscopic observable $O = \sum_{i=1}^N o_i$ of a system with $N$ components fluctuates thermally with variance $\propto\sqrt{N}$. This universal scaling follows directly from the statistical independence between different parts of the system, and the central limit theorem (assuming that the interface contribution is subdominant in the thermodynamic limit)~\cite{LANDAU19801}. At critical points, by contrast, fluctuations are bounded only by the size of the system, and therefore diverge in the thermodynamic $N\to\infty$ limit, and often display a universal critical scaling. % ( phenomena).

To further formalize the role of fluctuations, consider the probability of observing $o=O/N$ away from its average,
\begin{equation}\label{prob_x}
P(o)\propto e^{-N f(o)} \ ,
\end{equation}
where $f(o)$ is the large deviation function related to this probability, i.e., the intensive free energy of the system~\cite{touchette_large_2009}. Its global minimum, $o^*$, corresponds to the average, $\langle o \rangle = o^*$, in the thermodynamic limit $N\to\infty$. 
Given $f(o)$, the variance of fluctuations is straightforwardly given by the inverse of the second-order derivative (or, more generally, the Hessian) around $o^*$:
\begin{equation}\label{var_x}
\langle o^2 \rangle-\langle o \rangle^2 = \frac{1}{N\partial^2_o f(o)\big|_{o=o^*}} \ .
\end{equation}
For large enough $N$, the probability $P(o)$ can also be meaningfully approximated by a Gaussian distribution characterized by the aforementioned variance, because higher-order cumulants of the distribution grow with smaller powers of $N$. This regime of \textit{small fluctuations} describes typical fluctuations of extensive observables away from both spinodal points and 
phase transitions. 
%Let's recall that equilibrium fluctuations are profoundly connected to responses of the system by the fluctuation dissipation theorem,
%\begin{equation}
%\langle x^2 \rangle-\langle x \rangle^2 = T \partial_h\langle x \rangle = N^{-1}\partial^2_h \tilde{f}(h)
%\end{equation}
%where $h$ is a field conjugated to $x$ and $\tilde{f}$ is the Legendre transform of $f(x)$, or in more mathematical terms the cumulants' generating function.

If $o$ is a good order parameter, then a first-order phase transition corresponds to $o^*$ jumping from one state to another. In a mean-field description, this jump is accompanied by a barrier that diverges with $N$. For short-ranged models, the situation is more subtle because barriers between states scale subdominantly with $N$ with a power that depends on the nature of the interface between states. For simplicity, we thus here only consider the former case, and further exclude large deviation descriptions %in terms of $\exp(-N f(x))$ does not
that possess ambiguities resulting in the non-equivalence of ensembles~\cite{gross_microcanonical_2001}.

Mean-field disordered models (with either quenched or self-induced disorder) can exhibit a complex free energy landscape that contains a large number of distinct minima. More formally, the number of minima may grow exponentially with $N$ \cite{bray_metastable_1980,crisanti_thouless-anderson-palmer_1995}, while remaining separated by barriers that grow with a power of  $N$ \cite{ros_complexity_2019}. Given a \emph{good} local order parameter, $o_i$, which depends on some relevant microscopic degree of freedom, the macroscopic free energy $F(\{o_i\})$ %---also called TAP free energy, which corresponds to PEL at small temperature--- 
then exhibits a minimum for each metastable state \cite{thouless_solution_1977,plefka_convergence_1982,georges_how_1991}. At a given temperature, many of these metastable states have nearly equal Boltzmann weights, and each such state is deemed \textit{typical} if it belongs to the set of minima %, of given properties, 
that dominate the partition function. (For convenience, typical metastable states are denoted \textit{states} in the rest of the paper.) % hence the possibility of exploring the phase space of the system is exponentially suppressed.  
Given this structure, a (metastable) state can also be construed as the region of phase space visited by an equilibrium dynamics, such as that of a system in contact with a thermal bath of the same temperature or following Newtonian dynamics with the adequate kinetic energy. For large enough systems, the equilibrium dynamics thus mainly explores %corresponds to the exploration of 
a typical state with only very rare instantonic escapes \cite{ros_dynamical_2020,rizzo_path_2021}. %This picture is, however, significantly more complicated in short-range models, because barriers associated to metastable states then do not scale with $N$. The definition of an instantonic path in systems with many coexisting phases is also unclear.

While fluctuations in simple mean-field models without quenched disorder are typically introduced in undergraduate-level statistical physics, they have long been neglected %same cannot be said about 
in glassy models with or without disorder. 
Recently, however, it has been recognized that such fluctuations play an important role in several aspects of glassy physics, from dynamical heterogeneity~\cite{KT88,berthier_structure_2007,franz_field_2011,franz_static_2013} to metastable states~\cite{biroli_random-field-like_2014,berthier_self-induced_2021}. Furthermore, they are key ingredients to obtaining quantitative corrections to the infinite-dimensional mean-field theory of glasses~\cite{biroli_unifying_2020,biroli_local_2021}, and to understand non-perturbative relaxation processes~\cite{rizzo2015qualitative,rizzo2020solvable}.
In this work, we follow the approach of Ref.~\cite{KT88}, further developed a decade ago in Ref.~\cite{franz_field_2011},
to describe the critical scaling of fluctuations, and %use the replica method to 
extend its use to all equilibrium conditions %{ \it quantitative} grasp of fluctuations 
for the family of mean-field models--including the random Lorentz gas--that present a random first order transition (RFOT)~\cite{LW07,KT15}. 
Our analysis hence provides a quantitative grasp of the finite-size (or finite-dimensional) corrections to the mean-field treatment of these paradigmatic model glass formers.

The rest of the paper is structured as follows. Section~\ref{sec:def} introduces some basic ideas and definitions.
Section \ref{EF} details the general correspondence between overlap fluctuations and temporal averages. Section \ref{SMM} describes how the replica method can generally be used to evaluate the mass matrix and therefore quantify overlap fluctuations. Three specific models are then considered: spherical $p$-spin (Section~\ref{psp}), random orthogonal (Section~\ref{rom}), and random Lorentz gas (Section~\ref{rlg}). A brief conclusion follows in Section~\ref{conclusion}. Appendix \ref{AppendixA} describes the mass matrix diagonalization technique. Appendix \ref{AppendixB0} details the computation of fluctuations in a $p$-spin with homogeneous external field. Appendix \ref{AppendixB} evaluates fluctuations using the cumulant method, whose results agree with the mass matrix method presented in the main text.
Additional appendices provide technical details. %, when helpful. %to that of Sec.~\ref{SMM} for evaluating fluctuations.

\section{Definitions}
\label{sec:def}

\subsection{Fluctuations in RFOT Models}
\begin{figure}[t]    
    \centering
	\includegraphics[width=0.99\columnwidth]{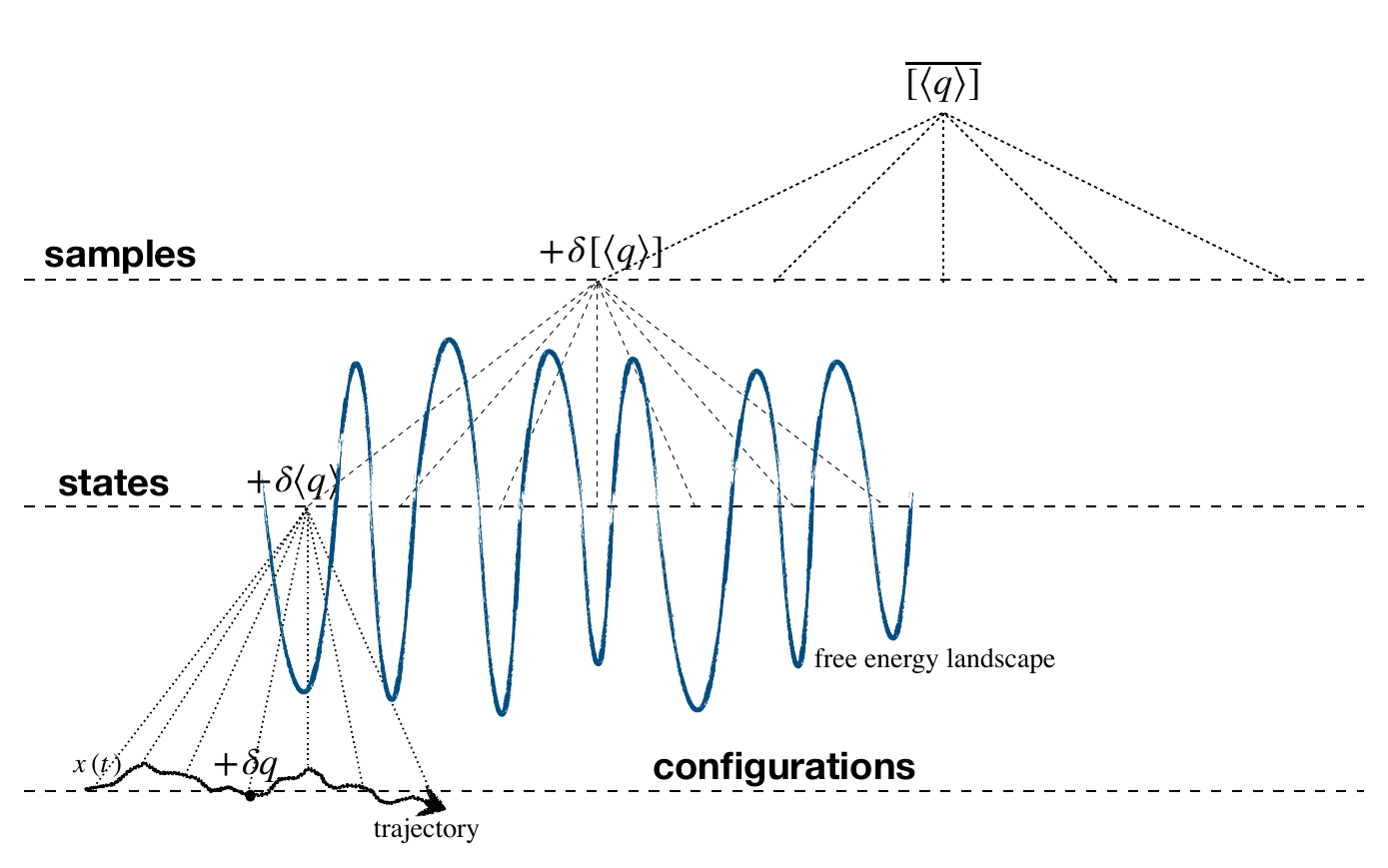}\\
	\caption{\footnotesize Hierarchy of fluctuations in disordered models exhibiting RFOT phenemenology.}
	\label{fig:sketch_hier}
\end{figure}

In order to sketch out this endeavor we first review the various physical contributions to fluctuations in these models. Denoting $\langle\bullet\rangle$ an average over equilibrium configurations within a given state,  $[\bullet]$ an average over the different states of a given realization of the quenched disorder, and $\overline{\bullet}$ an average over different quenched disorders, fluctuations for an \emph{appropriate} observable $q$, such as the overlap (or the distance) between two configurations, can be hierarchically decomposed as~\cite{berthier_structure_2007,franz_field_2011}:
\begin{itemize}
\item \textit{intra-state} fluctuations $\delta q$, which are related to the equilibrium exploration of a state;
\item \textit{inter-state} fluctuations $\delta \langle q \rangle$, which are related to the variability of states in a given system with fixed quenched disorder;
\item \textit{disorder} fluctuations $\delta [\langle q \rangle]$, which are related to the variability between different realizations of the quenched disorder.
\end{itemize}
This breakdown is generic for mean-field models with quenched disorder that exhibit RFOT phenomenology. (Comparable systems without quenched disorder, such as supercooled liquids in the high-dimensional $d\rightarrow\infty$ limit, lack disorder fluctuations.)
%A natural choice of $q$ is then . % is  in disordered systems, where there is no explicit symmetry. 
For the generic case sketched in Fig.~\ref{fig:sketch_hier}, the cumulative effect of these three fluctuations defines the observed $q$,
\begin{equation}
q - \overline{[\langle q \rangle]}= \delta [\langle q \rangle]+\delta \langle q \rangle+\delta q \ .
\end{equation}
%where $\overline{[\langle q \rangle]}$ is the expected value of $x$ when averaging over all samples $\overline{\bullet}$, states $[\bullet]$ and configurations $\langle \bullet \rangle$.
In the \textit{small-fluctuation} regime, each level of the hierarchy exhibits Gaussian fluctuations, and hence three susceptibilities (or variances) proportional to $1/N$ naturally emerge ~\cite{berthier_structure_2007,franz_field_2011}:
\begin{eqnarray}
\chi_{intra} = \overline{[\langle \delta q\delta q\rangle]} \ , \label{intra}\\
\chi_{inter} = \overline{[\delta \langle q \rangle\delta \langle q \rangle]} \ ,\label{inter}\\
\chi_{dis} = \overline{ \delta [\langle q \rangle]\delta [\langle q \rangle]} \ .\label{samples}
\end{eqnarray}
The total susceptibility is then $\chi_{tot} = \chi_{intra}+\chi_{inter}+\chi_{dis}$.

Furthermore, we can define a glass sample, or just a {\it sample}, as a single state in a single realization of the quenched disorder (if present).
This is because physically, a glass is confined into a single state.
Hence, we define the sample-to-sample susceptibility as $\chi_{sample} = \chi_{inter}+\chi_{dis}$.
Note that the intra-state susceptibility $\chi_{intra}$ is here averaged over all samples, but can be evaluated for a single state. Similarly, $\chi_{inter}$ does not need to be averaged over different realizations of quenched disorder.

Although these different contributions to fluctuations make their analysis somewhat more involved than for ordered system, the approach is nevertheless similar. By analogy to Eq.~\eqref{var_x}, we wish to define a large deviation function such that the Hessian evaluated at the saddle point provides the correct description of small fluctuations around the thermodynamic limit. We therefore consider a free energy built by averaging over configurations, states and samples. 
Given the free energy of a single sample $\alpha$,
\begin{equation}\label{free_en}
-\beta f_\alpha = \ln Z_\alpha = N^{-1}\ln \big [ \sum_{x} \exp(-\beta H_\alpha(x)) \big ] \ ,
\end{equation}
we compute the disorder average
\begin{equation}\label{rep_met}
\overline{f} = \sum_f P(f) f = \lim_{n\to0}\partial_n \overline{ \exp (n  f )} = \lim_{n\to0}\partial_n F(\mQ{n}{}) \ ,
\end{equation}
where $P(f)\propto\sum_\alpha \delta (f-f_\alpha)$ is the probability that a sample has free energy $f$. Note that the average over states is here implicit because Eq.~\eqref{free_en} contains a sum over all possible configurations of a given sample. 

\subsection{Replica Method for Computing Fluctuations}
The average over the quenched disorder can be evaluated by the replica method. %\cite{edwards_theory_1975}. 
(See e.g. Refs.~\cite{mezard_spin_1987,castellani_spin-glass_2005,parisi_theory_2020,folena_mixed_2020} for a pedagogical introduction.)  The replicated free energy $F(\mQ{}{})$ is a large deviation function for the $n\times n$ overlap matrix $\mQ{}{}$,
\begin{equation}\label{free_Q}
P(\mQ{}{})\propto e^{-N F(\mQ{}{})},
\end{equation}
where each element of the matrix $\mQ{}{}$ quantifies the overlap between two configurations, ${\mQ{}{}}_{a,b} =  q_{ab} = q(x_a, x _{b})$, for a given function $q(x,x')$ that measures similarity of configuration pairs $x,x'$ (defined more precisely below). 
The variable $n$ is a sort of temperature associated to the quenched disorder, and is conjugate to the free energy $f$. The operator $\lim_{n\to0}\partial_n$ then extracts the first cumulant (average) of $P(f)$.
In the thermodynamic limit, at fixed $n$, we expect the measure to concentrate on the most probable matrix $\mQ{*}{}$. At this minimum,  we have
\begin{equation}
{\mQ{*}{}}_{a,b} = \overline{[\langle q_{ab} \rangle]} = \overline{[\langle q(x_a, x_{b})\rangle]}.
\end{equation}
The standard way to surmount the difficulty of parameterizing the space of $n \times n$ matrices for performing an analytical continuation to non-integer $n$ % in full generality,identifying the global minimum commonly requires 
is to make an \textit{ansatz} on the structure of  $\mQ{}{}$ ---replica symmetric (RS) or $k$-times replica symmetry broken ($k$RSB)---  and of extremizing only in the subspace defined by this ansatz. %Historical advances that gave rise to this ansatz went more or less as follows. Sherrington and Kirkpatrick first applied the RS (or $k=0$) ansatz, which simply takes all system copies to have the same (typical) properties, to a fully-connected model of spin glasses~\cite{sherrington_solvable_1975}. De Almeida and Thouless~\cite{almeida_stability_1978} and Pytte and Rudnick~\cite{pytte1979scaling} later understood this solution to be unstable. Blandin~\cite{blandin1978theories} then formulated a replica symmetry broken ansatz, in which replicas where split in two identical groups ($k=1$ RSB with blocks of size $m=2$), and gave a physical interpretation of this breaking in terms of the Edwards-Anderson order parameter. Bray and Moore~\cite{bray1978replica} attempted a slightly different scheme in which replicas where split in two non-identical groups, but found inconsistencies.Finally, Parisi extended Blandin's ansatz to an arbitrary number $m$ of identical groups, and crucially realized than for non-integer $n<1$ one had to choose $n<m<1$ to significantly improve over the RS solution~\cite{parisi1979toward}; he then discovered that a full cascade of RSB steps with $k=\infty$ is needed to obtain the exact solution of some mean-field models~\cite{parisi_infinite_1979}. A complete physical interpretation of RSB in terms of a multiplicity of state was finally achieved a few years later~\cite{mezard1984nature,mezard_spin_1987}.

Assuming that the limits $n\to0$ and $N\to\infty$ commute, the resulting saddle point, $\mQ{*}{}$, encodes the probability of finding an overlap $q$ between two equilibrium configurations in a given state,
\begin{equation}
P(q) = \lim_{n\to0}\frac{2}{n(n-1)}\sum_{(a,b)}\delta(q_{ab}-q) \ ,
\end{equation}
after averaging over all possible states $i$ and samples $\alpha$, i.e., $P(q)\propto\sum_\alpha\sum_i P(q|i|\alpha)$. The ansatz in the structure of $\mQ{}{}$ therefore defines the distribution $P(q)$. In this work, we consider the simplest RS ansatz which corresponds to all non-diagonal elements of $\mQ{}{}$ being equal.
In terms of the free energy landscape $F(\{o_i\})$, this choice corresponds to a convex landscape, either because a single state is present (e.g., a paramagnetic state), or because multiple states are present but the system can be constrained into one of them by an external field (e.g., a ferromagnetic state)~\cite{parisi_theory_2020}.

\subsection{Relating Mass Matrix and Fluctuations}
In the context of the RS ansatz, fluctuations are obtained from the Hessian (or mass matrix) of $F(\mQ{}{})$ around the saddle point~$\mQ{*}{}$,
\begin{equation}\label{eq:Mabcddef}
\mathbbm{M}_{ab;cd} \equiv \partial_{q_{ab}}\partial_{q_{cd}} F(\mQ{}{})|_{\mQ{}{}=\mQ{*}{}} \ .
\end{equation}
Historically, the inverse 
$\mathbbm{G}_{ab;cd} = (\mathbbm{M}^{-1})_{ab;cd}$ 
of the mass matrix was mainly used to assess the stability of the saddle point $\mQ{*}{}$~\cite{almeida_stability_1978,de_dominicis_eigenvalues_1983,de_dominicis_ising_1985-1}, but it actually offers a much richer physical content. It notably encodes overlap fluctuations in the  \textit{small-fluctuation} regime,
\begin{equation}\label{var_Q}
\overline{[\langle q_{ab}q_{cd}\rangle]}-\overline{[\langle q_{ab}\rangle]}\overline{[\langle q_{cd}\rangle]} =\frac{1}{N} \mathbbm{G}_{ab;cd} \ ,
\end{equation}
%Equation~\eqref{var_Q} indeed 
and therefore can be formally related to the different susceptibilities in Eqs.~\eqref{intra}, \eqref{inter} and \eqref{samples}. In other words, the averages over configurations, states and samples are all embedded in the replicated free energy (and hence in the above variances), but need to be disentangled to be interpreted.
    
The authors of Ref.~\cite{franz_field_2011} (see also Refs.~\cite{franz_static_2013,franz_glassy_2013}) made fundamental advances on this problem by considering the behavior of overlap fluctuations around the mode-coupling theory (MCT) transition (where the RS ansatz becomes unstable). %By neglecting activated processes, 
By expressing the distinct intra-state, inter-state and sample-to-sample fluctuations in terms of the mass matrix, they extracted the critical scaling of the various contributions (in absence of instantonic escapes). Their analysis proposed that the initial (planted) configuration dominates fluctuations near the MCT transition temperature, $\TMCT$, as was then confirmed by numerical simulations (of the 3-spin Ising model), and was instrumental in better understanding earlier simulation results (of the random orthogonal model)~\cite{sarlat_predictive_2009-1}. %(The analysis of fluctuations using the replica method was further refined in Refs.~\cite{}.) 
The strength and magnitude of fluctuations outside the critical scaling regime, however, was not considered.
    
We here %explore further along this direction by first 
compute fluctuations far from the MCT transition and show how numerical time averages can be related to them.
In short, while expectation values of standard order parameters (say, magnetization in a ferromagnet) 
	  %(given a hypothetical pdf)
are obtained from one-time averages,
	 %$\mathbb{E}[f(\sigma)]  = \frac{1}{T} \int dt  f(\sigma(t))$
overlaps (between pairs of coupled systems) are obtained from two-time averages.
	 %$q=\sigma\cdot \sigma'$ represents and the associated expectation values are evaluated with two-time temporal averages $\mathbb{E}[f(q)] = \frac{1}{T^2} \int dt dt' f(q(t,t'))$.
In addition, when evaluating expectation values, different fluctuation types correspond to different time scales. Shorter time averages capture intra-state fluctuations, while longer time averages capture inter-state fluctuations. Overlap fluctuations are therefore directly related to two-time averages over specific time scales.

\subsection{Models studied in this work}
In order to validate our analytical findings, three different models are considered. (i) The $p$-spin spherical model is simulated as in Ref.~\cite{folena_gradient_2021} for $p=2$ with an external field, and for $p=3$ without. (ii) The random orthogonal model is simulated as in Refs.~\cite{cherrier_role_2003,sarlat_predictive_2009-1}. (iii) The random Lorentz gas is simulated as in Ref.~\cite{biroli_unifying_2020}, and its consideration builds on results presented in a companion letter~\cite{biroli_local_2021}.
All these models contain quenched disorder, and (except for the 2-spin) are studied in the glass phase, where multiple equilibria are possible. The glass transition that gives rise to this phase is generically of the random first order transition (RFOT) universality class \cite{castellani_spin-glass_2005,LW07,KT15}. In mean-field models, this corresponds to having a number of coexisting minima that grows exponentially with system size, and each minima being locally described by a RS ansatz. In other words, it is possible to use a Franz-Parisi~\cite{franz_recipes_1995} or a Monasson~\cite{monasson_structural_1995} potential (with a RS ansatz) to then select a single state. In all cases, analytical results for intra- and inter-state fluctuations are found to be in good agreement with numerical simulation results.

Additionally, we note that for certain models (finite-size) fluctuations can be evaluated by rigorous, albeit less universal methods. The simplest example is the random energy model, for which a detailed probabilistic 
analysis of fluctuations was presented in Ref.~\cite{mottishaw_finite_2015,derrida2018finite,derrida2021one}. 
Another example is the 2-spin spherical model, for which a mapping to random matrix theory offers the same level of rigor~\cite{baik_spherical_2021}. Reassuringly, our analysis of the latter with the replica method perfectly recapitulates these results.

%\medskip

\section{Equilibrium Fluctuations}\label{EF}

We here wish to quantify fluctuations within and between \emph{glassy} states of a disordered system at equilibrium. However, in order to do so the definition of such states needs to be clarified. From the dynamical point of view, we can write the correlation function between two different configurations visited by the equilibrium dynamics at two different times 
\begin{equation}\label{eq:Cdef}
C(t,t') = q(x(t),x(t')) \ ,
\end{equation}
where $q(x,x')$ is an overlap function. The precise definition of this function is not important as long as it: (i)~provides a measure of similarity of pairs of configurations, i.e., $q(x,x')=1$ if and only if $x=x'$, and (ii) otherwise lies between $0$ and $1$, with $q(x,x')\approx 0$ for pairs of configurations taken uniformly at random in phase space.  As a practical implementation for spin systems of size $N$, we consider the rescaled scalar product between configurations $q(x,x')=N^{-1}\sum_{i=1}^{N} x_i x'_i$.
We can then define a state as the collection of configurations explored by an equilibrium dynamics over a time larger than the microscopic relaxation time, and with a reciprocal overlap larger than a given threshold. 
In particular, we wish to distinguish fluctuations for configurations that belong to the \emph{same} glassy state (intra-state) from those between \emph{different} glassy states (inter-state), and eventually from those between samples. 
We here focus our analysis to temperatures (or analogous parameters) inside the ergodically broken phase (for $T<\TMCT$), where such distinction can be sharply formalized. The equilibrium dynamics is then indeed expected to exhibit two characteristic time scales: $\tau_\text{\tiny rel}$ to relax inside a state, and $\tau_\text{\tiny esc}$ to escape from a state by following an instantonic path. %, that can be energetically and/or entropically driven. 
In long-range models $\tau_\text{\tiny esc}$ is expected to increase with $N$ and $\tau_\text{\tiny rel}$ to remain finite, thus offering a natural separation between the two regimes: we will then here assume $\tau_\text{\tiny rel} \ll \tau_\text{\tiny esc}$. 

In the rest of this section we relate time-averages over the equilibrium dynamics to static averages over the replicated free energy. Hence \textit{small fluctuations} around the typical values of the time-correlations are put in correspondence with overlap fluctuations, allowing to disentangle the different levels of fluctuations embedded in the replicated action.

\subsection{A One-Dimensional Picture of Fluctuations}\label{oneD}

\begin{figure}[t]
    \centering
		\includegraphics[width=0.79\columnwidth]{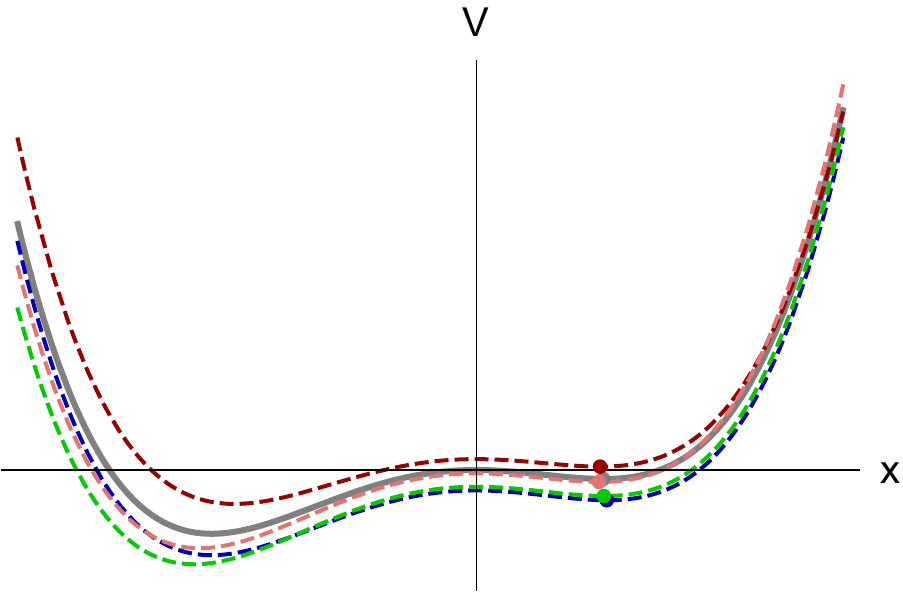}\\
		\caption{\footnotesize Intra-state and sample-to-sample fluctuations. Each color denotes a different potential with a different local minimum. The gray line is the average potential $\overline{V}(x)$. Intra-state fluctuations are inversely proportional to $V''(x)$, while sample-to-sample fluctuations correspond to fluctuations of the local minimum between different potentials.}\label{fig:randV}
	\end{figure}

Before diving into the general analysis of small fluctuations in disordered systems, % at equilibrium, 
we recall a simple picture of how sample-to-sample and intra-state fluctuations naturally arise due to randomness. This schematic was originally established in the context of the random field Ising model (for which the intra-state susceptibility is known as the ``connected susceptibility'') \cite{nattermann1998theory}. 
Let us consider a one-dimensional potential, $V(x)$, whose shape varies around the average $\overline{V}(x)$, and whose random contributions can be tuned to be arbitrarily small (see Fig.~\ref{fig:randV}). 
The variable $x$ here represents either a single degree of freedom, or a collective variable that provides a convenient order parameter for a many-body system.
	
Given a random realization $V(x)$, we can characterize the small thermal fluctuations experienced by a particle around a local minimum in the small temperature limit, $\beta \to \infty$, as
	\begin{equation*}
	    \langle (x - x^*)^2 \rangle \approx \frac{1}{\beta V''(x^*)} \ , \qquad x^*  \text{ s.t. }   V'(x^*)=0 \ .
	\end{equation*}
Average intra-state fluctuations are then 
	\begin{equation*}
	    \chi_{intra}=\overline{\langle (x - x^*)^2 \rangle} \approx \frac{1}{\beta \overline{V}''(\overline{x^*})} \ , \qquad \overline{x^*}  \text{ s.t. }   \overline{V}'(\overline{x^*})=0 \ ,
	\end{equation*}
where the equality holds in the limit of arbitrarily small randomness. 
	
Sample-to-sample fluctuations are then quantified by measuring the variance of the minimum $x^*$ for different realizations of the random potential.
These fluctuations are evaluated by decomposing the potential between the average and the random $\delta V(x)$ parts,
	\begin{equation*}
	    V(x) = \overline{V}(x)+\delta V(x),
	\end{equation*}
and rewriting the local minimum $x^*$ as the minimum of the average potential $\overline{x^*}$ plus a random shift $\delta x$,
    \begin{equation}
    \begin{split}
	    &V'(x^*) = 0 \\ 
	    &\Rightarrow \  \overline{V}'(\overline{x^*}+\delta x)+\delta V'(\overline{x^*}+\delta x) = 0 \\
	    &\Rightarrow \ \overline{V}'(\overline{x^*})+\overline{V}''(\overline{x^*})\delta x+\delta V'(\overline{x^*}) \\ &\qquad +\delta V''(\overline{x^*})\delta x = 0 \ .
	\end{split}
    \end{equation}
    
Excluding the last term, which is of second order in randomness, and noting that the first term is identically zero, we have
	\begin{equation*}
	   \delta x = -\frac{\delta V'(\overline{x^*})}{\overline{V}''(\overline{x^*})} \ .
    \end{equation*}
The variance of the sample-to-sample fluctuation finally reads
	\begin{equation*}
	\begin{aligned}
	    \chi_{dis} &=\overline{(x^* - \overline{x^*})^2} \approx \frac{\text{Var}[\delta V'(\overline{x^*})]}{\overline{V}''(\overline{x^*})^2} \\
	    &= \text{Var}[\delta V'(\overline{x^*})]\beta^2\chi_{intra}^2 \ .
	\end{aligned}
	\end{equation*}
	
This relationship is the characteristic signature of the presence of random-field--like disorder that tilts the potential close to the minimum. By applying this idea to the Franz-Parisi potential~\cite{franz_field_2011} one can further intuitively justify 
why this relationship should hold around a MCT transition as well. Sample-to-sample fluctuations of the minimum of a random potential near a spinodal point then diverge as the square of the average fluctuation around the minimum of each sample,
	\begin{equation}
	\chi_{dis} \propto \chi_{intra}^2\qquad \text{i.e.} \qquad \overline{\delta \langle  x \rangle \delta \langle  x \rangle}  \propto \overline{\langle  \delta x \delta x \rangle}^2.
	\end{equation}
	
A particular case of interest is when the randomness consists in just a linear tilt of the potential, i.e., $\delta V(x) = -\epsilon x$ with $\epsilon$ a random variable of zero mean that characterizes sample-to-sample fluctuations.
In that case, the minimum is the solution of
\begin{equation}
\overline{V}'(x^*_\epsilon) = \epsilon \ ,
\end{equation}
and because $V''(x) = \overline{V}''(x)$ intra-state fluctuations are given by
\begin{equation*}
	    \langle (x - x^*_\epsilon)^2 \rangle \approx \frac{1}{\beta \overline{V}''(x^*_\epsilon)} = \chi_{intra}(\overline{x^*}=x^*_\epsilon) \ .
	\end{equation*}
Therefore, atypical samples present fluctuations with the same variance of typical fluctuations at different external parameters $\epsilon$. In other words, a scatter plot over samples (i.e., over realizations of $\epsilon$) of the intra-state fluctuations versus the average $x^*_\epsilon$ collapses all points on a single line, which corresponds to the relation between the average intra-state fluctuations and the average order parameter. We report a concrete example of this phenomenon in Sec.~\ref{2spin} where we analyze the 2-spin spherical model with an external field. 

\subsection{Estimating Equilibrium Fluctuations through Time Kernels}\label{sec:timekernels}

In order to analyze equilibrium fluctuations we introduce two kernels for time averages, denoted as
``Franz-Parisi'' (FP) and ``Monasson'' (M) kernels, for reasons that will be explained in Sec.~\ref{SMM}.

\subsubsection{Franz-Parisi kernel}

The FP kernel is
a time-translationally invariant (TTI) kernel, ${\kappa_\FP(t-t')}$, which must select two configurations if they are far enough (in time) to be independent within a state ($t>\tau_\text{\tiny rel}$), but close enough to remain within the same state ($t<\tau_\text{\tiny esc}$). It must also be normalized such that $\int_{0}^{\infty}\kappa_\FP(s)ds=1$. Given two timescales, $\tau_{\kappa}$ and $\tau_\text{\tiny K}$, with $\tau_\text{\tiny rel} < \tau_{\kappa} \ll\tau_\text{\tiny K} < \tau_\text{\tiny esc}$, one such operator could be a flat function between time $\tau_{\kappa}$ and $\tau_\text{\tiny K}$ and zero outside,
%\begin{equation}
%    \kappa_{\theta}(s) \equiv  \frac{1}{\tau_\text{\tiny K}-\tau_{\kappa}} \theta(s- \tau_{\kappa}) \theta( \tau_\text{\tiny K}-s) \ .
%\end{equation}
but for practical reasons in numerical simulation we consider instead
\begin{equation}\label{Kd}
\kappa_\FP(s) \equiv \frac{1}{\tau_\text{\tiny K}/\tau_{\kappa}}\sum_{j=1}^{\tau_\text{\tiny K}/\tau_{\kappa}}\delta(s-j \tau_{\kappa}) \ ,
\end{equation}
which selects points that are equidistant in time.
Because $\tau_\text{\tiny esc}$ is an average quantity, in simulations the scale $\tau_\text{\tiny K}$ must be chosen adaptively so as to match the lifetime of a given state.
For example, $\tau_\text{\tiny K}$ can be fixed by the condition
\begin{equation}
\tau_\text{\tiny K} = \max_{t} [ q(x(t),x(0)) > q_\text{\tiny esc} ] \ ,
\end{equation}
which corresponds to taking the largest time such that the overlap is greater than a fixed threshold, $q_\text{\tiny esc}$, taking as the origin of time a point that belongs to a given state.
As is customary in statistical physics, we wish to eventually replace time averaging by averaging over static measures.
For a given two-time observable, e.g., $f(t,t')=q(x(t),x(t'))$, we thus define the kernel average over time, given a reference configuration at time~$t$,
\begin{equation}
\mathcal{T}_{\FP}[f](t)\equiv \int_{-\infty}^{+\infty} ds  \kappa_\FP(s) f (t+s,t).
\end{equation}

\subsubsection{Monasson kernel}

The Monasson kernel is instead a two-time kernel, $\kappa_\M(s,s')$, which homogeneously selects pairs of configurations that are inside a time window of length $\tau_K$ around a central configuration. Here again, many definitions are possible. A useful one is to take a two-time grid without the diagonal
\begin{equation}\label{Kd2}
\kappa_\M(s,s') \equiv \frac{\Sum{k=0}{K}\Sum{k'=0}{K}\delta(s-k \tau_{\kappa})\delta(s'-k' \tau_{\kappa})(1-\delta_{kk'})}{K(K+1)} \ ,
\end{equation}
where $K = \tau_\text{\tiny K}/\tau_{\kappa}$. The resulting kernel average around a reference configuration at time~$t$ reads
\begin{equation}
\mathcal{T}_{\M}[f](t)\equiv \int_{-\infty}^{+\infty} \int_{-\infty}^{+\infty} ds' ds  \kappa_\M(s,s') f (t+s,t+s') \ .
\end{equation}
Note that by TTI,
both these averages are independent of the specific value of the reference time. The $t$ dependence is hence dropped going forward. Note also that in the limit $K\to \infty$ the average becomes independent of the specific kernel definition
and corresponds to a flat measure up to time $\tau_{\text{\tiny K}}$.

\subsection{Time-Local Averages}

The kernels introduced in Sec.~\ref{sec:timekernels} enable the construction of a variety of time-local averages, as we now describe.

First, we introduce the time-local Franz-Parisi overlap
\begin{equation}
\begin{split}
q_\FP^{(0)} & \equiv \lim_{\tau_\text{\tiny K}/\tau_{\kappa}\to\infty}  \mathcal{T}_{\FP}[C](0) 
\\
&= \lim_{K\to\infty}\frac{1}{K}\sum_{k=1}^{K} q(x_0,x_k) \approx \langle q_{01}\rangle_{\FP} \ , 
\end{split}
\end{equation}
where $C$ is the correlation defined in Eq.~\eqref{eq:Cdef}, $x_0$ denotes the reference configuration (corresponding to the $(0)$ superscript),
and $x_k = x(k\tau_{\kappa})$ are the other configurations selected by the FP kernel.
Angle brackets $\langle q_{01} \rangle_{\FP}$ here indicate that the time average has been formally substituted by a probability measure related to an overlap action $F_\FP(\mQ{}{})$, which will be explicitly defined for mean-field models in Sec.~\ref{SMM}. The term $q_{01}$ denotes the overlap between the reference configuration $x_0$ and another typical equilibrium configuration that belongs to the same state.

A non-equivalent way to define the local overlap is the Monasson one
\begin{equation}\label{eq:qMdef}
\begin{aligned}
&q_{\M} \equiv \lim_{\tau_\text{\tiny K}/\tau_{\kappa}\to\infty}  \mathcal{T}_{\M}[C] \\
&= \lim_{K\to\infty}\frac{1}{(K+1)K}\sum_{k'=0}^{K}\sum_{\substack{k=0\\k\neq k'}}^{K} q(x_{k'},x_k) \approx \langle q_{12}\rangle_{\M} \ .
\end{aligned}
\end{equation}
where we have again substituted the time average with an average over a probability measure $\langle\bullet \rangle_{\M}$ related to an overlap action $F_{\M}(\mQ{}{})$. 
While the averages of $q_\M$ and $q_{\FP}$ are expected to be the same in a stable enough state (for $K\to\infty$) and in the thermodynamic limit, their fluctuations generally differ in finite-size systems.

Assuming that the equilibrium dynamics is confined to within a well-defined state, averaging the one-time kernel $\kappa_\FP$ over different reference configurations is equivalent to considering the two-time kernel $\kappa_\M$. Therefore we can define the Edwards-Anderson overlap of the state as
 \begin{equation}\label{eq_q}
 q_\text{\tiny EA} \equiv 
 \lim_{K\to\infty}\frac{1}{K}\sum_{k=0}^{K-1} q_\FP^{(k)} = q_{\M}.
 \end{equation}
In the following, in order to lighten the notation we will implicitly assume the $K\to\infty$ limit.
Two different kinds of time-local \emph{intra-state} susceptibilities for the overlap can then be introduced.

\subsubsection{Intra-state Franz-Parisi Susceptibility}

The intra-state Franz-Parisi susceptibility, which is computed with respect to a reference configuration at $t=0$, is
	\begin{equation}\label{chith}
	\begin{split}
	\chi_{intra}^{\FP,(0)} &\equiv \mathcal{T}_{\FP}[C^2]-\mathcal{T}_{\FP}[C]^2 = \mathcal{T}_{\FP}[(C-q^{(0)}_\FP)^2] \\
	&=  \frac{1}{K}\sum_{k=1}^{K} q(x_0,x_k)^2 -  \left[\frac{1}{K}\sum_{k=1}^{K} q(x_0,x_k) \right]^2 \\
	&\approx \langle q_{01}^2\rangle_{\FP} - \langle q_{01}\rangle_{\FP}^2 \ ,
	\end{split}\end{equation}
where again, in the $K\to \infty$ limit, the time average over configurations has been substituted with the appropriate measure.
The subscript $0$ refers to the reference configuration and $1$ is a typical equilibrium configuration distinct from $0$ (i.e. distant in time by at least $\tau_{\kappa}$).
Analogously to Eq.~\eqref{eq_q}, if the state is \emph{well defined}, we can take an arbitrary reference configuration $k$ and redefine the intra-state FP susceptibility as
\begin{equation}\label{eq_chis}
 \begin{aligned}
&  \chi_{intra}^\FP \equiv \frac{1}{K}\sum_{k=0}^{K-1} \chi_{intra}^{\FP,(k)}\\ 
	&= \frac{1}{K^2}\Sum{k=0}{K-1}\Sum{k'\neq k}{K} q(x_{k},x_{k'})^2 - \frac{1}{K^3}\Sum{k=0}{K-1}\left[\Sum{k'\neq k}{K} q(x_{k},x_{k'})\right]^2  \\
	&= \Mean{k\neq k'}[q_{kk'}^2] - \Mean{k\neq k'\neq k''}[q_{kk'}q_{k k''}] + O(K^{-1}) \ .
	\end{aligned}
 \end{equation}
As in Eq.~\eqref{eq:qMdef}, we can assume that there exists a probability measure associated to the two-time average,
such that
\begin{equation}\label{thk2}
 	\chi_{intra}^\FP \approx \langle q_{12}^2\rangle_{\M} - \langle q_{12}q_{13}\rangle_{\M} \ .
\end{equation}
	
\subsubsection{Intra-state Monasson Susceptibility} 

The intra-state two-time susceptibility describes total fluctuations inside a state,
	\begin{equation}\label{DYN}
	\chi_{intra}^{\M} \equiv  \mathcal{T}_{\M}[C^2]-\mathcal{T}_{\M}[C]^2 =\mathcal{T}_{\M}[(C-q_\text{\tiny M})^2] \ .
	\end{equation}
We can again consider the specific kernel 
in Eq.~\eqref{Kd2}, which gives
	\begin{equation}
	\begin{aligned}
	&\chi_{intra}^{\M} = \frac{\Sum{k'=0}{K}\Sum{k\neq k'}{K} q(x_{k'},x_k)^2}{(K+1)K} - \left[\frac{\Sum{k'=0}{K}\Sum{k\neq k'}{K} q(x_{k'},x_k)}{(K+1)K} \right]^2\\
	=&\quad \Mean{k\neq k'}[q_{kk'}^2] - \Mean{k\neq k'\neq k''\neq k'''}[q_{kk'}q_{k''k'''}] 
	+O(K^{-1}) \ .
	\end{aligned}
	\end{equation}
As for $\chi_{intra}^\FP$, the correction in powers of $K^{-1}$  depends on the specific choice of kernel. % $\kappa$.
Again, we can rewrite the fluctuations as an average over a probability measure,%Although the limit $K\to\infty$ is independent of that choice, it still depends on the chosen state,
	\begin{equation}\label{chiintraM}
	\chi_{intra}^{\M} \approx \langle q_{12}^2\rangle_{\M} - \langle q_{12}q_{34}\rangle_{\M} \ ,
	\end{equation}
where, as before, the four subscripts $1,2,3,4$ refer to four typical yet distinct equilibrium configurations.
We thus have $\langle q_{12}\rangle_{\M}  \equiv \langle q_{23}\rangle_{\M}$.
	
Therefore, $\mathcal{T}_{\FP}$ and $\mathcal{T}_{\M}$ define two classes of local-time fluctuations around the reference configuration $x_0$, which implement two different ways of evaluating the variance of intra-state fluctuations. Roughly speaking,
	\begin{equation}
	\begin{aligned}
	\chi_{intra}^\FP & \approx \text{Mean}_{s'}[\text{Var}_{s}[C(s,s')]] \ , \\
	\chi_{intra}^{\M} &\approx \text{Var}_{s,s'}[C(s,s')] \ ,
	\end{aligned}
	\end{equation}
where the times $s,s'$ are taken such that the dynamics does not permit escaping the state. % defined by the reference $x_0$.
We then obtain the following inequality
 	\begin{equation}
 	\chi_{intra}^\FP < \chi_{intra}^{\M} \ .
 	\end{equation}
 	
 	%Moreover we observe that $\chi^{\kappa}_{intra}^\text{\tiny FP}$ is well defined for every equilibrium dynamics independently on the nature of the "caging", since it consider only one time index. However $\chi^{\kappa}_{intra}^\text{\tiny M}$ needs to 
	
	    \begin{figure}[t]    
		\centering
		\includegraphics[width=0.89\columnwidth]{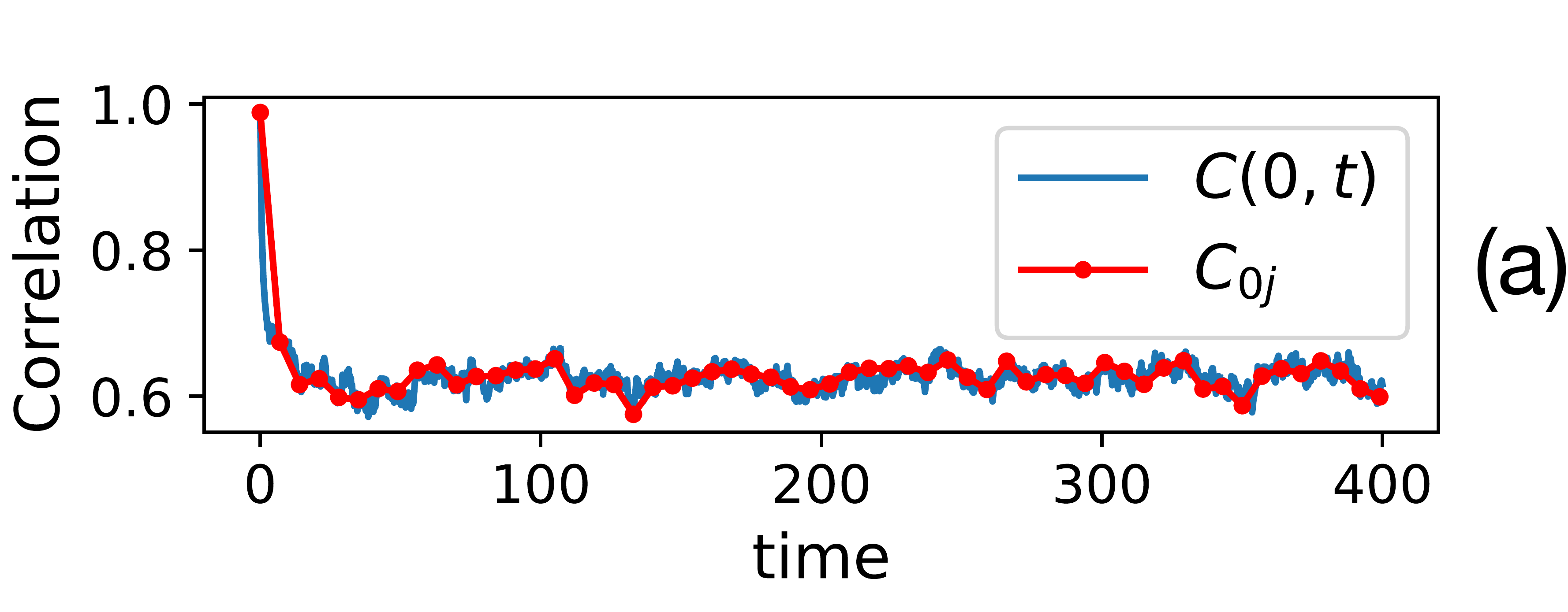}
		\includegraphics[width=0.89\columnwidth]{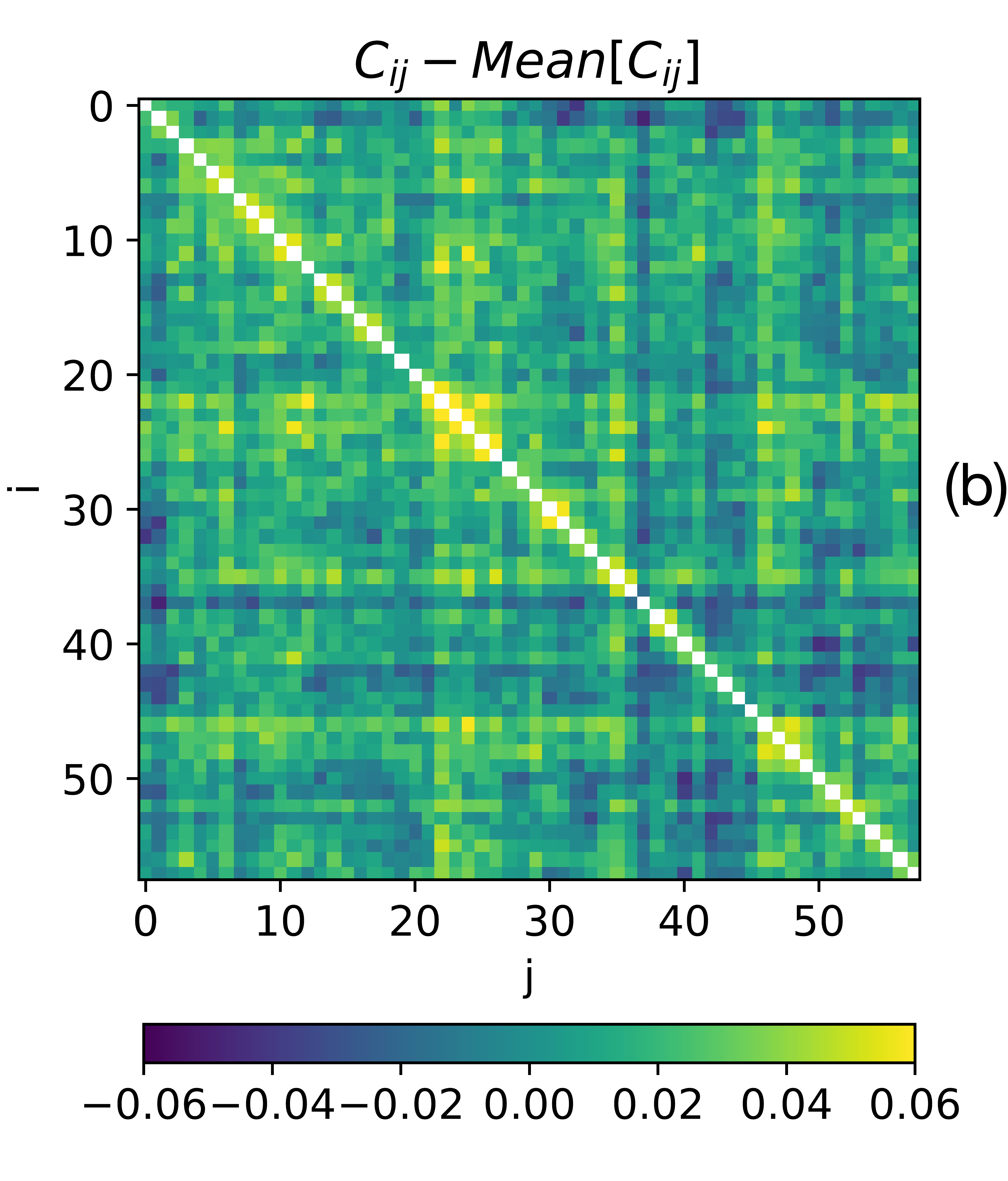}
		\caption{Intra-state fluctuations in the 3-spin spherical model of size $N=3200$ at $T=0.59<\TMCT$ (with $\tau_{\kappa} = 7$).
			 \textbf{(a)}: correlation to a given reference configuration inside the state. The kernel $\kappa_\FP(s)$ selects one configuration every $\tau_{\kappa}$ (red dots).
			 \textbf{(b)}: correlation matrix $C_{ij}$ for configurations $x_i,x_j$ inside a state. Here, we have
			$q_\text{\tiny EA} = 0.656, \hat{\chi}_{intra}^\FP = 0.5056, \hat{\chi}_{intra}^{\M} = 0.688$. See Sec.~\ref{pspinsimul} for numerical details.}
		\label{fig:SingleTraj}
	\end{figure}
	
\subsection{Numerical Implementations of Kernels %$\mathcal{T}_{\kappa}$ and $\mathcal{T}_{\kappa^2}$}
}\label{numSusc}
 Given these kernel definitions, we now provide a numerical implementation in order to clarify and validate their analysis. We select a reference configuration at equilibrium at $t=0$ (sample) and consider the correlations $C_{ij} = q(x(i \tau_{\kappa}),x(j\tau_{\kappa} ))$, with $i,j\in [0,K]$. 
From this correlation we can estimate the overlap of the cage,
	\begin{equation*}
	q_\text{\tiny EA} =\langle q_{12}\rangle_{\M} \approx \text{Mean}_{i\neq j}[C_{ij}] \ ,
	\end{equation*}
as well as two susceptibilities,
	\begin{equation*}
		\begin{aligned}
		\chi_{intra}^\FP  &=\langle q_{12}^2\rangle_{\M} - \langle q_{12}q_{13}\rangle_{\M} \approx \text{Mean}_{i}[\text{Var}_{j}[C_{ij}]] \ , \\
		\chi_{intra}^{\M}  &= \langle q_{12}^2\rangle_{\M} - \langle q_{12}q_{34}\rangle_{\M} \approx \text{Var}_{i\neq j}[C_{ij}]\ .
		\end{aligned}
	\end{equation*}
	Fluctuations around $q_\text{\tiny EA}$ scale as $1/N$ in the thermodynamic limit, i.e. the susceptibilities $\chi$ scale as $N^{-1}$. We will thus also use rescaled quantities, such that a finite value is obtained in the thermodynamic limit, $\hat{\chi} = N \chi$.
	For the sake of illustration, Fig.~\ref{fig:SingleTraj} shows the two-time correlation matrix $C_{ij}$ for a fully-connected model of size $N=3200$. (See Sec.~\ref{pspinsimul} for numerical details.)

	\subsection{Average over Samples (States and Disorder)}\label{secAv}
Up to this point, we have considered time-local susceptibilities $\chi_{intra}^\FP,\chi_{intra}^{\M}$, inside one glassy state, i.e., close enough to a reference configuration $x_0$. This same $x_0$ can be thought as being drawn from an equilibrium distribution. At very long times, $t\gg \tau_\text{\tiny esc}$, we expect the system to explore different states and therefore to present local overlap and susceptibilities that fluctuate depending on $x_0$. To average over different states, or equivalently over different reference configurations,
we define the kernel $\mathcal{S}(t)$ that selects independent reference configurations homogeneously in time. Two configurations are deemed independent if their mutual overlap is approximately zero, i.e., $q(x(t),x(s))\approx 0$. (In RFOT systems, it is assumed that two configurations have vanishing mutual overlap %$\approx 0$ 
for $t\gg\tau_\text{\tiny esc}$.) One possible implementation is
	\begin{equation}\label{Sd}
	\mathcal{S}(t) \equiv \frac{1}{S}\sum_{j=1}^{S}\delta(t-j \tau_{\mathcal{S}}) \ ,
	\end{equation}
	such that $\tau_\mathcal{S} \gg \tau_\text{\tiny esc}$, where $S$ is the total number of averaged states. %so to select independent configurations 
 In small systems, however, the time to escape a state and reach another one fluctuates broadly. A more convenient choice is then a variable time step, such as
\begin{equation}
\tau_{j+1} = \min_{\tau} q[x(\tau_{j}+\tau),x(\tau_{j})]<0 \ .
\end{equation}
\begin{figure}
    \includegraphics[width=1.1\columnwidth]{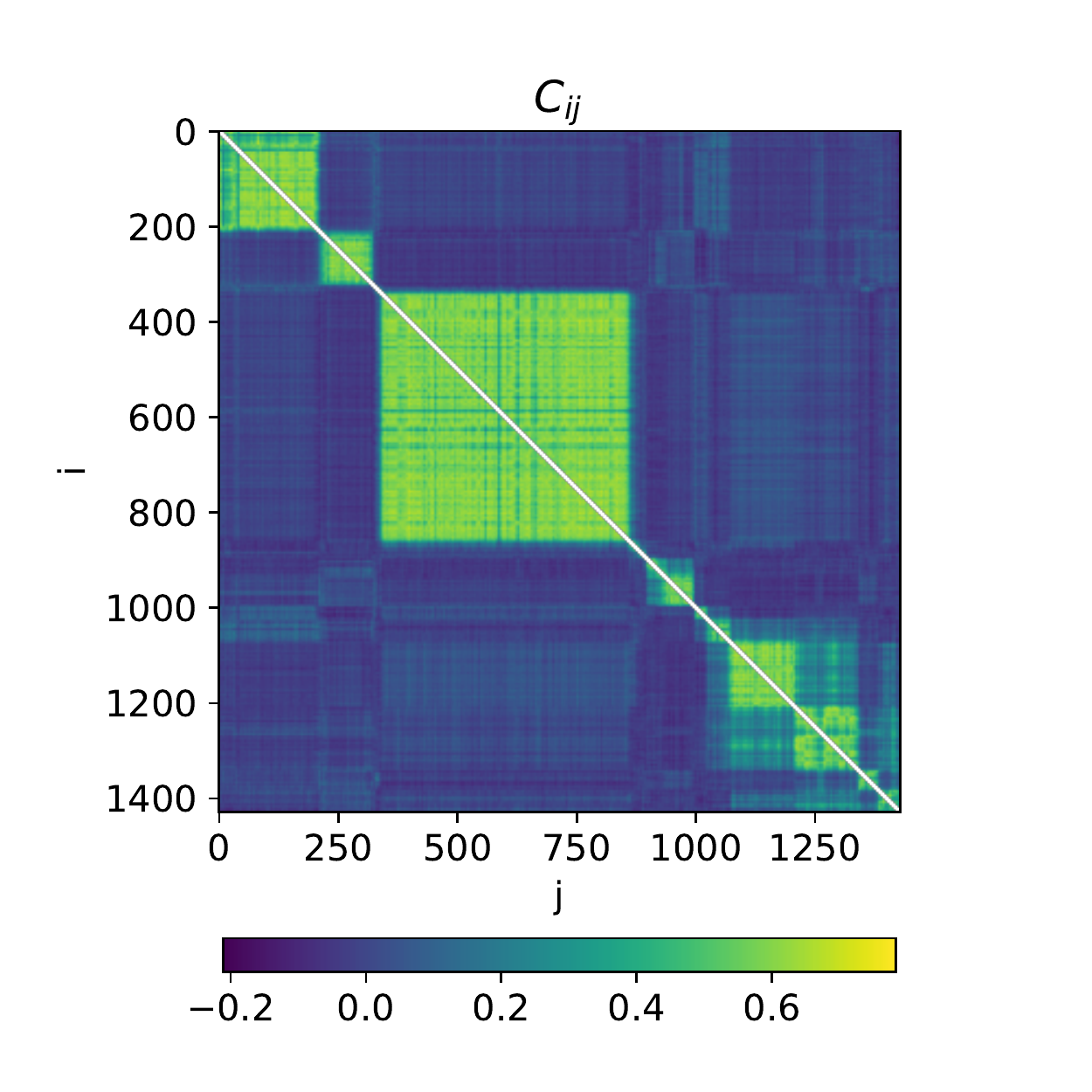}
    \caption{\footnotesize Inter-state fluctuations in the 3-spin spherical model of size $N=300$ at $T=0.59<\TMCT$ ($\tau_{\kappa} = 7$).
	Instantonic paths between different states take typical times larger than $\tau_\text{\tiny esc}$.}
    \label{fig:Inst300}
\end{figure}
Given the kernel $\mathcal{S}(t)$, we define an operator that averages over different states
	\begin{equation}
		\mathcal{T}_{\mathcal{S}}[f](t)\equiv \int_{-\infty}^{+\infty} ds \mathcal{S}(s) f (t+s) \ .
	\end{equation}
In long-range models the escape time $\tau_\text{\tiny esc}$ grows exponentially with $N$. Sufficiently long simulations thus rapidly become computationally prohibitive. It is common practice to instead directly consider different samples;
recall that in our notation, a sample is an equilibrium configuration inside a glass state, for a given quenched disorder (if present).
For certain models, a \textit{planting} procedure \cite{krzakala_hiding_2009}--as is used here--is possible, and analogous preparation protocols can be used in others. For instance, the \textit{swap algorithm} very efficiently prepares equilibrium configurations of polydisperse systems of spheres~\cite{ninarello_models_2017}.
Note that states and samples coincides in systems with self-induced disorder, i.e. without quenched disorder. For systems with quenched disorder the two averages play different roles, but averaging over the latter implies averaging over the former, i.e., $\overline{[\bullet]} = \overline{\bullet}$. 

For numerical convenience, we thus here average over samples directly, and define sample-averaged quantities using bold fonts:
	\begin{equation}
	\begin{aligned}
	\bm{q}_\text{\tiny EA} &\equiv \overline{q_\text{\tiny EA}} =\lim_{S\to\infty} \frac{1}{S} \sum_{s=1}^S q^{(s)}_\text{\tiny EA} = \lim_{\substack{S\to\infty\\K\to\infty}} \mathcal{T}_{\mathcal{S}}\big [ \mathcal{T}_{\FP}[C]\big ] \ ,
	%\\
	 %&=  \overline{q^{(0)}_{intra}^\text{\tiny M}} =\lim_{S\to\infty} \frac{1}{S} \sum_{s=1}^S q^{s}_{intra}^\text{\tiny M} = \lim_{\substack{S\to\infty\\K\to\infty}} \mathcal{T}^{+}_{\mathcal{S}}\big [ \mathcal{T}^{+}_{\kappa^2}[C]\big ] 
	 \end{aligned}
	\end{equation}
	 \begin{equation}
	 \begin{aligned}
	\bm{\chi}_{intra}^\FP &\equiv \overline{\chi_{intra}^{\FP} }= \lim_{S\to\infty} \frac{1}{S} \sum_{s=1}^S \chi^{(s)}_\FP \\
	&= \lim_{\substack{S\to\infty\\K\to\infty}} \mathcal{T}_{\mathcal{S}}\big [\mathcal{T}_{\FP}[C^2]-\mathcal{T}_{\FP}[C]^2 \big ]  \ ,
	 \end{aligned}
	\end{equation}
	\begin{equation}
	\begin{aligned}
	\bm{\chi}_{intra}^{\M} &\equiv \overline{ \chi_{intra}^{\M} }= \lim_{S\to\infty} \frac{1}{S} \sum_{s=1}^S \chi_{intra}^{\M,(s)} \\
	&= \lim_{\substack{S\to\infty\\K\to\infty}} \mathcal{T}_{\mathcal{S}}\big [\mathcal{T}_{\M}[C^2]-\mathcal{T}_{\M}[C]^2 \big ] \ .
	\end{aligned}
	\end{equation}

	\subsection{Sample-to-sample Fluctuations}\label{sampleSusc}

In order to quantify fluctuations of the local overlap $q^\FP_{intra}$ between different samples (different states and different disorders), we consider the sample-to-sample susceptibility, given by the sum of inter-state and disorder susceptibilities,  
\begin{equation}
\bm{\chi}_{sample}=\bm{\chi}_{inter}+\bm{\chi}_{dis} \ .
\end{equation}

\subsubsection{Franz-Parisi scheme}

The sample-to-sample Franz-Parisi susceptibility
	\begin{equation}
	\label{hetSusc}
	\begin{aligned}
	\bm{\chi}_{sample}^\FP &\equiv \overline{(q^{(s)}_\FP-\overline{q^{(s)}_\FP})^2} \\
	&= \lim_{S\to\infty} \frac{1}{S} \sum_{s=1}^S {q^{(s)}_\FP}^2- \left[ \frac{1}{S} \sum_{s=1}^S q^{(s)}_\FP \right]^2\\
	&= \lim_{\substack{S\to\infty\\K\to\infty}} \mathcal{T}_{\mathcal{S}}\big [ \mathcal{T}_{\FP}[C]^2\big ]-\mathcal{T}_{\mathcal{S}}\big [ \mathcal{T}_{\FP}[C]\big ]^2,
	\end{aligned}
	\end{equation}
 can be rewritten in terms of the specific kernels $\kappa_{\FP}$ and $\mathcal{S}$, defined in Eqs.~\eqref{Kd} and \eqref{Sd}, respectively. 
 Using two encapsulated indexes, i.e. $s$ to denote a state (reference configuration) and $k$ to denote an equilibrium configuration in the state that includes $s$, we can expand the sums to get
	\begin{equation}\label{chihet}
	\begin{aligned}
	&\bm{\chi}^\FP_{sample}= \\
	&=  \frac{1}{S} \sum_{s=1}^S\big ( \frac{1}{K}\sum_{k=1}^{K} q_{s,s+k} \big )^2 -  \left[ \frac{1}{S} \sum_{s=1}^S\big ( \frac{1}{K}\sum_{k=1}^{K} q_{s,s+k}  \big ) \right]^2\\
	%&= \frac{1}{S} \sum_{s}^S\big ( \frac{1}{K^2}\sum_{k_s\neq {k'}_s}^{K^2}  q_{sk_s}q_{sk_s'} +\frac{1}{K^2}\sum_{k_s=1}^{K}  q_{sk}^2 \big ) \\
	%&\quad -  \frac{1}{S^2} \sum_{s}^S\big ( \frac{1}{K}\sum_{k_s=1}^{K} q_{sk_s} \big )^2 - \frac{1}{S^2}\frac{1}{K^2} \sum_{s\neq s'}^{S^2}\sum_{k_s}^{K}\sum_{k_s'}^{K}q_{sk_s}q_{s'k_s'}  \\
	&= \frac{S-1}{S} \Big ( \Mean{s,k\neq k'}[q_{sk}q_{sk'}]-\Mean{s\neq s',k\neq k'}[q_{sk}q_{s'k'}]\Big ) \\
	&\approx \overline { \langle q_{01}\rangle^2_{\FP}} -  \overline { \langle q_{01}\rangle_{\FP}}^2 \ ,
	\end{aligned}
	\end{equation}
	where the last line holds in the limit $K\to\infty$ and $S\to \infty$.
Unlike the previous susceptibilities, this one is \emph{time global} and its value depends on two encapsulated averages, over samples and equilibrium configurations.
Like for the other susceptibilities, we expect this limit to be independent of the specific definition of the kernels $\kappa_\FP$ and $\mathcal{S}$.
Following the same reasoning that led to Eq.~\eqref{thk2}, we can also express the FP sample-to-sample susceptibility
in terms of the kernel $\kappa_\M$, as follows:
\begin{equation}\label{het2t}
\bm{\chi}_{sample}^\FP \approx \overline { \langle q_{12}q_{13}\rangle_{\M}} -  \overline { \langle q_{12}\rangle_{\M}}^2
\end{equation}
	
Finally, we can define a total susceptibility that captures the fluctuations of the overlap over all possible sources of randomness. It can thus be expressed as the sum of the intra-state and sample-to-sample susceptibilities,
	\begin{equation}\begin{split}
	\bm{\chi}_{tot} &\equiv \lim_{\substack{S\to\infty\\K\to\infty}} \mathcal{T}_{\mathcal{S}}\big [ \mathcal{T}_{\FP}[C^2]\big ]-\mathcal{T}_{\mathcal{S}}\big [ \mathcal{T}_{\FP}[C]\big ]^2 \\
	&= \bm{\chi}_{intra}^\FP +\bm{\chi}_{sample}^\FP \\
	&\approx \overline { \langle q_{01}^2\rangle_{\FP}}  -  \overline { \langle q_{01}\rangle_{\FP}}^2 =  
	   \overline { \langle q_{12}^2\rangle_{\M}} - \overline { \langle q_{12}\rangle_{\M}}^2 \ .
	\end{split}\end{equation}
The last line gives two alternative decompositions of the total susceptibility, either in terms of the Franz-Parisi measure, Eqs.~\eqref{chith} (averaged over samples) 
and \eqref{chihet}, or 
in terms of the Monasson measure, Eqs.~\eqref{thk2} (again, averaged over samples) and \eqref{het2t}.
	
\subsubsection{Monasson scheme}
	
As before, we can also introduce a second way of quantifying sample-to-sample fluctuations by means of the Monasson two-time susceptibility,
	\begin{equation}
	\begin{aligned}
	\bm{\chi}_{sample}^{\M} &\equiv \overline{(q_{\M}-\overline{q_{\M}})^2} \\
	&= \lim_{S\to\infty} \frac{1}{S} \sum_{s=1}^S {q^{s}_\M}^2-(\frac{1}{S} \sum_{s=1}^S q^{s}_\M)^2 \\
	&=  \lim_{\substack{S\to\infty\\K\to\infty}} \mathcal{T}_{\mathcal{S}}\big [ \mathcal{T}_{\M}[C]^2\big ]-\mathcal{T}_{\mathcal{S}}\big [ \mathcal{T}_{\M}[C]\big ]^2 \ .
	\end{aligned}
	\end{equation}
Again we can write this average in terms of the static measures as
\begin{equation}\label{eq:chiMsample}
\bm{\chi}_{sample}^{\M} \approx \overline { \langle q_{12}q_{34} \rangle_{\M}} -  \overline { \langle q_{12}\rangle_{\M}}^2 \ ,
\end{equation}
which leads, together with Eq.~\eqref{chiintraM}, to the following decomposition of the total fluctuations:
	\begin{equation}\begin{split}
	\bm{\chi}_{tot} &\equiv \lim_{\substack{S\to\infty\\K\to\infty}} \mathcal{T}_{\mathcal{S}}\big [ \mathcal{T}_{\M}[C^2]\big ]-\mathcal{T}_{\mathcal{S}}\big [ \mathcal{T}_{\M}[C]\big ]^2 \\
	&
	= \bm{\chi}_{intra}^{\M} +\bm{\chi}_{sample}^{\M} \approx   
	   \overline { \langle q_{12}\rangle^2_{\M}} - \overline { \langle q_{12}\rangle_{\M}}^2 \ .
	\end{split}\end{equation}
In summary, we have seen that the total overlap fluctuations can be decomposed into intra-state and sample-to-sample fluctuations
(keeping in mind that here {\it sample} indicates a single glass state of a given quenched disorder, if present) in many different ways,
depending on the chosen definition of overlap and of the choice of time averages. 
Because these different decompositions have previously been used confusingly, we provide in Appendix~\ref{appendixH} a summary of the different susceptibilities used in the literature, and how they map on our notation.

\subsection{From Time Averages to Probability Measures}\label{fromAvtoProb}
 
We now discuss how different time-averaging schemes can
be recast into different probability measures.
The averages over $S$ and over $K$ play different roles. The first corresponds (for $S\to\infty$) to the standard time average
 	\begin{equation}\begin{split}
 	&\frac{1}{S} \sum_{s=1}^S f(x_s)\approx \frac{1}{T}\int_0^{T}ds  f(x(s)) \\ &\approx  Z^{-1}\int\mathcal{D}x e^{-\beta H(x)}  f(x) \ .
 	\end{split}
 	\end{equation}
If the system is ergodic, it thus coincides with the static average given by the Gibbs measure.
The second, by contrast,  corresponds to a local time average within a single glass sample. For the one-time Franz-Parisi kernel $\kappa_\FP$, the equilibrium probability measure of finding a configuration at a given overlap $p=q_{0 k}$ with a specific $x_0$ is
 	 \begin{equation}\label{FPmeasure}
 	 \begin{aligned}
 	 \frac{1}{K}&\sum_{k=1}^K f(q_{0k})\approx \frac{1}{\tau_\text{\tiny K}}\int_0^{\tau_\text{\tiny K}}dt f[q(x(0),x(t)] \\
 	 &\approx  Z^{-1}\int dp e^{-N\beta V_{\text{\tiny FP}}(p)} f(p) =  \langle f\rangle_{\text{\tiny FP}} \ ,
 	 \end{aligned}
 	 \end{equation}
where $V_{\text{\tiny FP}}(p)$ is the Franz-Parisi (free energy) potential~\cite{franz_recipes_1995}, which explains our notational choice. 
This quantity generally depends on the reference configuration $x_0$ and the given state. The time average given by the two-time Monasson kernel $\kappa_\M$, by contrast, corresponds to a free energy that measures the probability of observing the overlap $q$ between any pairs of equilibrium configurations in a given state, i.e.,
  	\begin{equation}\label{Mmeasure}
  	\begin{aligned}
  	\frac{\sum_{k\neq k'}^{1,K} f(q_{k k'})}{K(K-1)}&\approx \frac{1}{\tau_\text{\tiny K}^2}\int_0^{\tau_\text{\tiny K}}dt\int_0^{\tau_\text{\tiny K}}dt' f[q(x(t),x(t')] \\
 	 &\approx  Z^{-1}\int dq e^{-N\beta V_\text{\tiny  M}(q)} f(q) = \langle f\rangle_\text{\tiny  M} \ ,
  	\end{aligned}
  	\end{equation}
where $V_\text{\tiny  M}(q)$ is the Monasson (free energy) potential~\cite{monasson_structural_1995}. 
(Section~\ref{SMM} describes how these two potentials can be evaluated using the replica method.)
We therefore have a direct correspondence between different time averages and different potentials
\begin{equation}
\mathcal{T}_{\FP}[\bullet] \approx \langle \bullet \rangle_\text{\tiny FP} \ , \
\mathcal{T}_{\M}[\bullet] \approx \langle \bullet \rangle_\text{\tiny  M} \ .
\end{equation}

\section{Susceptibilities in Mean-field Models}\label{SMM}
In this section we relate the mass matrix to equilibrium fluctuations extracted from dynamical correlations. As mentioned in the introduction, the replicated free energy $F(\mathbbm{Q})$ encodes three different averages: over configurations (thermal), over states and over the quenched disorder. In order to disentangle them, the physical meaning of the overlap fluctuations in Eq.~\eqref{var_Q} must be teased out. To this end, we mainly follow the approach of Refs.~\cite{franz_field_2011,franz_static_2013}.
We here restrict our analysis to models and parameter regimes that present a simple RS structure of the saddle point $\mathbbm{Q}^*$. Such a mathematical structure can describe two quite distinct physical situations.

In the first case (an example is given in Sec.~\ref{2spin}),
the free energy landscape $F(\{o_i\})$ has a single minimum and is locally convex around it. 
As a result, any two equilibrium configurations present the same typical overlap, defined by $\mathbbm{Q}^*_\text{\tiny RS}$. 
Because there is a single state for each quenched disorder,
inter-state fluctuations are absent and only intra-state and sample-to-sample fluctuations persist.  Our goal is then to relate, in the small-fluctuation regime, the susceptibilities in Eqs.~\eqref{intra} and \eqref{samples} to the RS mass matrix,
$\mathbbm{M}^\text{\tiny RS}_{ab;cd}$, defined in Eq.~\eqref{eq:Mabcddef}.

The second case we study corresponds to mean-field models that belong to the RFOT class~\cite{LW07,KT15}, 
and thus present a genuine MCT transition accompanied by an underlying 1RSB transition. Because below the MCT transition phase space gets ergodically broken in a large number of states (that diverges exponentially with $N$), dynamical correlations correspondingly develop a plateau of diverging length. Despite this abundance of states, we nevertheless only consider the sample-to-sample susceptibility that includes both inter-state and disorder fluctuations. 
An important practical reason is that 
for large systems it is numerically near impossible to study an equilibrium dynamics that requires jumping between states. 
The only known technique to produce equilibrium samples is the planting technique, which gives a single sample for each given disorder, 
and hence does not allow one to disentangle sample and disorder fluctuations.
An additional reason is that the calculation of sample-to-sample fluctuations within the RS ansatz is analytically simpler. We will make use
of two distinct approaches.

The first approach, called the Franz-Parisi (FP) potential calculation~\cite{franz_recipes_1995}, selects an equilibrium configuration $x_0$ and defines the free energy as a function of the overlap $p$ with this reference configuration, 
\begin{equation}
\begin{aligned}
-N\beta \bm{V_{\text{\tiny FP}}}(p) &\equiv \overline{ [ \ln \left\{   \Tr_{x} e^{-\beta H(x)} \delta [p-q(x,x_0)]\right\} ]}\\
&= N  \lim_{n\to 0} \partial_n \big [\text{Ext}_{\mQ{n}{RS}} F(\mQ{n+1}{RS}) \big ] \ ,
\end{aligned}
\end{equation}
where 
\begin{equation}
[ \bullet ] = \Tr_{x_0} \left[ \bullet \frac{e^{-\beta H(x_0)}}Z \right]
\end{equation}
denotes the equilibrium average over $x_0$, and
$\text{Ext}_{\mQ{n}{RS}}$ refers to the extremization with the respect to the parameters of the RS overlap matrix: the overlap $q_{ab}=q(x_a,x_b)$
for $a,b\in[1,n]$, and eventually the self-overlap $q_d=q(x,x)$, i.e. the norm of the configuration, with $q_{0a} = q(x_0,x_a)=p$ being fixed.
As described in Sec.~\ref{fromAvtoProb} the FP potential induces an overlap measure $\langle \bullet  \rangle_{\text{\tiny FP}}$ which corresponds to the one-time average $\mathcal{T}_{\FP}[ \bullet  ]$. Notice that $\bm{V_{\text{\tiny FP}}}(p)=\overline{[V_{\text{\tiny FP}}(p)]}$ is the averaged potential over $x_0$ and the quenched disorder (when not self-induced). 
By analogy with Eq.~\eqref{rep_met}, the replica method can be used to compute the overlap action $F(\mQ{n+1}{})$, which, when extremized using a RS ansatz, corresponds to a locally convex free energy landscape around $x_0$. (The notation $\mathbbm{Q}^{n+1}_\text{\tiny RS}$ describes a RS matrix with one special replica, $x_0$, that breaks replica symmetry and is fixed to have the same overlap $p$ with all other replicas. The first row and column of the matrix are thus identically $p$.)

The second approach couples $m$ real copies of the system so as to energetically favor configurations in the same state. The resulting Monasson (M) potential \cite{monasson_structural_1995} gives the free energy
\begin{equation}
\begin{aligned}
-N\beta &V_\M^m(q)= \\
&\ln \left\{  \Tr_{x^{\otimes m}} e^{-\sum_{k=1}^m\beta H(x_k)} \prod^m_{k\neq k'}\delta [q-q(x_k,x_{k'})]\right\} \ , \\
  \bm{V_\text{\tiny  M}}(q) &\equiv \lim_{m\to 1}\partial_m \overline{  V_\M^m(q) } \\
&=\lim_{m\to 1}\partial_m \big [ \lim_{n\to 0} \partial_n [\text{Ext}_{\mQ{m}{RS}^{\otimes n}} F_{intra}^\text{\tiny  M}({\mathbbm{Q}^m_\text{\tiny RS}}^{\otimes n})] \big ] \\
&=\lim_{m\to 1}\partial_m  \big [ \text{Ext}_{\mQ{m}{RS}} F(\mathbbm{Q}^m_\text{\tiny RS}) \big ]\ ,
\end{aligned}
\end{equation}
where ${\mathbbm{Q}^m_\text{\tiny RS}}^{\otimes n}$ stands for an RS matrix of dimension $m$ that is repeated $n$ times with all off-diagonal blocks equal to zero, so as to encode the orthogonality between different states. This structure is mirrored in long-time equilibrium simulations (see Fig.~\ref{fig:Inst300}). (Interestingly, an analogous structure has been reported for small systems of the supercooled Kob-Andersen binary Lennard-Jones liquid~\cite{appignanesi_democratic_2006}, which is part of the same mean-field universality class.)
Analogously to the FP potential the M potential induces an overlap measure $\langle \bullet  \rangle_\text{\tiny  M}$ which corresponds to the two-time average $\mathcal{T}_{\M}[ \bullet  ]$.

Notice that in the two potentials introduced above, the large deviation function of the overlaps $F(\mQ{}{})$ is the same (since it depends only on the Hamiltonian and on the space of configurations). What changes is the ansatz chosen for the matrix $\mQ{}{}$ which reflects the imposed constraints, i.e., the fixed overlap $p$ with the reference configuration for the FP potential and the overlap $q$ between two different configurations in the M potential. 
In either case, even if the free energy landscape $F(\{o_i\})$ presents a large number of equilibrium states, a RS structure is obtained for the saddle point matrix $\mQ{*}{RS}$.  Also, at \textit{equilibrium} ($p=\qEA=q$) the two saddle points are equivalent,
\begin{equation}
\lim_{n\to0}F(\mQ{* n+1}{RS}) = \lim_{m\to1}F(\mQ{* m}{RS}) \ .
\end{equation} 
Said differently, if the overlap with the reference configuration is taken to be the equilibrium value ($p=\qEA$) in the FP potential, and if the constraint between two configurations is also taken to be the equilibrium value ($q=\qEA$) in the M free energy, the two expressions are equal. The resulting saddle point $\mathbbm{Q}^*$ is an RS matrix of dimension $1$ that we will denote as the RFOT saddle in order to distinguish it from the standard RS saddle obtained from a matrix of dimension going to $0$.
(See Ref.~\cite{folena_mixed_2020} for a detailed discussion of the two potentials.)

Studying RFOT models with a RS ansatz by means of a potential that constrains the available phase space results in  the quenched disorder and the disorder induced by the reference configurations being averaged concurrently. In other words, inter-state and disorder fluctuations are then absorbed into sample-to-sample fluctuations. 

In order to emphasize the role of the total average over the overlap action we define the expectation value associated with a curly bracket with a subscript denoting the ansatz (FP or M or RS or RFOT) under consideration,
\begin{equation}\label{curlyRS}
\begin{aligned}
\avRS{ A }{ansatz} \equiv \lim_{N\to\infty} \frac{\int \mathcal{D} \mathbbm{Q} e^{ -N F(\mathbbm{Q})}A(\mathbbm{Q})}{\int \mathcal{D} \mathbbm{Q} e^{ -N F(\mathbbm{Q})} }\approx A(\mQ{*}{ansatz}) \ .
\end{aligned}
\end{equation}
The FP/M/RFOT ansatzes corresponds to different ways of averaging over samples, states and configurations, e.g., $\avRS{ A }{FP}  = \overline{ [\langle A \rangle_{\text{\tiny FP}} ] }$.
The RFOT scenario is by definition the optimal (with respect to the overlap $\qEA$) RS matrix of dimension $1$. It coincides with the FP and M ansatz for $p=q=\qEA$, therefore
\begin{equation}
\avRS{\bullet}{FP}|_{p=\qEA} = \avRS{\bullet}{M}|_{q=\qEA} = \avRS{\bullet}{RFOT} \ .
\end{equation}
In light of the above considerations, the rest of this section focuses on the structure of the RS ($n\to0$) and RFOT mass matrix ($n=1$) and the evaluations of the respective susceptibilities.

\subsection{Matrices of RS Fluctuations and Correlations}
The study of fluctuations around the RS saddle-point (with an external field) was first reported by de Almeida et al.~\cite{almeida_stability_1978}. Their study of the stability of the RS solution of the Sherrington-Kirkpatrick (SK) model \cite{sherrington_solvable_1975} revealed that the RS solution of this model is stable---the landscape is convex---at high external magnetic fields or high temperatures only, i.e., above the (now-called) de Almeida-Thouless line. As the field or the temperature are lowered, one of the eigenmodes (the replicon) of the RS mass matrix (RSMM) vanishes, thus giving rise to a transition. 

Here, we consider the structure of the RSMM in the simplified case of zero external field and with non-fluctuating self-overlap (diagonal entries).
Whenever the saddle point corresponding to the metastable minimum of the Franz-Parisi or Monasson potential is replica symmetric, the matrix of fluctuations around the saddle point can be described by three independent parameters (or masses), $m_1$, $m_2$ and $m_3$, as
\begin{equation}\label{RSMM}
\begin{aligned}
&\mathbbm{M}^\text{\tiny RS}_{a\neq b;c\neq d} \equiv\partial_{q_{ab}}\partial_{q_{cd}} F(\mathbbm{Q})\big |_{\mathbbm{Q} = \mathbbm{Q}_\text{\tiny RS}} \\
&= \frac{m_1}{2} (\delta_{a c}\delta_{ b d}+\delta_{a d}\delta_{ b c}) +  \frac{m_2}{4} (\delta_{a c}+\delta_{a d}+\delta_{ b c}+\delta_{ b d}) + m_3 \ .
\end{aligned}
\end{equation}
Note that the more general case for which diagonal overlaps fluctuate has seven independent parameters, but the form in Eq.~\eqref{RSMM} suffices for the models considered here.

The RSMM can be rewritten in the basis of its three distinct set of degenerate eigenmodes (see Appendix~\ref{AppendixA}) as
\begin{equation}
\mathbbm{M}^\text{\tiny RS} = 
\lambda_\text{\tiny R}\sum_{i=1}^{\mu_\text{\tiny R}}|v^{\text{\tiny R}}_i\rangle\langle v^{\text{\tiny R}}_i|
+\lambda_\text{\tiny A}\sum_{i=1}^{\mu_\text{\tiny A}}|v^{\text{\tiny A}}_i\rangle\langle v^{\text{\tiny A}}_i|
+\lambda_\text{\tiny L}\sum_{i=1}^{\mu_\text{\tiny L}}|v^{\text{\tiny L}}_i\rangle\langle v^{\text{\tiny L}}_i| \ ,
\end{equation}
where  the three eigenvalues $\lambda$ with relative multiplicity $\mu$ are
\begin{eqnarray}\label{REP}
\begin{aligned}
\lambda_\text{\tiny R} &= m_1 \ , \qquad \qquad \qquad \mu_\text{\tiny R} = \frac{n(n-1)}{2} -n \ , \label{lR} \\
\lambda_\text{\tiny A} &= m_1+ \frac{n-2}{2}m_2 \ ,  \qquad \qquad \mu_\text{\tiny A} = n-1 \ , \label{lA}  \\
\lambda_\text{\tiny L} &= m_1 +(n-1)m_2 + n(n-1)m_3 \ , \qquad \mu_\text{\tiny L} = 1 \ . \label{lL} 
\end{aligned}
\end{eqnarray}
The first (or replicon) eigenvalue describes the instability of the chosen ansatz for the overlap matrix. Upon approaching a marginal phase the replicon vanishes, thus indicating that the minima associated with typical states are getting flatter along certain directions. 
The second (or anomalous) eigenvalue describes eigenmodes that break RS symmetry along only one direction. 
Finally, the third (or longitudinal) eigenvalue corresponds to a shift of the average overlap in the system, i.e., $\delta \avRS{q}{}$.
This nomenclature for the eigenmodes of the Hessian was first introduced in \cite{bray_replica_1979} and is carefully discussed in~\cite{de_dominicis_ising_1985-1}.

%These are fluctuations in directions for which the average overlap of any configuration with all the others does not change and the probability distribution of overlaps splits.

Another representation of the RSMM by three elements (as originally formulated in Ref.~\cite{almeida_stability_1978}) is possible
\begin{equation}
\begin{aligned}\label{EqM}
\mathbbm{M}^\text{\tiny RS}_{12;12} &= \frac{m_1}{2} + \frac{m_2}{2} + m_3 \ , \\
\mathbbm{M}^\text{\tiny RS}_{12;13} &= \frac{m_2}{4} + m_3 \ , \\
\mathbbm{M}^\text{\tiny RS}_{12;34} &=  m_3\ .
\end{aligned}
\end{equation}
%This representation more naturally lends itself to computation, because it corresponds to second-order derivatives in the basis of replica couples, i.e. $\mathbbm{M}^\text{\tiny RS}_{12;12} = \partial_{q_{12}}\partial_{q_{12}} F(\mQ{}{})$,
%$\mathbbm{M}^\text{\tiny RS}_{12;13} = \partial_{q_{12}}\partial_{q_{13}} F(\mQ{}{})$,
%$\mathbbm{M}^\text{\tiny RS}_{12;34} = \partial_{q_{12}}\partial_{q_{34}} F(\mQ{}{})$.
It is also possible to define a symmetrized version of this representation,
$$\mathbbm{M}^\text{\tiny RS}_{(ab);(cd)} \equiv \mathbbm{M}^\text{\tiny RS}_{a\neq b;c\neq d}+\mathbbm{M}^\text{\tiny RS}_{a\neq b;d\neq c}+\mathbbm{M}^\text{\tiny RS}_{b\neq a;c\neq d}+\mathbbm{M}^\text{\tiny RS}_{b\neq a;d\neq c}$$
where each pair of distinct indices is counted only once.

%\subsection{Matrix of Correlations}

The matrix of correlations around the RS saddle point is then obtained by inverting $\mathbbm{M}^\text{\tiny RS}$.
In the diagonal basis, the operation straightforwardly gives
\begin{equation}
\mathbbm{G}^\text{\tiny RS} = 
\frac1{\lambda_\text{\tiny R}}\sum_{i=1}^{\mu_\text{\tiny R}}|v^{\text{\tiny R}}_i\rangle\langle v^{\text{\tiny R}}_i|
+\frac1{\lambda_\text{\tiny A}}\sum_{i=1}^{\mu_\text{\tiny A}}|v^{\text{\tiny A}}_i\rangle\langle v^{\text{\tiny A}}_i|
+\frac1{\lambda_\text{\tiny L}}\sum_{i=1}^{\mu_\text{\tiny L}}|v^{\text{\tiny L}}_i\rangle\langle v^{\text{\tiny L}}_i| \ .
\end{equation}
More explicitly, taking advantage of the fact that matrices of the form in Eq.~\eqref{RSMM} form a closed algebra,
the inverse has the same form
\begin{equation}\label{RSCM}
\begin{aligned}
&\mathbbm{G}^\text{\tiny RS}_{a\neq b;c\neq d} \equiv\avRS{\delta \hat{q}_{ab}\delta \hat{q}_{cd}}{RS/RFOT}\\
&= \frac{g_1}{2} (\delta_{a c}\delta_{ b d}+\delta_{a d}\delta_{ b c}) +  \frac{g_2}{4} (\delta_{a c}+\delta_{a d}+\delta_{ b c}+\delta_{ b d}) + g_3 \ , 
\end{aligned}
\end{equation}
where we introduced a new notation for the re-scaled fluctuation of the overlap
\begin{equation}
\delta \hat{q}_{ab} = N^{\frac{1}{2}}\big (q_{ab}-\avRS{ q_{ab}}{RS/RFOT}\big ) \ .
\end{equation}
Imposing that the product of $\mathbbm{M}^\text{\tiny RS}$ and $\mathbbm{G}^\text{\tiny RS}$ is the identity
leads to the conditions
\begin{equation}\label{eq:genericinverse}
\begin{split}
&g_1 = \frac{1}{m_1} \ ,\\
&g_2 = -\frac{2 m_2}{m_1  [2 m_1 +m_2 (n-2)]} \ ,\\
&g_3 = \frac{-2 m_1  m_3+m_2^2+m_2 m_3 n}{m_1  [2 m_1 +m_2 (n-2)] [m_1 +(n-1) (m_2+m_3 n)]} \ ,
\end{split}
\end{equation}
which specify the three parameters that describe the matrix of correlations between overlaps around a RS (or RFOT) saddle point.
If the RS saddle point globally minimizes the free energy, then $n=0$; if it is but a local minimum of the Franz-Parisi or Monasson potential (as in the RFOT case), then  $n=1$. 
In either case, the matrix $\mathbbm{G}^\text{\tiny RS}_{a\neq b;c\neq d}$ can be used to express the various susceptibilities defined in Sec.~\ref{intro}, as described in the following subsections.

\subsection{Total Susceptibility}
For both simple RS ($n=0$) and RFOT ($n=1$) phases, the total susceptibility around the RS saddle point is given by the trace of the matrix,
\begin{equation}
\begin{aligned}
\bm{\hat{\chi}}_{tot}&\equiv \avRS{ \delta \hat{q}_{12}\delta \hat{q}_{12} }{RS/RFOT}\\
&= 4\mathbbm{G}^\text{\tiny RS}_{12;12}
=\frac{4}{n(n-1)}\sum_{a b}\mathbbm{G}^\text{\tiny RS}_{a\neq b;a \neq b}\\
&=\frac{4}{n(n-1)}(\mu\text{\tiny R}\lambda^g_\text{\tiny R}+\mu\text{\tiny A}\lambda^g_\text{\tiny A}+\mu\text{\tiny L}\lambda^g_\text{\tiny L}) \\
&= 2g_1 + 2g_2 + 4g_3 \ .
\end{aligned}
\end{equation}
The associated fluctuations are those expected from averaging over all sources of fluctuations. In other words, the total susceptibility corresponds to what is obtained by considering very long timescales, such that configurations are \emph{ergodically} sampled.

\subsection{Franz-Parisi Susceptibilities}\label{secth}

The intra-state FP susceptibility introduced in Eq.~\eqref{thk2} describes system fluctuations inside a state corresponding to the two-time kernel (Monasson potential), here also averaged over states:
\begin{equation}
\begin{aligned}\label{atypicalth}
\hat{\bm{\chi}}_{intra}^\text{\tiny FP} &=  \overline{\langle \delta \hat{q}_{12}^2\rangle_\text{\tiny  M} - \langle \delta \hat{q}_{12}\delta \hat{q}_{13}\rangle_\text{\tiny  M}}\\
&= \avRS{  \delta \hat{q}_{12}^2 }{M} - \avRS{  \delta \hat{q}_{12}\delta \hat{q}_{13} }{M} \\
&= 4 (\mathbbm{G}^\text{\tiny RS}_{12;12} -\mathbbm{G}^\text{\tiny RS}_{12;13}) = 2g_1 +  g_2 \\
&=\frac{2}{m_1}\left[1-\frac{m_2}{2m_1-m_2(2-n)} \right] \ .
\end{aligned}
\end{equation}
Note that the average given by the RS ansatz for the overlap matrix, $\avRS{ }{M}$, can also be evaluated for overlaps that are not typical in the thermodynamic limit $N\to\infty$, i.e., $q\neq\qEA$. This average is thus expected to be valid for every overlap such that the RS solution remains stable, that is as long as all the RSMM eigenvalues (given in Eqs.~\eqref{lR}) remain positive. In the typical case $q = \qEA$ we have $\avRS{ }{M}$ = $\avRS{ }{RS/RFOT}$. 

The complement of the intra-state FP susceptibility is the sample-to-sample FP susceptibility (Eq.~\eqref{het2t} in the Monasson form),
\begin{equation}\label{eqhet}
\begin{aligned}
&\bm{\hat{\chi}}_{sample}^\text{\tiny FP} = \overline{\langle \delta \hat{q}_{12}\delta \hat{q}_{13}\rangle_\text{\tiny  M}} \big |_{q=\qEA}\\
&= \avRS{\delta \hat{q}_{12}\delta \hat{q}_{13}}{RS/RFOT}= 4 \mathbbm{G}^\text{\tiny RS}_{12;13}= g_2 + 4 g_3\\
&=-\frac{2 [m_1(m_2+4 m_3)-m_2(3-n) (m_2+m_3n)]}{m_1[2 m_1-m_2(2-n)] [m_1-(1-n) (m_2+m_3n)]}\ .
\end{aligned}
\end{equation}
Interestingly, these two susceptibilities have (at $q=\qEA$) a static interpretation in terms of the Franz-Parisi potential $V_{\text{\tiny FP}}(p)$ (see Eqs.~\eqref{chith} and ~\eqref{chihet}). The intra-state FP susceptibility is equivalent to the inverse of the second derivative of that potential with respect to the overlap with the reference configuration $x_0$, $p\equiv q_{01}$,
\begin{equation}\label{chithFP}
	\bm{\hat{\chi}}_{intra}^{\text{\tiny FP}} = \frac{1}{\beta V''_{\text{\tiny FP}}(p)} =  N\big ( \langle  p^2\rangle_{\text{\tiny FP}}-\langle p\rangle^2_{\text{\tiny FP}} \big )\ .
\end{equation}
Note that here the average over samples is not needed, but can be taken afterwards.
The sample-to-sample FP susceptibility can be expressed in terms of the fluctuations of the derivatives of the Franz-Parisi potential,
\begin{equation}\label{chihetFP}
	\bm{\hat{\chi}}_{sample}^\text{\tiny FP}  = \frac{\overline{\delta  V'_{\text{\tiny FP}}(p)^2}}{ \bm{V''_{\text{\tiny FP}}}(p)^2 }\bigg |_{p=\qEA} = N\big( \overline{\langle  p \rangle_{\text{\tiny FP}}^2}-\overline{\langle  p\rangle_{\text{\tiny FP}}}^2 \big ).
\end{equation}

The FP potential is therefore a \textit{proper one-dimensional random potential} (as in Sec.~\ref{oneD}), for which randomness arises from the reference configuration (and the quenched disorder, if present). 
Details about this equivalence are provided in Appendix~\ref{AppendixB}. 
Figure~\ref{fig:3spin} depicts the average FP potential. Its second derivative around $\qEA$ gives $[\bm{\hat{\chi}}_{intra}^\text{\tiny FP}]^{-1}$.

\subsection{Monasson Susceptibilities}\label{secdyn}
Alternatively, the total susceptibility can be subdivided between the intra-state M susceptibility introduced in Eq.~\eqref{DYN},
\begin{equation}
\begin{aligned}\label{atypicaldyn}
\hat{\bm{\chi}}_{intra}^\text{\tiny M} &= \langle \delta \hat{q}_{12}^2\rangle_\text{\tiny  M}- \langle \delta \hat{q}_{12}\delta \hat{q}_{34}\rangle_\text{\tiny  M}  \big|_{q=\qEA}\\
&= \avRS{ \delta q_{12}^2 }{M} - \avRS{ q_{12}q_{34} }{M}  \\
&= 4 (\mathbbm{G}^\text{\tiny RS}_{12;12} -\mathbbm{G}^\text{\tiny RS}_{12;34})= 2g_1 +  2g_2 \\
&= \frac{2}{m_1}\big (1-\frac{2m_2}{2m_1-m_2(2-n)} \big )\ ,
\end{aligned}
\end{equation}
and the complementary sample-to-sample M susceptibility given in Eq.~\eqref{eq:chiMsample},
\begin{equation}\label{eqvar}
\begin{aligned}
&\bm{\hat{\chi}}_{sample}^\text{\tiny M} = \overline{\langle \delta \hat{q}_{12}\delta \hat{q}_{34}\rangle_\text{\tiny  M}} \big|_{q=\qEA}\\
&= \avRS{\delta \hat{q}_{12}\delta \hat{q}_{34} }{RS/RFOT} = 4 \mathbbm{G}^\text{\tiny RS}_{12;34}=  4 g_3\\
&=\frac{4 \left(-2 m_1 m_3+m_2^2+m_2 m_3 n\right)}{m_1 [2 m_1-m_2 (2-n)] [m_1-(1-n) (m_2+m_3 n)]} \ .
\end{aligned}
\end{equation}
While the FP potential is a one-dimensional potential and intra-state susceptibilities can evaluated by second order total derivatives of the potential, the Monasson potential instead has a two-dimensional nature and there is no direct way to evaluate intra-state susceptibilities by mean of total derivatives.

\subsection{About Sample-to-Sample Fluctuations in Mean-field Models}\label{Discussion}
We have seen how to extract sample-to-sample fluctuations from the replica action. In the case of systems that admits an RS global solution there exists only one state in the system, therefore inter-state fluctuations are absent and $\bm{\hat{\chi}}_{sample}=\bm{\hat{\chi}}_{dis}$. This will be the case in the $2$-spin spherical model discussed in Sec.~\ref{2spin}. 

On the other extreme if a system presents a RFOT phase and there is no external field one can argue, based on the replica method (see \cite{franz_field_2011}), that the fluctuations of the quenched disorder (at the Gaussian level) are absent, therefore $\bm{\hat{\chi}}_{sample}=\bm{\hat{\chi}}_{inter}$. This will be the case in the $3$-spin model, in the ROM and in the RLG, respectively in Secs.~\ref{3spinsec},\ref{rom},\ref{rlg}.

We will not discuss the case of a RFOT system with a global external field (see for example \cite{cavagna_1999}), in which both inter-state and disorder fluctuations are present. 
In this case the replica analysis become more complicated, since the FP or M potential do not factorize on independent blocks (with reciprocal zero overlap) and therefore the mass matrix of small fluctuations becomes 1RSB. This remains an open problem for future investigations.

\subsection{Single State Replicon and Diverging Susceptibilities}

Eq.~\eqref{REP}, in addition to enabling susceptibility calculations, indicates how far a RS state is from breaking that symmetry. The \emph{replicon} of a state, which is related to the inner stability of a given state, can be expressed as 
\begin{equation}\label{rep}
\lambda_\text{\tiny R}=m_1=\frac{2}{2 \hat{\chi}_{intra}^\text{\tiny FP} - \hat{\chi}_{intra}^\text{\tiny M} } \ ,
\end{equation}
which may provide a practical way to measure the replicon in numerical simulations.

Furthermore, for a RFOT ($n=1$) phase, the leading divergence of the local and non-local susceptibilities at the MCT transition, at which the replicon vanishes, therefore reads (see Eqs.~\eqref{atypicalth}, \eqref{atypicaldyn}, \eqref{eqhet}, and \eqref{eqvar})
\begin{equation}\begin{split}
&\hat{\bm{\chi}}_{intra}^\text{\tiny FP}\sim \hat{\bm{\chi}}_{intra}^\text{\tiny M} \sim m_1^{-1} \ , \\
&\bm{\hat{\chi}}_{sample}^\text{\tiny FP}\sim \bm{\hat{\chi}}_{sample}^\text{\tiny M} \sim  m_1^{-2} \ .
\end{split}\end{equation}
As anticipated from our consideration of a simple random potential in Sec.~\ref{oneD}, non-local susceptibilities therefore diverge as the square of the local ones~\cite{franz_field_2011}.

\section{Fluctuations in the $p$-spin Spherical Model}\label{psp}
The fully-connected $p$-spin spherical model is a good candidate to verify mean-field suspectibility predictions, because of its ease of analysis and general versatility. For instance, it is possible to tune it, so as to obtain systems with very different phases and transitions~\cite{crisanti_spherical_2006,crisanti_amorphous-amorphous_2007}, including a RFOT phase. We here specifically consider the 2-spin model with external field $h$ and the 3-spin model with $h=0$ (see \cite{franz_field_2011}). (The 2-spin with zero field presents a marginal spin-glass state for $T<1$ \cite{kosterlitz_spherical_1976}, and any $O(1)$ (in $N$) external field $h$ convexifies the free energy landscape \cite{baik_spherical_2021}.) The former offers a single global minimum, and hence instantonic escapes are of no concern. The latter is one of the simplest models in the RFOT universality class. In both cases, we find a robust agreement between our analytical description of susceptibilities and direct simulations of large $N$ systems. 

The Hamiltonians of interest read respectively as
\begin{equation}\begin{split}
H_2 &= -\sum^N_{ij} J_{ij}s_is_j-\sum_i h_i s_i \ , \\
H_3 &= -\sum^N_{ijk} J_{ijk}s_is_js_k \ ,
\end{split}\end{equation}
where the quenched coupling constants $J_{ij}$ are Gaussian distributed with zero mean and variance $\frac{1}{2}N/\binom{N}{p}$. 
For the external fields $h_i$, we consider both a Gaussian distributed field $h_i$ with zero mean and variance $\overline{h}^2$ and a homogeneous field $h_i \equiv h$.
Note that $p$-spin spherical models are uniquely determined by the covariance between the Hamiltonians of systems with different quenched disorders
\begin{equation}\label{covH}
\overline{H[s]H[s']}-\overline{H[s]}\ \overline{H[s']}=N f\big (\frac{s\cdot s'}{N}\big )\ ,
\end{equation}
where $\overline{\bullet}$ corresponds to averaging over $J$ and $f(q)$ is a polynomial. 
In particular, for the 2-spin with Gaussian field and the 3-spin models we have
\begin{eqnarray}
\label{eq:fq}
f(q) =f_2(q) &\equiv q^2/2 + \overline{h}^2 q\ ,\\
\label{eq:fq2}
f(q) =f_3(q) &\equiv q^3/2\ ,
\end{eqnarray}
respectively.
In the presence of a homogeneous external field $h$, the average over the disorder of the Hamiltonian 
does not vanish, and one has
\begin{equation}\label{avH}
\overline{H[s]}=-N h \big (\frac{\bm{1}\cdot s}{N}\big ) \ ,
\end{equation}
where $\bm{1}$ is a vector of ones $(1,1,...,1)$. This last case is treated in details in Appendix~\ref{AppendixB0}.

\begin{figure}[t]
	\centering
	\includegraphics[width=1\columnwidth]{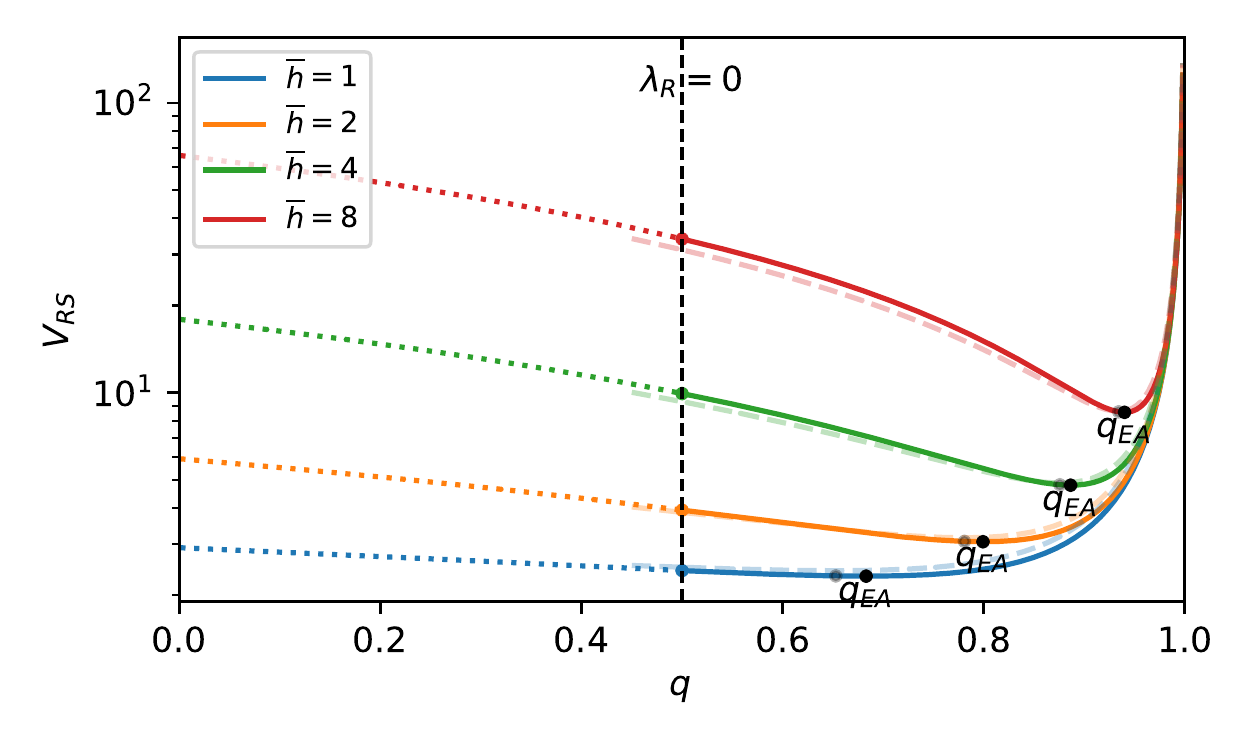}
	\caption{Overlap dependence of the RS free energy in the 2-spin spherical model at $T=0.5$ for Gaussian external fields $\overline{h}=1,2,4,8$ (solid and dotted lines). Colored points denote the overlap at which the RS solution becomes unstable ($\lambda_\text{\tiny R}=0$); black points denote the typical overlap, $\qEA$, given by the saddle point of the RS free energy. The RSMM and related parameters $m_1,m_2,m_3$ can be evaluated for any value of the overlap in the RS phase (solid line). The shadowed lines show the change in free energy upon increasing temperature ($T=0.55$). }
	\label{fig:2spin}
\end{figure}

\subsection{Free Energy}

\begin{figure}[t]
	\includegraphics[width=1\columnwidth]{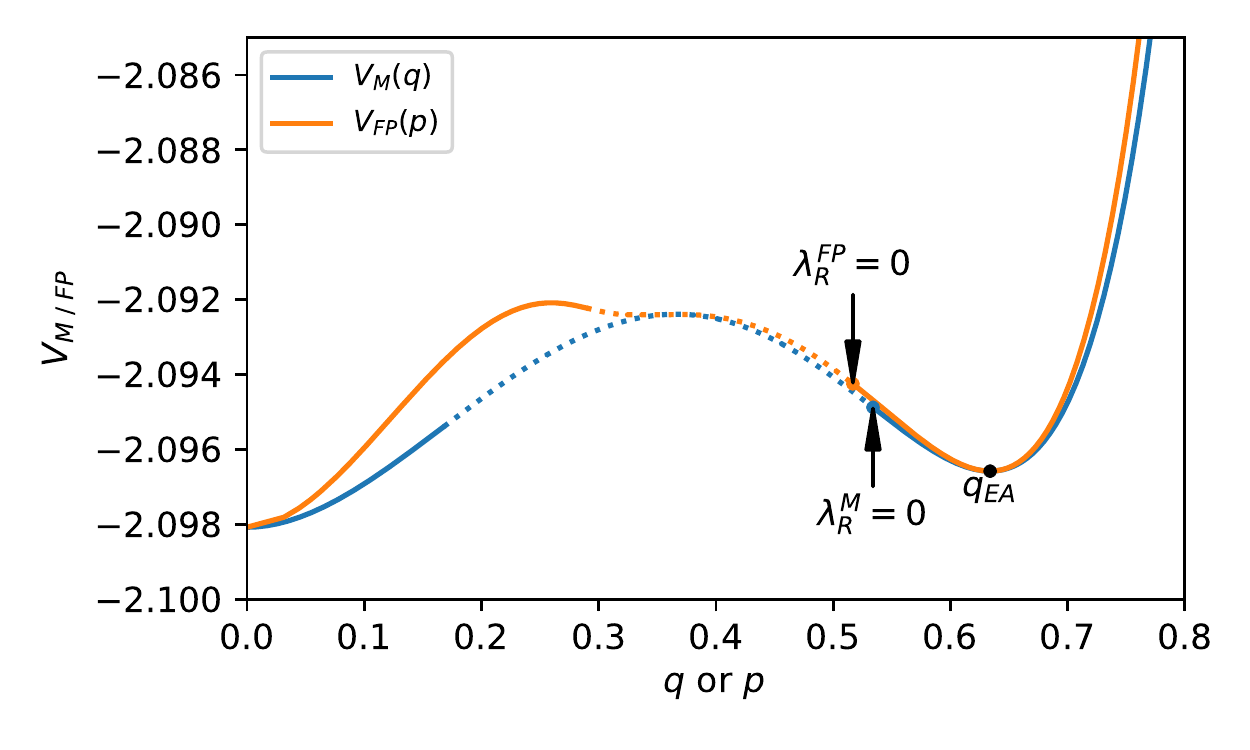}
	\caption{\footnotesize FP and M potentials for the 3-spin model at $T=0.59\lesssim \TMCT$. Colored points denote the value of the overlap at which the RS solutions become unstable, i.e.,  $\lambda_\text{\tiny R}=0$. The saddle point is the same for the two potentials and corresponds to the typical overlap $\qEA$. The RSMM at this point, which is equivalent for the two potentials, is used to obtain the intra-state and sample-to-sample susceptibilities. The second derivative of the FP potential at $\qEA$ directly provides the inverse of $\bm{\hat{\chi}}_{intra}^\text{\tiny FP}$. Note that the M potential is plotted as a function of the overlap $q$ between two typical equilibrium configurations, while the FP is a function of the overlap $p$ between an equilibrium configuration and the reference one (planted at equilibrium).}\label{fig:3spin}
\end{figure}

% We now consider the explicit case of the p-spin model, where 
Following Eq.~\eqref{rep_met}, in order to calculate the free energy averaged over the disorder we replicate the system $n$ times and obtain
\begin{equation}
 \begin{aligned}
& \overline{\exp (-\beta n f)}= \overline{\Tr_{s^n} e^{-\beta \sum_{a=1}^{n} H[s_a]} } =\\
=&\Tr_{s^n} e^{-\beta \sum_{a=1}^{n} \overline{H[s_a]} + \frac{1}{2}\beta^2 \sum^{n}_{a,b} \big (\overline{H[s_a]H[s_b]}-\overline{H[s_a]}\,\overline{H[s_b]} \big )} \ .
 \end{aligned}
\end{equation}
Because of its Gaussian nature, this expression gives the second-order cumulant expansion of the disorder. Substituting Eqs.~\eqref{covH} and \eqref{avH}, we obtain the overlap action (up to an irrelevant constant term), 
\begin{equation}\label{overlapAc}
F(\mQ{n}{}) =\frac{1}{2}\Big [ \ln \det(\mQ{}{}) + \sum_{ab} \beta^2f(q_{ab}) \Big ] \ ,%- \frac{\beta^4 h^4}{2}(\sum_{ab} q_{ab})^2
\end{equation}
 where the determinant of the Jacobian, $\det(\mQ{}{})$, accompanies the change of variables from spins $s$ to overlaps $q_{ab}$.  Recalling that for a generic square matrix $\mathbbm{A} $, $\partial_{\mathbbm{A}_{ab}}\ln \det \mathbbm{A}  = [\mathbbm{A} ^{-1}]_{ba}$, the saddle point equation is
 \begin{equation}
 \partial_{q_{ab}}F(\mQ{n}{}) \Longrightarrow \beta^2f'(q_{ab}) = -[\mQ{}{}^{-1}]_{ab} \ .%-\beta^4h^4(\sum_{cd} q_{cd})
 \end{equation}
 For the RS ansatz $q_{ab} = \delta_{ab}(1-q)+q$, the inverse matrix reads
 \begin{equation}\label{Qinv}
 [\mQ{}{}^{-1}]_{ab} = \frac{1}{1-q}\left[\delta_{ab}-\frac{q}{1+(n-1)q}\right] \ ,
 \end{equation}
 hence the general RS saddle point solution is 
 \begin{equation}\label{RSsaddle}
\beta^2 =\frac{\qEA}{f'(\qEA)(1-\qEA) [1+(n-1) \qEA]} \ .
\end{equation}
Plugging this ansatz in Eq.~\eqref{overlapAc} the overlap action reads
\begin{equation}\label{overRS}
\begin{aligned}
-\beta\bm{V}_\text{\tiny RS}(q)&\equiv \partial_n F(\mQ{n}{RS})\\&=\frac{1}{2} \Big\{ \ln (1-q)+\frac{q}{1+(n-1) q}\\
&+\beta ^2 [f(1)+(2 n-1) f(q)]\Big\}\ .
\end{aligned}
\end{equation}
Note that in the RFOT case ($n=1$), we recover the Monasson potential, which, as discussed in Sec.~\ref{SMM}, has the same RS saddle point (and $\qEA$) as the Franz-Parisi potential (see Fig.~\ref{fig:3spin}).

\subsection{RS Mass Matrix and Susceptibilities} \label{Susc3spin}

For convenience, we subdivide the total mass matrix into entropic (ent) and energetic (ene) contributions. 
 %we 
% we evaluate the entropic contribution of mass matrix
%\footnote{For a generic square matrix $\hat{M}$: $\partial_{M_{cd}}{ \sum_k M_{ka}[\hat{M}^{-1}]_{ab}} =0 \Longrightarrow \partial_{M_{cd}}[\hat{M}^{-1}]_{ab} = -[\hat{M}^{-1}]_{ac}[\hat{M}^{-1}]_{db}$}.
%\begin{equation}
%$\ln[\det(Q)]$
%\end{equation}
%by first considering a generic matrix $Q$, and then imposing the RS ansatz.
Because $\mQ{}{}$ is symmetric, we can write $\partial_{q_{ab}} \to (\partial_{q_{ab}}+\partial_{q_{ba}})$, and hence
\begin{equation}\label{ent1}
\begin{aligned}
-2 \mathbbm{M}^{_\text{\tiny ent}}_{ab;cd} &\equiv \partial_{q_{cd}}\partial_{q_{ab}}\ln \det(\mQ{}{}) \\
&= -2 \big ([\mQ{}{}^{-1}]_{ac}[\mQ{}{}^{-1}]_{db}+[\mQ{}{}^{-1}]_{ad}[\mQ{}{}^{-1}]_{cb}\big ).
\end{aligned}
\end{equation}
%where $_\text{\tiny ent}$ stands for entropic. 
Using Eq.~\eqref{Qinv} then gives
\begin{equation}
\begin{aligned}
\partial_{q_{cd}}&\partial_{q_{ab}}\ln \det(\mQ{}{}) =\\
&= - \frac{2}{(1-q)^2}\big \{[\delta_{ac}-\frac{q}{1+(n-1)q}][\delta_{bd}-\frac{q}{1+(n-1)q}]\\
&+[\delta_{ad}-\frac{q}{1+(n-1)q}][\delta_{bc}-\frac{q}{1+(n-1)q}]\big \} \ .
\end{aligned}
\end{equation}
We can thus identify
\begin{equation}\label{eq:ent}
\begin{aligned}
m_1^{_\text{\tiny ent}} &= \frac{2}{(1-q)^2} \ , \\
m_2^{_\text{\tiny ent}} &= -\frac{2}{(1-q)^2}\frac{2q}{1+(n-1)q} \ , \\
m_3^{_\text{\tiny ent}} &= \frac{2}{(1-q)^2}\frac{q^2}{[1+(n-1)q]^2} \ .
\end{aligned}
\end{equation}
The energetic contribution is then
\begin{equation}\label{eq:ene}
\begin{aligned}
-2\mathbbm{M}^{_\text{\tiny ene}}_{ab;cd} & \equiv \partial_{q_{cd}}\partial_{q_{ab}} \big [ \sum_{ef} \beta^2f(q_{ef}) \big ] \\ 
&= 2 \big (\delta_{ac}\delta_{bd}+\delta_{ad}\delta_{bc})\beta^2 f''(q)
\end{aligned}
\end{equation}
for $a\neq b$ and $c\neq d$, and hence 
\begin{equation}
m_1^{_\text{\tiny ene}} = -\frac{1}{2} 4 \beta^2f''(q)  \ ,
\end{equation}
while $m_2^{_\text{\tiny ene}}=m_3^{_\text{\tiny ene}}=0$.

The total mass matrix is obtained by summing the contributions in Eqs.~\eqref{eq:ent} and \eqref{eq:ene}, 
\begin{equation}\label{MassM}
\begin{aligned}
m_1(q) &= m^{_\text{\tiny ent}}_1 + m^{_\text{\tiny ene}}_1 =  \frac{2}{(1-q)^2} - 2\beta^2f''(q)  = \lambda\text{\tiny R}(q) \ , \\
m_2(q)  &= m^{_\text{\tiny ent}}_2  =  -\frac{4q}{(1-q)^2[1+(n-1)q]} \ , \\
m_3(q)  &= m^{_\text{\tiny ent}}_3 + m^{_\text{\tiny ene}}_3  = \frac{2q^2}{(1-q)^2[1+(n-1)q]^2} \ .
\end{aligned}
\end{equation}
As shown in Appendix~\ref{AppendixB0} in the case of a homogeneous external field an additional factor of $2\beta^4 h^4$ appears in $m_3(q)$.

Note that these results are only valid if the RS solution is stable, which is here (minimally) checked by ensuring that the replicon eigenvalue (in Eq.~\eqref{lR}) is positive, and therefore
\begin{equation}\label{lRpspin}
\lambda\text{\tiny R}>0 \quad \Longrightarrow \quad \beta^2 < \frac{1}{(1-q)^2f''(q)} \ .
\end{equation}
Given the mass parameters in Eq.~\eqref{MassM}, the various susceptibilities defined in Sec.~\ref{SMM} can be computed. The result are explicitly reported in Appendix~\ref{AppendixC}.

\subsection{Simulation Details}\label{pspinsimul}
We have simulated the equilibrium dynamics of the 2-spin spherical model with an  external field and the 3-spin spherical model without, implementing the Langevin over-damped dynamics on the $N$-dimensional sphere (following the simulations details presented in~\cite{folena_gradient_2021}). Because the free energy landscape of the 2-spin in a field is convex, the system is initialized on a random configuration and equilibration is fast. By contrast, the 3-spin landscape is complex. To overcome equilibration difficulties, the starting configuration is planted. More specifically, couplings are biased to emulate equilibrium around the random initial configuration (see Refs.~\cite{folena_gradient_2021} and \cite[Sec.~2.2.2]{folena_mixed_2020}).  For the 3-spin model the number of couplings grows as $N^3$. In order to reach large system sizes we dilute the couplings~\cite{folena_gradient_2021} by a factor of $6/N$. For every model and set of parameters, many different system sizes $N$ are considered. For each of the $N_\mathrm{sample}$ samples (different disorder and/or reference configuration) an equilibrium dynamics is simulated and configurations are sampled every time $\tau_{\kappa}$  (taken long enough to decorrelate, see Fig.~\ref{fig:SingleTraj}), for a total of $K$ configurations. The resulting correlation matrix of $K\times K$ overlaps is used to evaluate the local susceptibilities of each sample. Specific system sizes, number of samples and relative $K$ are reported in Tables~\ref{table2spin} and \ref{table3spin}.

\begin{table}[t]
	\centering
	\begin{tabular}{|c|c|c|c|}
		%\multicolumn{4}{c}{2-spin; } \\
		\hline 
		$N$ & 50 & 100 & 200 \\ 
		\hline 
		$N_\mathrm{sample}$ & 100 & 100 & 100 \\ 
		\hline 
		$K$ & 700 & 700 & 700 \\ 
		\hline 
	\end{tabular} 	
		\caption{Simulation parameters for the (RS) 2-spin at $T=0.5$ and $h=1,2,4,8$ for $\tau_{\kappa}=7$.}\label{table2spin}
\end{table}

\begin{table}[t]
	\centering
	\begin{tabular}{|c|c|c|c|c|c|c|}
		\hline 
		$N$ & 400 & 800 & 1600 & 3200 & 6400 & 12800 \\ 
		\hline 
		$N_\mathrm{sample}$ & 378 & 84 & 95 & 98 & 48 & 48 \\ 
		\hline 
		$K$ & 100 & 50 & 20 & 20 & 100 & 20 \\ 
		\hline 
	\end{tabular}
	\caption{Simulation parameters for the (RFOT) 3-spin at $T=0.59$ for $\tau_{\kappa}=21$.}\label{table3spin}
	\end{table}

\subsection{$2$-spin Spherical Model (RS) Results}\label{2spin}

\begin{figure}[t]
	\centering
	\includegraphics[width=1\columnwidth]{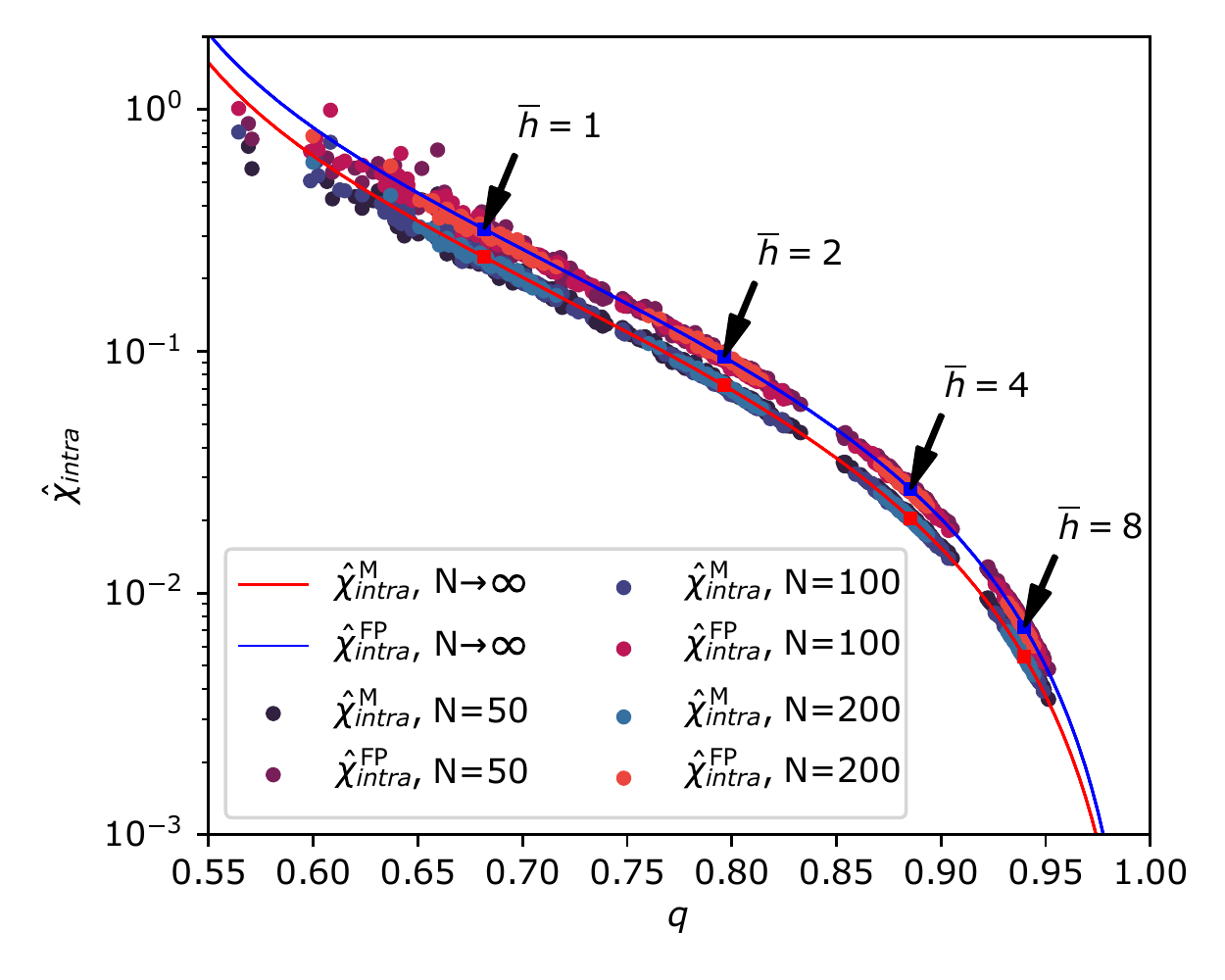}
	\caption{Scatter plot of the intra-state FP (blue points) and M (red points) susceptibilities versus the overlap in the $2$-spin spherical model at temperature $T=0.5$ with external field $\overline{h}=1,2,4,8$ for $N=50,100,200$. For every field and system size, 100 points (corresponding to $N_\mathrm{sample}=100$) are shown. Squares denote the expected value in the thermodynamic limit, given the relative external field $h$. Non-typical samples align with the theoretical line of susceptibilities, i.e., with typical samples at different fields. This effect is a consequence of Eq.~\eqref{eqthdyn}, as described in the text.}\label{fig:2spinScat}
\end{figure}

\begin{figure}[t]
	\centering
	\includegraphics[width=1\columnwidth]{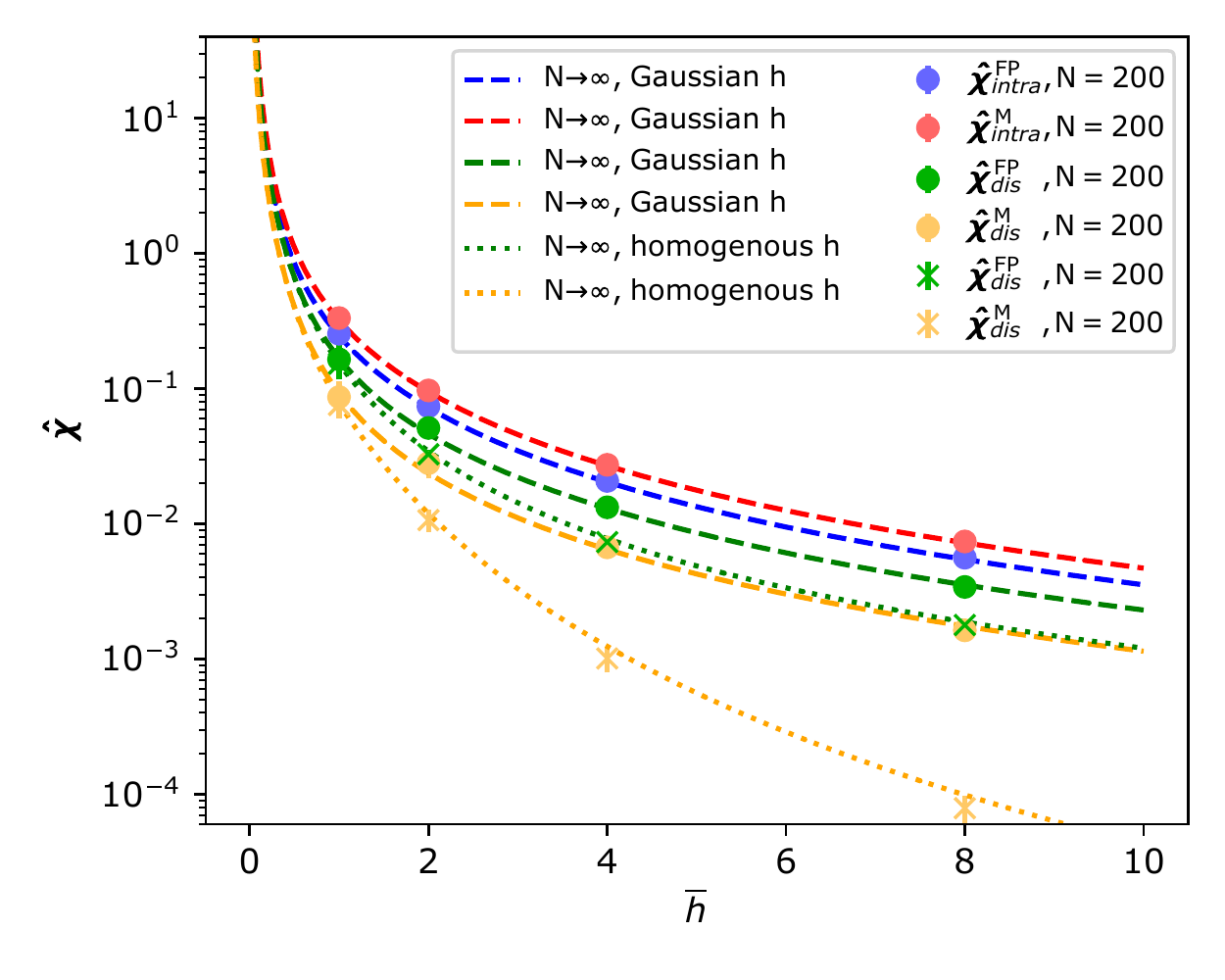}
	\caption{Four susceptibilities for the 2-spin spherical model as a function of the standard deviation $\overline{h}$ of the Gaussian distributed external field. The dotted lines and associated crosses denote sample-to-sample fluctuations for a homogeneous external field. Already at small systems sizes (and with as few as $N_\mathrm{sample}=100$ per point) the agreement with the thermodynamic result is very good. Such fast convergence follows from the steepness of the free energy in Fig.~\ref{fig:2spin}.}\label{fig:localsusc}
\end{figure}

Because the $2$-spin model possesses a single global minimum,  heterogeneity between samples arises only from the quenched disorder of the couplings,  not from different reference configurations. %Each sample corresponds to a different choice of couplings $J$.
The RS solution in Eq.~\eqref{RSsaddle} for $n=0$ is then
\begin{equation}
\beta^2=\frac{\qEA}{(1-\qEA)^2f_2'(\qEA)} \ ,
\end{equation}
with $f_2(q)$ as in Eq.~\eqref{eq:fq}. %\equiv q^2/2+h^2 q$ in the 2-spin with field.
This equation implicitly defines the typical overlap $\qEA(\beta,h)$.
For the 2-spin at any inverse-temperature $\beta$, an arbitrary small $h$ results in a RS stable saddle point at $\qEA(\beta,h)$. 
However, the possibility that atypical $q$ have a negative $\lambda\text{\tiny R}$ must be considered. 
We then have the condition (see Eqs.~\eqref{lRpspin} and \eqref{eq:fq}),
\begin{equation}
\lambda\text{\tiny R}>0 \quad \Longrightarrow \quad q > 1-\frac{1}{\beta} \ , \quad \forall h \ .
\end{equation}
For this system the overlap is expected to follow the large deviation function given by the RS overlap action in Eq.~\eqref{overRS}
\begin{equation}
-\beta\bm{V}_\text{\tiny RS} = \frac{1}{2} \left[\beta ^2 \big(\frac{1-q^2}{2} + \overline{h}^2(1-q)\big )+\frac{q}{1-q}+\ln (1-q)\right].
\end{equation}
Figure~\ref{fig:2spin} shows the RS overlap action for different external fields at inverse temperature $\beta=2$, for which the condition of positive replicon is $q>0.5$.
For any $q > 1-\frac{1}{\beta}$, the intra-state susceptibilities (Eqs.~\eqref{Chth} and \eqref{Chdyn}) read
\begin{equation}\label{eqthdyn}
\begin{aligned}
\hat{\bm{\chi}}_{intra}^\text{\tiny FP}(q) &= \frac{(1-q)^2 \left[1+2q-\beta ^2 (1-q)^3\right]}
{1+q-2 \beta ^2 (1-q)^2+\beta ^4 (1-q)^5} \ , \\
\hat{\bm{\chi}}_{intra}^\text{\tiny M}(q) &=\frac{(1-q)^2 \left[1+3q-\beta ^2 (1-q)^3\right]}
{1+q-2 \beta ^2 (1-q)^2+\beta ^4 (1-q)^5} \ .
\end{aligned}
\end{equation}
Note that $\hat{\bm{\chi}}^{\text{\tiny RS},\text{\tiny M}}_{intra}(q)$ is the same as \cite[Eq.~(10.19)]{baik_spherical_2021} 
(with a different notation, see Appendix~\ref{appendixH}), 
which was obtained using random matrix theory. This equivalence is an important validation of our analysis.
These two local susceptibilities are also perfectly recovered in numerical simulations of small systems (Figs.~\ref{fig:2spinScat} and~\ref{fig:localsusc}). Remarkably, this correspondence not only holds around the thermodynamic saddle point, $\qEA$, but also for $q$ well beyond %the Gaussian fluctuations regime 
(in a large deviation sense). The fact that atypical overlaps, at given external field $h$, also lie on the same curve defined by typical overlaps at different $h$ (squares), is a consequence of the independence of intra-state susceptibilities (given by Eq.~\eqref{eqthdyn}) on the external field. This property applies generally to RS systems with an external field, and follows from the equivalence to a linearly tilted one-dimensional potential, as discussed in Sec.~\ref{oneD}. 

The Gaussian sample-to-sample fluctuations of $q$ around $\qEA$ are described by
Eqs.~\eqref{Chihet} and \eqref{Chivar},
\begin{equation}\label{eqhetvar}
\begin{aligned}
\bm{\hat{\chi}}^\text{\tiny FP}_{dis} &= \frac{ 6 \qEA^2/[1 - \beta^2(1 - \qEA)^2 ] + \qEA - 3 \qEA^2 %- 2 \beta^4h^4 (1 - q)^4
}
       { (1 - \qEA)^{-2}[1 + \qEA - \beta^2(1 - \qEA)^3 ]^2} \ , \\
\bm{\hat{\chi}}^\text{\tiny M}_{dis} &=\frac{ 4 \qEA^2/[1 - \beta^2(1 - \qEA)^2 ] - 2 \qEA^2 %- 2 \beta^4h^4 (1 - q)^4
}
       { (1 - \qEA)^{-2}[1 + \qEA - \beta^2(1 - \qEA)^3 ]^2} \ ,
\end{aligned}
\end{equation}
and are compared with numerical simulations in Fig.~\ref{fig:localsusc}.
Once again, our ${\hat{\chi}}^\text{\tiny M}_{dis}$ prediction is the same as Ref.~\cite[Eq.~(10.18)]{baik_spherical_2021}.

Figure~\ref{fig:localsusc} reports the susceptibility for a homogeneous external $h$ with intensity equal to the standard deviation of the Gaussian case, i.e., $h=\overline{h}$. In this case, an extra factor $2\beta^4 h^4$ appears in $m_3$ of Eq.~\eqref{MassM} (see Appendix \ref{AppendixB0} for the derivation), which results in a factor $- 2 \beta^4h^4 (1 - \qEA)^4$ appearing in the numerator of both Eqs.~\eqref{eqhetvar}. 
Although sample-to-sample fluctuations then change with the external field (green and yellow lines), intra-state fluctuations are unaffected (blue and red lines).

\begin{figure}[t]
	\includegraphics[width=1\columnwidth]{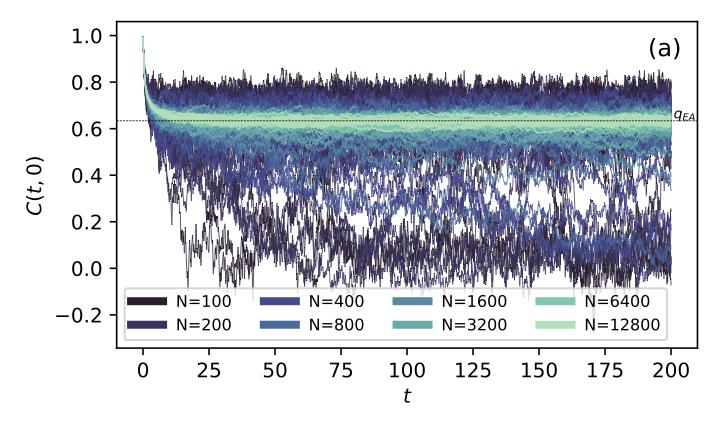}
	\includegraphics[width=1\columnwidth]{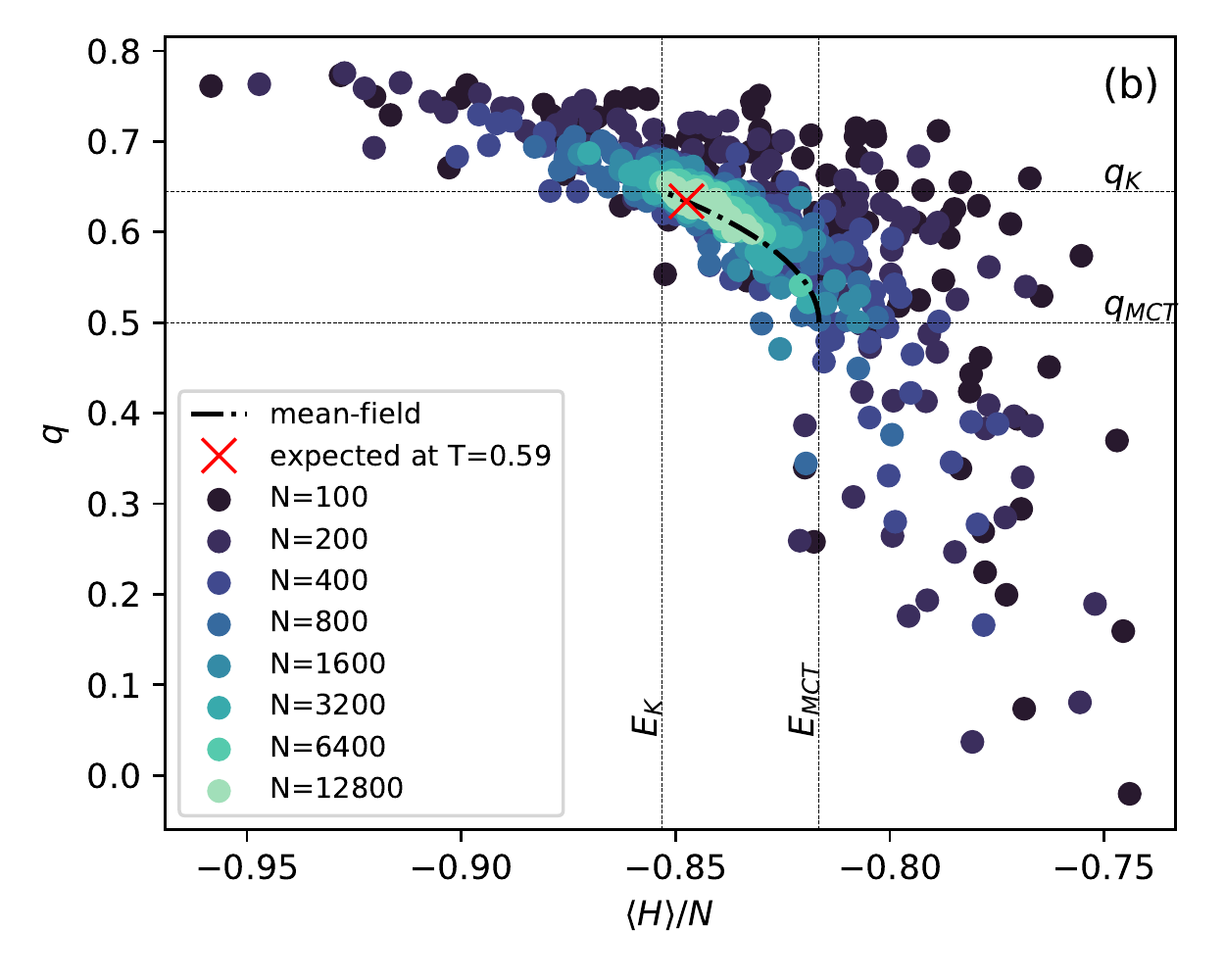}
	\caption{Equilibrium dynamics in the $3$-spin spherical model at $T=0.59 \lesssim \TMCT$, for  $N=100,200,400,800,1600,3200,6400,12800$. 
		\textbf{(a):} Time evolution of $C(t,0)=\sum_{i=1}^Ns_i(t)s_i(0)/N$.
		\textbf{(b):} Scatter plot of the equilibrium overlap vs equilibrium energy below $\TMCT$. Each point represents a different sample (and state) for a single equilibrium trajectory. The red cross gives the thermodynamic expectation. The dashed-dotted line corresponds to the typical overlap $\qEA$ (Eq.~\eqref{3spinEquil}) vs the typical energy $\langle H \rangle=-\beta/2$, at different temperatures between $T_\text{\tiny K}$ and $\TMCT$.}
		\label{fig:3spin-1}%\label{ENergyVSCorr}
\end{figure}

\begin{figure}[th!]
	\includegraphics[width=0.95\columnwidth]{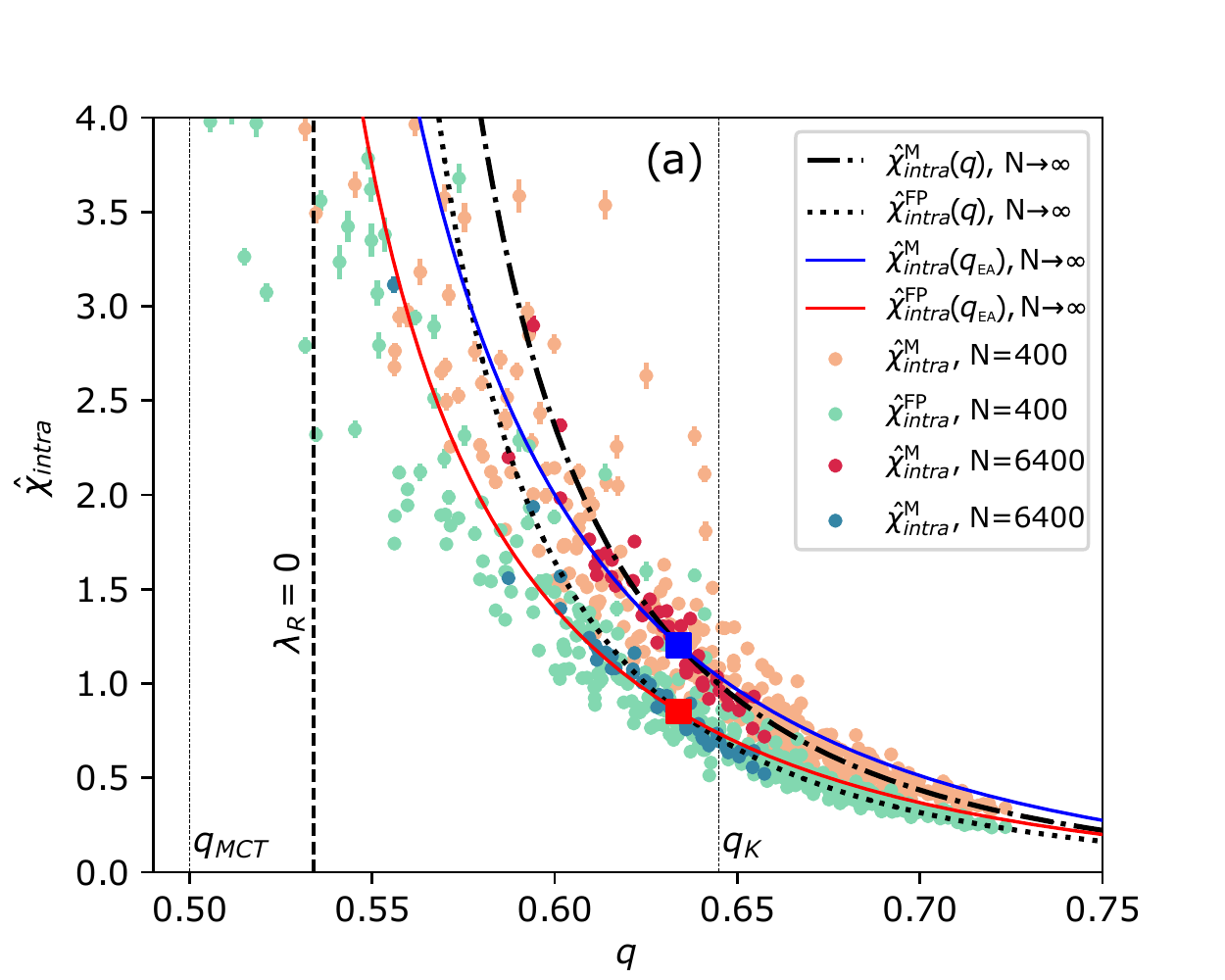}
	
	\includegraphics[width=0.95\columnwidth]{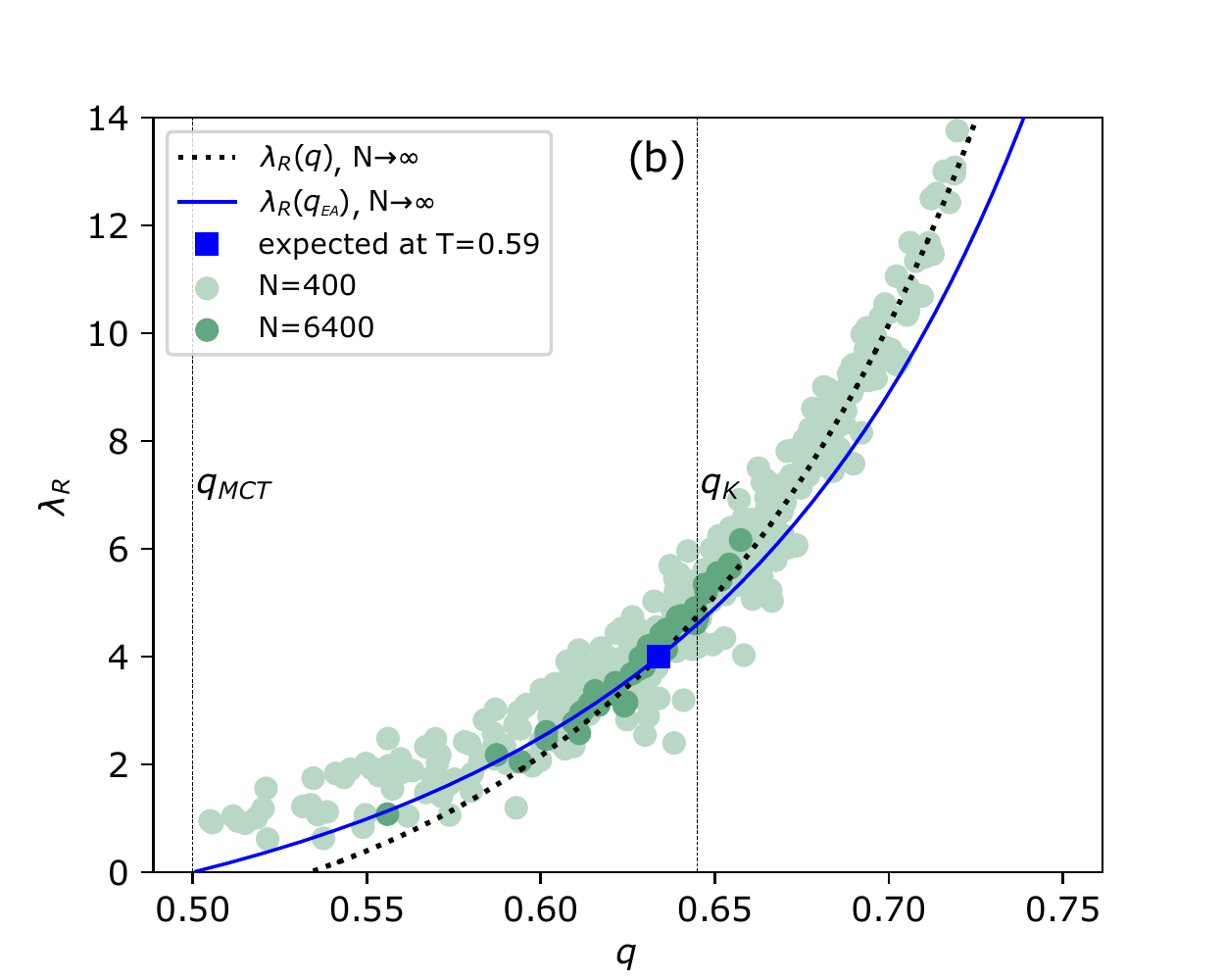}
	\caption{{\bf (a):} Scatter plot of the intra-state susceptibilities versus the overlap in the $3$-spin spherical model at $T=0.59$. On each sample the bar indicates the estimate of the error in evaluating the intra-state susceptibility. The dotted line represents $\bm{\hat{\chi}}_{intra}^\text{\tiny FP}(q)$ at fixed temperature $T=0.59$ (Eq.~\eqref{Chth}), and the dashed-dotted line represents $\bm{\hat{\chi}}_{intra}^\text{\tiny M}(q)$ (Eq.~\eqref{Chdyn}). The red and blue  squares give the typical value in the thermodynamic limit at that temperature, i.e., $\bm{\hat{\chi}}_{intra}^\text{\tiny FP}(\qEA)$ and $\bm{\hat{\chi}}_{intra}^\text{\tiny M}(\qEA)$, and the red and blue lines give these typical values at different temperatures. The samples seem to follow the atypical branch (dotted and  dashed-dotted lines) for overlaps greater than $\qEA$ and the typical branch (red and blues lines) for overlaps below $\qEA$.
	{\bf (b):}
	Same plot for the replicon eigenvalue $\lambda\text{\tiny R}(q)=2/[2\bm{\hat{\chi}}_{intra}^\text{\tiny FP}(q)-\bm{\hat{\chi}}_{intra}^\text{\tiny M}(q)]$, Eq.~\eqref{rep}, which measures how far a state is from breaking into RSB ($\lambda\text{\tiny R}=0$). The dotted line is $\lambda\text{\tiny R}(q)$ in the thermodynamic limit at that temperature. The blue line is the $\lambda\text{\tiny R}(\qEA)$ for different temperatures and it intersects the dynamical transition point $(0,q_\text{\tiny MCT}=0.5)$. Also in this case samples follow the atypical branch (dotted line) above $\qEA$ and the typical branch below (blue line).
    }\label{3Susc}
\end{figure}

\begin{figure}[t]
	\includegraphics[width=1\columnwidth]{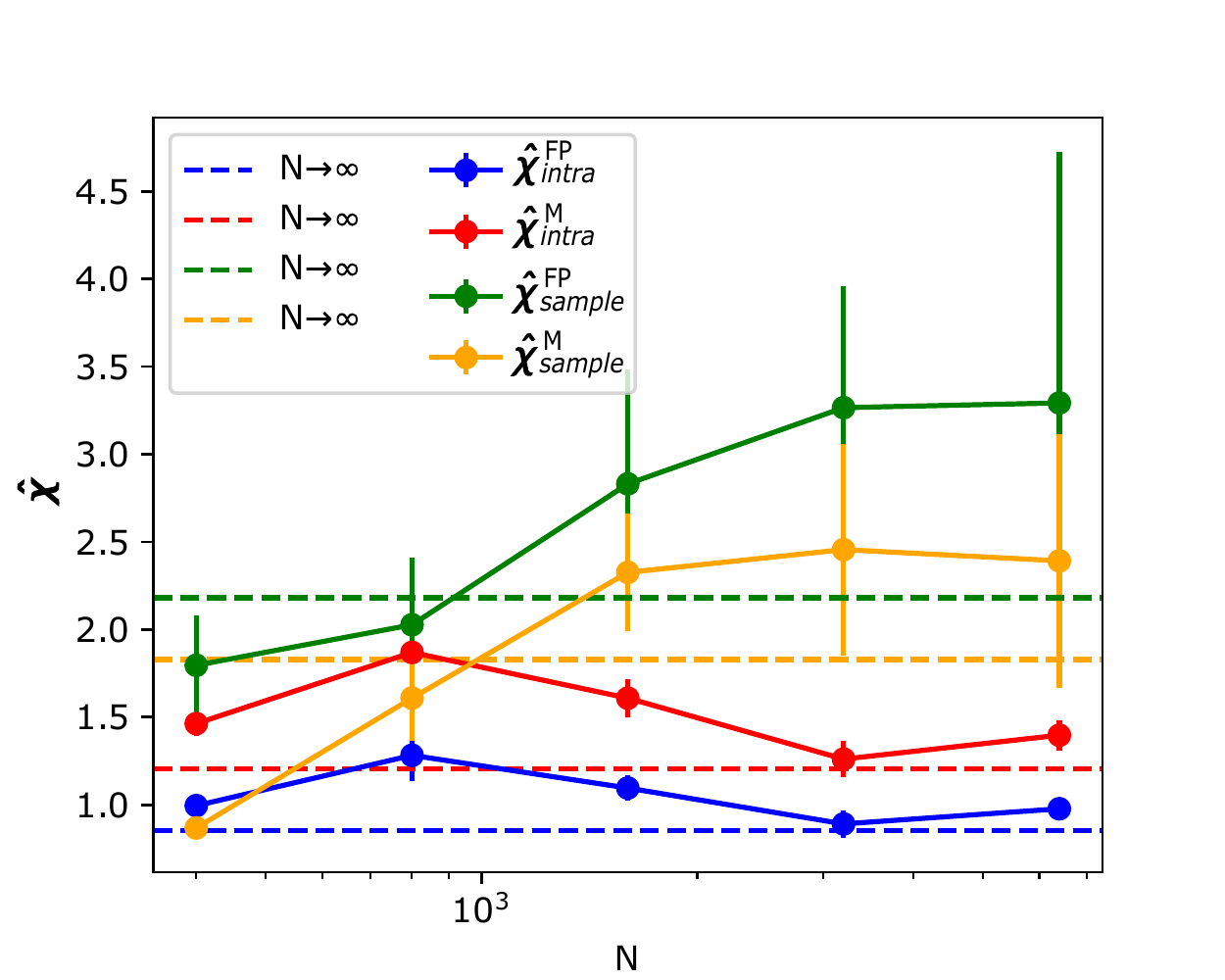}
	\caption{Size scaling of the intra-state susceptibilities $\bm{\hat{\chi}}_{intra}^\text{\tiny FP}=\overline{\hat{\chi}_{intra}^\text{\tiny FP}}$, $\bm{\hat{\chi}}_{intra}^\text{\tiny M}=\overline{\hat{\chi}_{intra}^\text{\tiny M}}$ and of the sample-to-sample susceptibilities $\bm{\hat{\chi}}_{sample}^\text{\tiny FP}$, $\bm{\hat{\chi}}_{sample}^\text{\tiny M}$ in the 3-spin spherical model at $T=0.59$. $N=400,800,1600,3200,6400$. The large error bars for the sample-to-sample susceptibility follow from the relatively small $N_\mathrm{sample}$ (see Table~\ref{table3spin}).
	}
	\label{fig:3spin-3}
\end{figure}

\subsection{3-spin Spherical Model (RFOT) Results}\label{3spinsec}
%This is a simple model that that presents RFOT in temperature. 
The RS solution for the 3-spin model is given by Eq.~\eqref{RSsaddle} for the RFOT phase ($n=1$),
\begin{equation}\label{3spinEquil}
\beta^2=\frac{\qEA}{(1-\qEA)f_3'(\qEA)} \ ,
\end{equation}
with $f_3(q)$ given in Eq.~\eqref{eq:fq2}.
Here again, this equation implicitly defines the typical overlap $\qEA(\beta)$, which graphically corresponds to the local minimum (saddle point) of the RS action (see Fig.~\ref{fig:3spin}). For each temperature, overlaps away from the saddle correspond to atypical states, but only if the corresponding replicon eigenvalue is positive, i.e., RS stable (see Eqs.~\eqref{lRpspin} and \eqref{eq:fq2}),
\begin{equation}\label{3spinReplicon}
m_1=\lambda_\text{\tiny R}>0\qquad \Longrightarrow \qquad \beta^2<\frac{1}{3(1-q)^2 q} \ .
\end{equation}
This condition is valid for both the Monasson and Franz-Parisi potentials and gives different low-$q$ limits of RS stability as shown in Fig.~\ref{fig:3spin}. Note that the M potential is a lower bound to the FP potential, because the constrained reference configuration of the latter is relaxed in the former.

In order to assess the validity of the analysis of RFOT susceptibilities, we consider the equilibrium dynamics of the 3-spin model for $T=0.59$ (see Appendix \ref{AppendixC2} for $T=0.6$ results), which is below the MCT temperature $T_\text{\tiny MCT} = \sqrt{3/8} \approx 0.612$ and above the Kauzmann temperature $T_\text{\tiny K} \approx 0.586$~\cite{folena_mixed_2020}. At the MCT transition, the overlap is $q_\text{\tiny MCT}=1/2$ (from $\lambda_\text{\tiny R}(\qEA)=0$). Figure~\ref{fig:3spin-1} shows the equilibrium time correlation with the reference (planted) configuration for different samples (quenched disorder and reference configuration) and different $N$. Upon increasing the system size, fluctuations decrease and the typical correlations concentrate towards the expected thermodynamic value $\qEA \approx 0.634$ at long times. We note that, in the considered time window, escape processes (instantons) from the initial state are very rare ($<1/100$) for systems larger than $N>800$.  To build intuition on how the equilibrium observables concentrate around the thermodynamic limit upon increasing $N$, Fig.~\ref{fig:3spin-1} presents a scatter plot of the average energy vs the average overlap with the reference configuration for the same conditions. %All the samples are at equilibrium at the same temperature $T=0.59$ and 
%The corresponding typical energy and overlap in the thermodynamic limit is shown by a red cross. 
It is interesting to observe that, for fixed $N$, different samples follow quite strictly the mean-field expectation at other temperatures (dashed-dotted line). Therefore the most frequently observed atypical samples are those that would be typical at another temperature. 
%This consideration suggests that the most suitable way to study the convergence to the thermodynamic limit of finite size systems is in terms of extensive observables rather than intensive one \cite{gross_microcanonical_2001}, thus it would be better to select samples based on their average energy rather then their preparation temperature.
%Rough definition of instantons: if $C(t,0)<\qEA/2$

 We next compare predictions for the equilibrium susceptibilities in the thermodynamic limit  (Sec.~\ref{Susc3spin}) with the numerical simulations at finite $N$ (following Sec.~\ref{numSusc}). A scatter plot (Fig.~\ref{3Susc}a) of the intra-state susceptibilities $\hat{\chi}_{intra}^\text{\tiny FP},\hat{\chi}_{intra}^\text{\tiny M}$ vs the intra-state overlap $q^{\text{\tiny M}}_{intra}$ for different samples shows that both intra-state susceptibilities concentrate around the theoretical expectation. Atypical samples have susceptibilities that follow the atypical value of the susceptibility given by Eqs.~\eqref{atypicalth} and~\eqref{atypicaldyn}. They therefore correspond to the RSMM away from the saddle along the Monasson potential (see Fig.~\ref{fig:3spin}). (Similarly, one can look at the RSMM along the Franz-Parisi potential, with almost identical results.) Both lines diverge when the replicon approaches zero at an overlap larger than $q_\text{\tiny MCT}$. %The blue and red line are show for comparison and correspond to the typical susceptibilities at different temperatures, i.e. $\hat\chi_{intra}^\text{\tiny M}^{(0)}(\qEA(T))$ and $\hat{\chi}^{(0)}_{intra}^\text{\tiny FP}(\qEA(T))$, where $\qEA(T)$ is the typical overlap at temperature $T$.
In order to further investigate the behavior of the susceptibilities at $T=0.59$ we use Eq.~\eqref{rep} and define the replicon eigenvalue associated to each sample. The corresponding scatter plot (Fig.~\ref{3Susc}b) shows that while for $q>\qEA$ the replicon follows the atypical line (dotted), for $q<\qEA$ the behavior is less clear.

To conclude this section, we consider the finite-size scaling of the intra-state susceptibilities averaged over all the samples and of the sample-to-sample susceptibilities $\bm{\chi}_{sample}^\text{\tiny FP},\bm{\chi}_{sample}^\text{\tiny M}$ introduced in Sec.~\ref{sampleSusc} (Fig.~\ref{fig:3spin-3}). Due to the small number of samples and the small time windows considered (see Table~\ref{table3spin}) the error bars for sample-to-sample susceptibilities are too large to provide a stringent test of the analytical results.

\section{Fluctuations in the Random Orthogonal Model}\label{rom}

%\subsection{Overlap Free Energy}
We next consider the random orthogonal model (ROM), which can be construed as a generalization of the SK model. The ROM Hamiltonian is
\begin{equation}
H = \sum_{ij}J_{ij}s_is_j \ , \qquad J_{ij} = [O^{T}DO]_{ij} \ ,
\end{equation}
for Ising spins, $s_{i} = \pm 1$, where $O$ is a random orthogonal matrix and $D$ is a diagonal matrix with entries that are sampled from a given distribution $P(d_{ii})$. Taking the Wigner semicircle law as distribution recovers the SK model, because the $J_{ij}$ couplings are then Gaussian distributed. %Thus ROM is a generalization of the SK model for general distribution of eigenvalues.

The motivation for considering this model is two-fold. First, the model has only two-spin interactions, and hence simulations have a computational complexity of $N^2$. Second, $P(d)$ can be tuned so as to obtain a strong RFOT model, with well-separated $\TMCT$ and $T_\text{\tiny K}$ transitions (see Sec.~\ref{sec:simuROM}). This feature is particularly useful because we have then a wide range of temperatures
$T_\text{\tiny K} < T < \TMCT$ at which the model can be equilibrated via quiet planting with arrested dynamics.
For comparison, the $p$-spin model offers either a fragile RFOT behavior with $T_\text{\tiny K} \approx \TMCT$ at $p=3$, or a broader regime at larger $p$, but at a markedly increased computational cost~\cite{parisi2004fragility}. Note that a prior analysis of the regime of small fluctuations for this model was made in Ref.~\cite{sarlat_predictive_2009-1}, but that work preceded the full appreciation of the role of intra-state and disordered susceptibilities.  

\subsection{Free Energy}

From Ref.~\cite[Eq.~(43)]{cherrier_role_2003}, we know that the ROM overlap free energy is
\begin{equation}\label{LL}
\begin{split}
S[\mQ{}{},\Lambda] & \equiv \frac{1}{2}\Tr G(\beta \mQ{}{}) -\frac{1}{2}\Tr \mQ{}{} \Lambda \\
&+\ln \Big ( \Tr_{s} e^{\frac{1}{2}\sum_{a,b} \Lambda_{ab}s_a s_b} \Big ) \ ,
\end{split}\end{equation}
where the trace is  over replica indexes, 
$\Tr \mQ{}{} = \sum^n_{a=1} q_{aa}$.
Evaluating the saddle point in $\mQ{}{}$ gives
\begin{equation}\label{Lambda}
\beta[G'(\beta \mQ{}{})]_{ab}=\Lambda_{ab} \ ,
\end{equation}
which provides the replicated free energy as a functional of the overlap matrix (see Ref.~\cite[Eq.~(51)]{cherrier_role_2003}),
\begin{equation}\label{SQ}
S[\mQ{}{}]  \equiv S^\text{\tiny I}[\mQ{}{}]+S^\text{\tiny II}[\mQ{}{}] \ ,
\end{equation}
where
\begin{equation}
 S^\text{\tiny I}[\mQ{}{}] = \frac{1}{2}\Tr G(\beta \mQ{}{}) - \frac{\beta}{2}\Tr \mQ{}{} G'(\beta \mQ{}{})
\end{equation}
and
\begin{equation}
S^\text{\tiny II}[\mQ{}{}] = \ln \Big [ \Tr_s \exp(\frac{\beta}{2} \sum_{a,b} [G'(\beta \mQ{}{})]_{ab}s_a s_b) \Big ] \ .
\end{equation}

We now focus on the RS ansatz, which corresponds to the overlap matrix $\mQ{}{RS} = (1-q)\mathbb{I}+q\mathbb{J}$, where $\mathbb{I}$ is the identity matrix and $\mathbb{J}$ is a matrix with all entries set to unity. 
Given an arbitrary function $f$, we then have
\begin{equation}\label{RS}
\begin{split}
f(\mQ{}{RS}) &= f(1-q)\mathbb{I} +  \big [ f(1-q+nq)-f(1-q) \big ]\frac{\mathbb{J}}n \\ &= \gamma_{f}\mathbb{I} + \lambda_{f}\mathbb{J} \ , \\
\Tr f(\mQ{}{RS} ) &=  n(\gamma_{f} + \lambda_{f}) \ .
\end{split}
\end{equation}
These identities will be particularly helpful in subsequent calculations. 

For example, using $f(q)=G(\beta q)$ and $g(q) = q G'(\beta q)$, 
the RS ansatz for the free energy gives 
\begin{equation}
n S^\text{\tiny I}[\mQ{}{RS} ] = \frac{1}{2} n[\gamma_{f} + \lambda_{f}] - \frac{\beta}{2} n[\gamma_{g} + \lambda_{g}] \ ,
\end{equation}
and, for $\ell(q) = \beta G'(\beta q)$,
\begin{equation}
\begin{aligned}
&n S^\text{\tiny II}[\mQ{}{RS} ] = \\
&=\ln \Big [ \exp(\frac{1}{2}\sum_{c,d}[\gamma_{\ell}\mathbb{I}+\lambda_{\ell}\mathbb{J}]_{cd}\partial_{h_c}\partial_{h_d})\times\\&\qquad\times \prod_a 2 \cosh(h_a) \Big |_{h_a=0}\Big]\\
&=\ln \Big\{  \exp\big (\frac{1}{2}\lambda_{\ell}\partial_{h}^2\big )  \big [ \exp(\frac{1}{2}\gamma_{\ell}) 2 \cosh(h) \big ]^n \Big |_{h=0}\Big\} \\
&=\frac{n}{2}\gamma_{\ell}+n\ln(2)+\ln \Big [ \int \frac{dz}{\sqrt{2\pi}} e^{-\frac{z^2}{2}} \cosh(\sqrt{\lambda_{\ell}}z)^n \Big] \ .
\end{aligned}
\end{equation}
The resulting total $S[\mQ{}{RS} ]$ is equivalent to that of Ref.~\cite[Eq.~(65)]{cherrier_role_2003}, except for an irrelevant constant.

\begin{figure}[t]
	\centering
	\includegraphics[width=1\columnwidth]{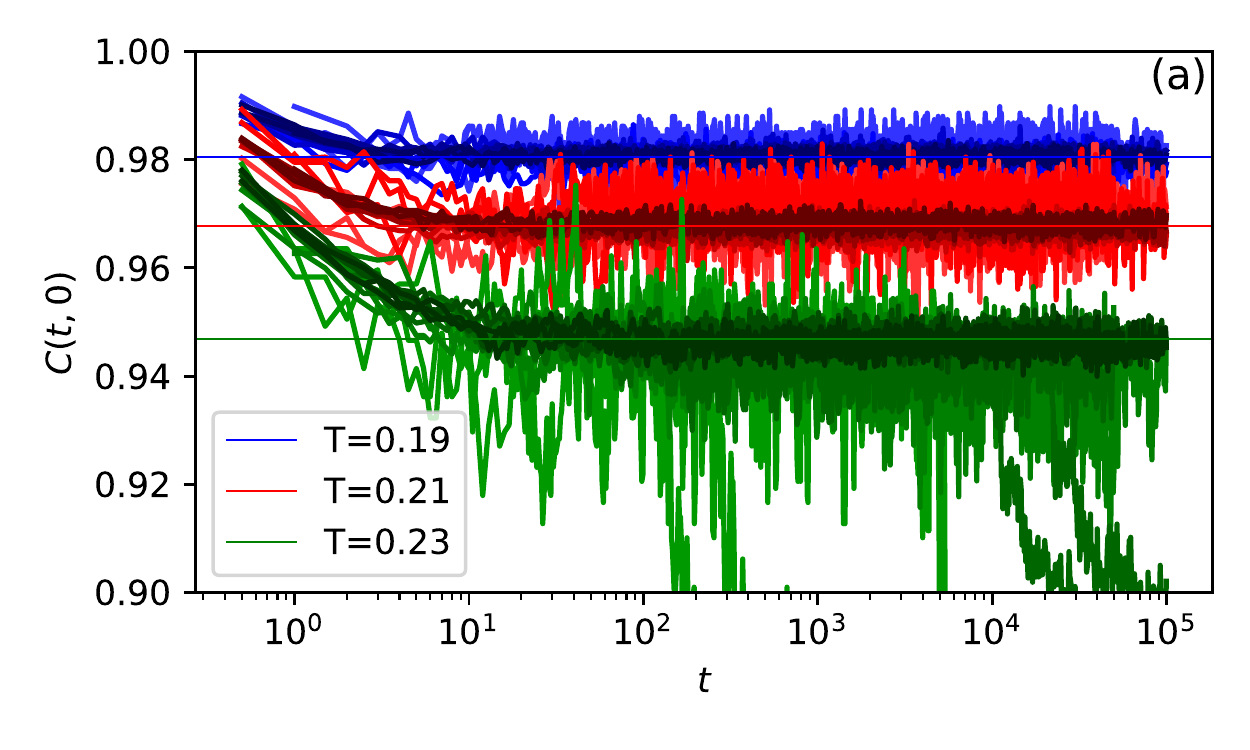}
	\includegraphics[width=1\columnwidth]{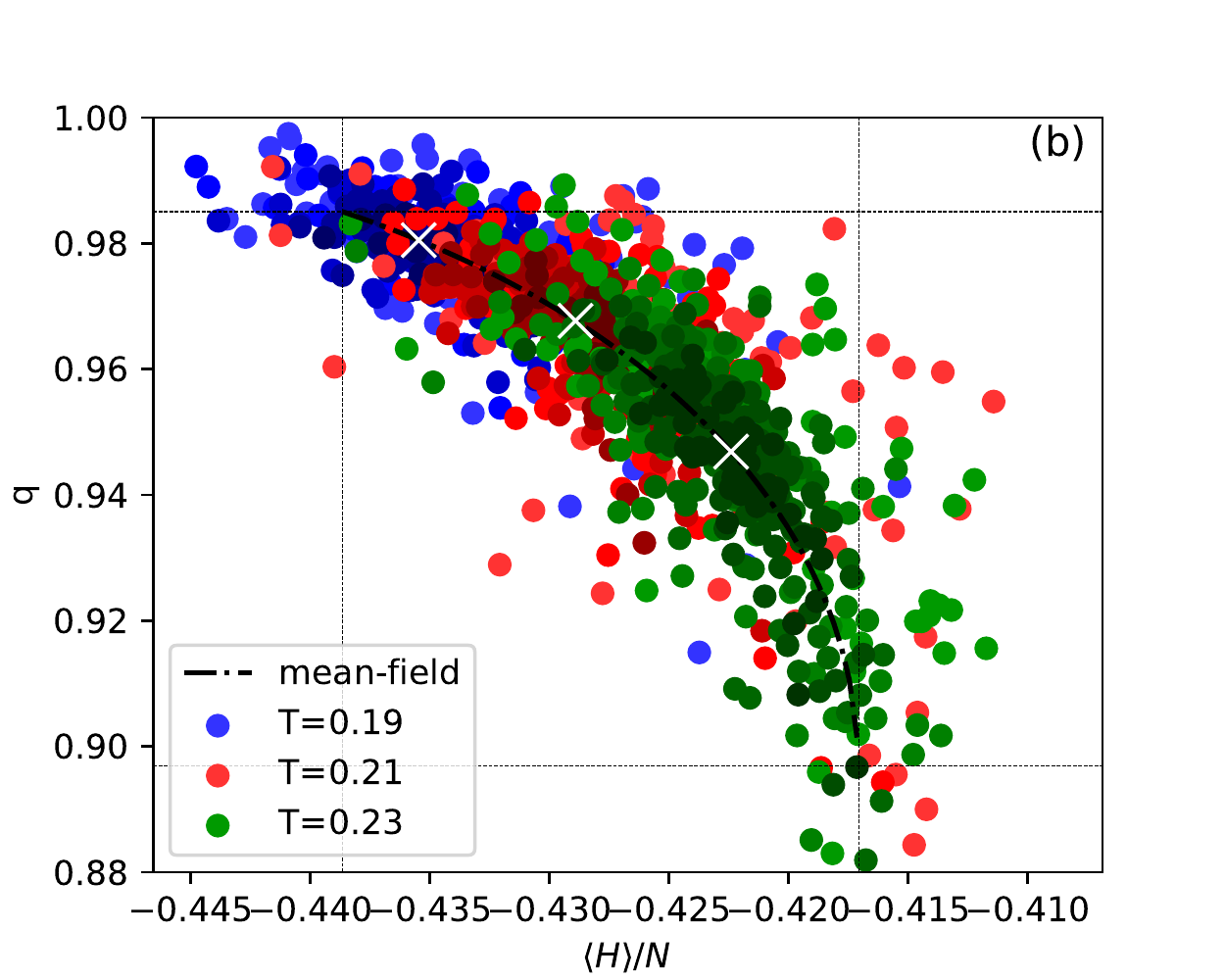}
	\caption{Equilibrium dynamics in the ROM with $\alpha=0.3$, for $T=0.19,0.21,0.23$ between $T_\text{\tiny K}$ and $\TMCT$, and $N=256,512,1024,2048,4096$. \textbf{(a):} Time evolution of the equilibrium correlation $C(t)$. Straight lines denote the thermodynamic overlap, $\qEA$. \textbf{(b):} Scatter plot of the overlap versus the energy for different samples at the same three temperatures. The black dashed-dotted line shows the mean-field average result, plotted parametrically with temperature; the white crosses indicate the specific values at these three temperatures.}\label{FI1}
\end{figure}

\begin{figure}[t]
	\centering
	\includegraphics[width=1\columnwidth]{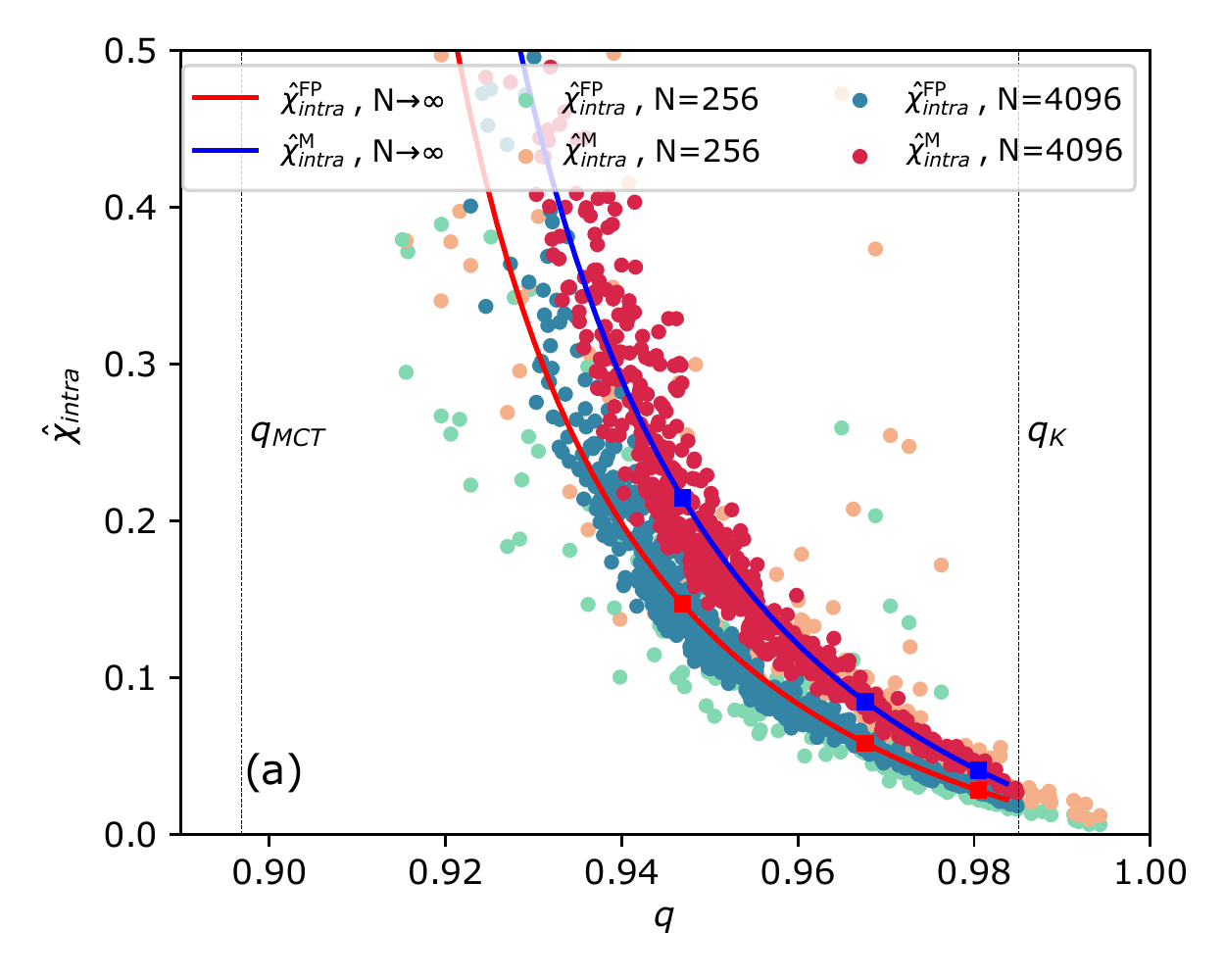}
	\includegraphics[width=1\columnwidth]{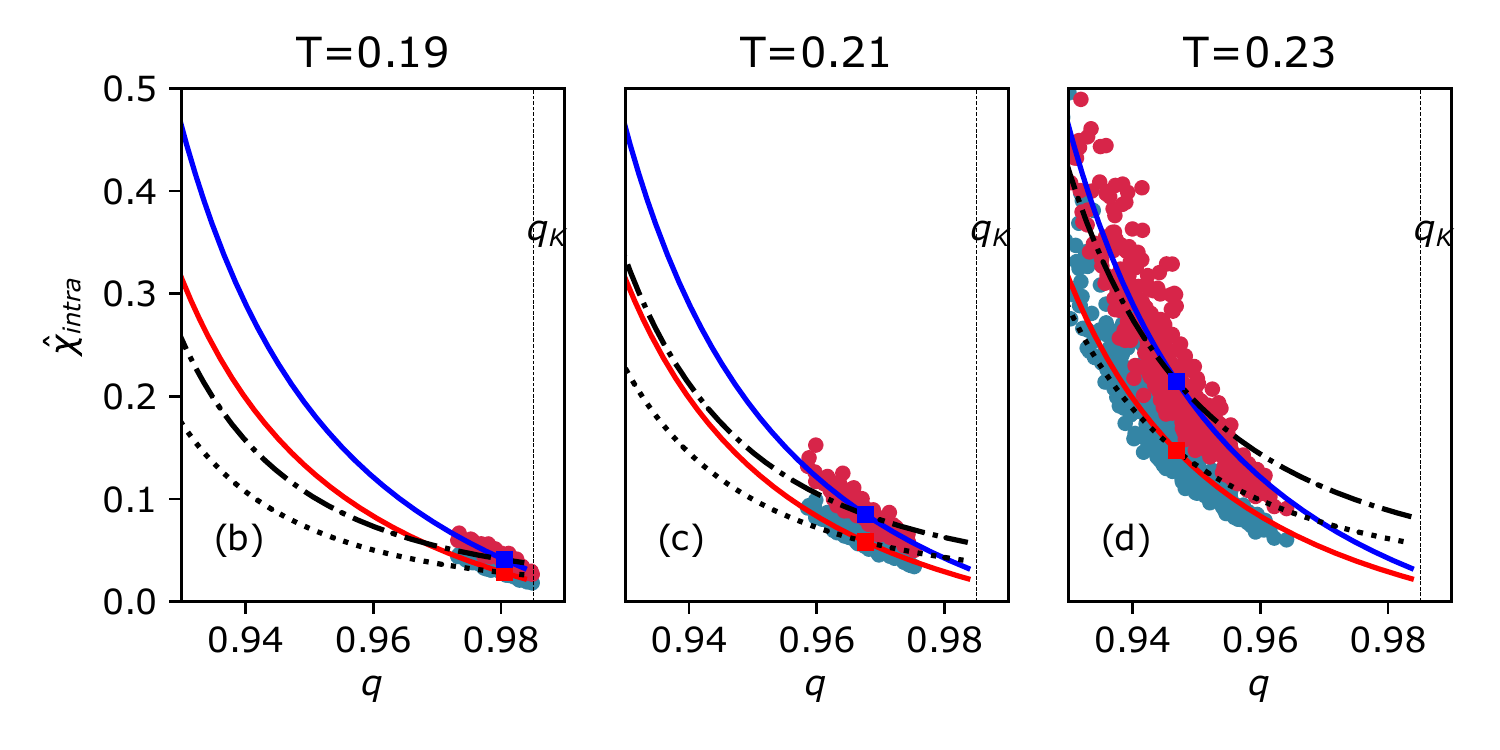}
\caption{\textbf{(a):} Scatter plot of intra-state susceptibilities in the ROM  with $\alpha=0.3$ for $T=0.19,0.21,0.23$ between $T_\text{\tiny K}$ and $T_\text{\tiny MCT}$. Each point denotes a different sample. The red and blue lines show the thermodynamic (or typical value) result for different temperatures. 
\textbf{(b),(c),(d):}
Same samples, now separated for different temperature $T=0.19,0.21,0.23$. Black lines (dotted and dashed-dotted) denote atypical branches of susceptibilities, while red and blue lines represent the typical fluctuations upon changing temperature. 
Unlike for the $p$-spin model (see  Fig.~\ref{3Susc}), ROM samples follow the typical branch of susceptibilities. The underlying reason behind the discrepancy between ROM and $p$-spin results is unclear.
}
\label{scatRom}
\end{figure}

\subsection{RS Mass Matrix and Susceptibilities}

We now repeat the calculation of fluctuations by evaluating the mass matrix $\partial_{q_{ab}}\partial_{q_{cd}} S[\mQ{}{}]$. In order to simplify the expressions, we consider the free energy with an explicit Lagrange multiplier $\Lambda$, as defined in Eq.~\eqref{LL}. The total mass matrix then reads
\begin{equation}\label{M}
\mathcal{M} = 
\begin{pmatrix}
\mathbbm{M} & -\mathbb{I}\\
-\mathbb{I} & \mathbbm{L}\\
\end{pmatrix}
= 
\begin{pmatrix}
\partial_q\partial_qS[\mQ{}{},\Lambda]  & \partial_q\partial_\lambda S[\mQ{}{},\Lambda]\\
\partial_\lambda\partial_q S[\mQ{}{},\Lambda] & \partial_\lambda\partial_\lambda S[\mQ{}{},\Lambda]
\end{pmatrix} \ .
\end{equation}
In order to evaluate the  RSMM %matrix $\mathbbm{M}$ matrix, 
we need to evaluate the matrix derivative of Eq.~\eqref{Lambda}. Following Ref.~\cite[Appendix]{marinari_replica_1994}, we first consider the derivative of a generic power $k$ of the overlap matrix and then infer the matrix derivative for a generic function $f(\mQ{}{})$, the sum of powers of $\mQ{}{}$. The details of the calculation are reported in Appendix \ref{appendixD}.
Given the result in Eq.~\eqref{fQ}, considering $f(\mQ{}{})=\partial_{q_{ab}}\frac{1}{2}\Tr G(\beta \mQ{}{}) = \frac{\beta}{2}G'(\beta \mQ{}{})$ with $n=1$, and setting $u=\beta$ and $d=\beta(1-q)$, we obtain the elements
\begin{equation}
\begin{aligned}
&\mathbbm{M}_{12,34} =  \frac{\beta^2}{2} \Big [ 2 \Psi_\beta(q) -4\Phi_\beta(q)  \Big ]\ ,\\
&\mathbbm{M}_{12,13} =  \frac{\beta^2}{2} \Big [ 2 \Psi_\beta(q) -3\Phi_\beta(q)  \Big ]\ ,\\
&\mathbbm{M}_{12,12} =  \frac{\beta^2}{2} \Big [ 2 \Psi_\beta(q) -2\Phi_\beta(q)  \Big ]\ ,
\end{aligned}
\end{equation}
where the $\Psi_\beta(q) = G''(\beta) + G''(\beta(1-q))$ and $\Phi_\beta(q) = G'(\beta)/(\beta q)-G'(\beta(1-q))/(\beta q)$.
Using the change of parametrization in Eq.~\eqref{EqM}, the previous equations can be further recast as
\begin{equation}
\begin{aligned}
m_1 & = \beta^2 G''(\beta(1-q)) \ , \\
m_2 
%&=  2\beta^2\Big [ \frac{G'(\beta)-G'(\beta(1-q))}{\beta-\beta(1-q)}-G''(\beta(1-q))\Big ] \\
&=  - 2\beta^2G''(\beta(1-q))+2\frac{\lambda^*}{q} \ , \\
m_3 
%&= \beta^2 \Big [G''(\beta) + G''(\beta(1-q))-2\frac{G'(\beta)-G'(\beta(1-q))}{\beta-\beta(1-q)} \Big]\\
&=  \beta^2 \Big [G''(\beta) + G''(\beta(1-q))\Big ]-2\frac{\lambda^*}{q}\ ,
\end{aligned}
\end{equation}
where $\lambda^* = \lambda_{\ell}$ is  the saddle point value of
the RS ansatz for $\Lambda_{ab}=-\lambda\delta_{ab}+\lambda$, see Eqs.~\eqref{Lambda} and~\eqref{RS}.
We now turn to the term $\mathbbm{L}=\partial_{\lambda}\partial_{\lambda}S[\mQ{}{},\Lambda]$. The mass matrix $\mathbbm{L}$ is evaluated along the same lines as for the SK model, see, for example, Ref.~\cite[Sec.~3.1.1]{nishimori_statistical_2001}. We thus here only briefly review the derivation. Given that
\begin{equation}
\Tr_s e^{\frac{1}{2} \sum_{ab}^n \Lambda_{ab}s_a s_b} = \Big [ e^{\frac{1}{2}  \sum_{ab} \Lambda_{ab}\partial_{h_a} \partial_{h_b}} \prod_c 2\cosh(h_c) \Big ]_{h_*=0}\ ,
\end{equation}
inserting the RS ansatz we have
\begin{equation}
\begin{aligned}
\ln \Big [ &e^{\frac{1}{2}  \sum_{ab} \Lambda_{ab}\partial_{h_a} \partial_{h_b}} \prod_c 2\cosh(h_c)\Big ]_{h_*=0} =\\
&= \ln \Big [e^{\frac{\lambda}{2} \partial_{h}^2} \big (e^{-\frac{\lambda}{2} \partial_{h}^2} 2 \cosh(h)\big )^n \Big ]_{h=0} \\
&= n (2-\lambda)+\ln \Big [e^{\frac{\lambda}{2} \partial_{h}^2} \cosh(h)^n \Big ]_{h=0} \ .
\end{aligned}
\end{equation}
The first derivative then reads
\begin{equation}
\begin{aligned}
\partial_{\lambda_{ab}} &\Tr_s e^{\frac{1}{2} \sum_{cd} \Lambda_{cd}s_c s_d} =\\
&= \frac{  e^{\frac{\lambda}{2} \partial_{h}^2} \Big [\big ( \frac{1}{2}\partial_{h_a} \partial_{h_b}+\frac{1}{2}\partial_{h_b} \partial_{h_a} \big )  \prod_c \cosh(h_c) \Big ]_{h_*=h}}{e^{\frac{\lambda}{2} \partial_{h}^2} \cosh(h)^n }\Big|_{h=0}\\
 &= \frac{ e^{\frac{\lambda}{2} \partial_{h}^2} \sinh(h)^2 \cosh(h)^{n-2}}{e^{\frac{\lambda}{2} \partial_{h}^2} \cosh(h)^n }\Big|_{h=0} \equiv \langle\tanh^2 \rangle \ ,
\end{aligned}
\end{equation}
where we have defined the average
\begin{equation}
\begin{aligned}
\langle A \rangle &= \frac{ e^{\frac{\lambda}{2} \partial_{h}^2} A(h) \cosh(h)^n}{e^{\frac{\lambda}{2} \partial_{h}^2} \cosh(h)^n }\Big|_{h=0} \\
&= \frac{\int \frac{dz}{\sqrt{2\pi\lambda}} e^{-\frac{z^2}{2\lambda}} A(z) \cosh(z)^n}{\int \frac{dz}{\sqrt{2\pi\lambda}} e^{-\frac{z^2}{2\lambda}} \cosh(z)^n} \ .
\end{aligned}
\end{equation}
Finally, the mass matrix is
\begin{equation}
\begin{aligned}
\mathbbm{L}_{ab,cd} &=   \langle\tanh^4 \rangle -  \langle\tanh^2 \rangle^2 \ , \\
\mathbbm{L}_{ab,ad} &=   \langle\tanh^2 \rangle -  \langle\tanh^2 \rangle^2 \ , \\
\mathbbm{L}_{ab,ab} &=  1 -  \langle\tanh^2 \rangle^2 \ ,
\end{aligned}
\end{equation}
and in the other parametrization
\begin{equation}
\begin{aligned}
\lambda_1 &=  2 \langle[1-\tanh^2]^2 \rangle \ , \\
\lambda_2 &=  4 \langle [1-\tanh^2]\tanh^2 \rangle \ ,\\
\lambda_3 &=  \langle\tanh^4 \rangle -  \langle\tanh^2 \rangle^2\ .
\end{aligned}
\end{equation}
We thus have all the terms that define the RSMM in Eq.~\eqref{M}.

To obtain the fluctuations of the overlap values, we need to invert $\mathcal{M}$. For convenience, we define the generic inverse $\mathcal{M}\mathcal{G} = \mathbb{I}$, %of $\mathcal{M}$:
\begin{equation}
\mathcal{M} = 
\begin{pmatrix}
\mathbbm{M} & -\mathbb{I}\\
-\mathbb{I} & \mathbbm{L}\\
\end{pmatrix} \ , \qquad
\mathcal{G} = 
\begin{pmatrix}
\mathbbm{G} & \mathbbm{A}\\
\mathbbm{A} & \mathbbm{H}\\
\end{pmatrix} \ ,
\end{equation}
where $\mathbbm{A}$ has the same symmetry between replica indices as $\mathbbm{G}$ and $\mathbbm{H}$.
%From the definition of inverse $\mathcal{M}\mathcal{G} = \mathbb{I}$, 
We then obtain
\begin{equation}\label{invA}
\begin{aligned}
 \mathbbm{M}  \mathbbm{G} &= \mathbb{I}+ \mathbbm{A} \ , \qquad
 \mathbbm{M}  \mathbbm{A} =  \mathbbm{H} \ , \\
 \mathbbm{L} \mathbbm{A} &= \mathbbm{G} \ , \qquad\quad
 \mathbbm{L} \mathbbm{H}  = \mathbb{I}+\mathbbm{A} \ , \\
\end{aligned}
\end{equation}
but the second and third expressions are equivalent. Using the RS ansatz, and considering that each of these matrices then has the form of Eq.~\eqref{RSMM}, we can use the inversion formula
in Eq.~\eqref{eq:genericinverse} to obtain for the RFOT ($n=1$) case
\begin{equation}
\begin{aligned}
g_1&=\frac{\lambda_1}{\lambda_1 m_1-1} \ , \\
g_2&=\frac{2 \lambda_1}{\lambda_1 m_1-1}-\frac{4 (2\lambda_1- \lambda_2)}{(2\lambda_1 - \lambda_2) (2 m_1- m_2)-4} \ , \\
g_3&=\frac{4 (2 \lambda_1-\lambda_2)}{(2 \lambda_1-\lambda_2) (2 m_1-m_2)-4} \\
		&-\frac{\lambda_1 [\lambda_1 (2 m_1+m_2+m_3)-2]+\lambda_2+\lambda_3}{(\lambda_1 m_1-1)^2}\ .
\end{aligned}
\end{equation}
Therefore, given $m_1,m_2,m_3,\lambda_1,\lambda_2,\lambda_3$, we can evaluate $g_1,g_2,g_3$ and thus calculate the various susceptibilities defined in Sec.~\ref{secdyn} and Sec.~\ref{secth} as
\begin{equation}\label{CHI}
\begin{aligned}
\bm{\hat{\chi}}_{intra}^\text{\tiny FP} &= 2g_1+g_2 \ , \qquad
\bm{\hat{\chi}}_{intra}^\text{\tiny M} = 2g_1+2g_2 \ , \\
\bm{\hat{\chi}}_{sample}^\text{\tiny FP} &= g_2+4g_3 \ , \qquad
\bm{\hat{\chi}}_{sample}^\text{\tiny M} = 4g_3 \ . \\
\end{aligned}
\end{equation}

\begin{figure}[t]
	\centering
	\includegraphics[width=1\columnwidth]{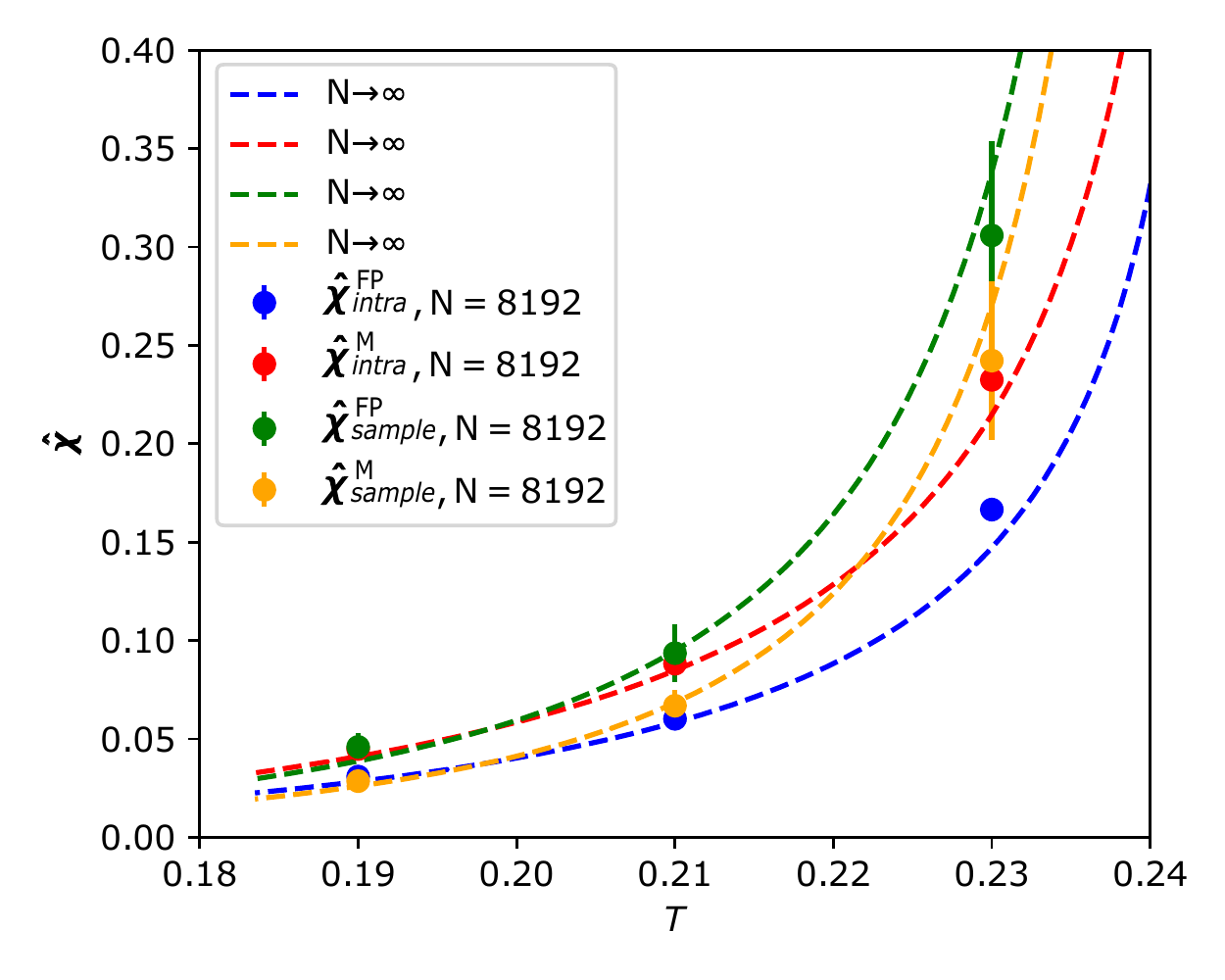}
	\caption{Average susceptibilities for the ROM with $N=8192$ compared with the analytical results for the thermodynamic limit. Note that $N_\mathrm{sample}=100$ is too small to display reasonable error bars for $T=0.23$.  }\label{FI2}
\end{figure}

\subsection{Simulations Details}
\label{sec:simuROM}

\begin{table}[b]
	\centering
	\begin{tabular}{|c|c|c|c|c|c|c|}
		%\multicolumn{6}{c}{ROM; } \\
		\hline 
		$N$ & 256 & 512 & 1024 & 2048 & 4096 & 8192\\ 
		\hline 
		$N_\mathrm{sample}$ & 100 & 100 & 100 & 100 & 100 & 100\\ 
		\hline 
		$K$ & 500 & 500 & 500 & 500 & 500 & 500\\ 
		\hline 
	\end{tabular} 	
	\caption{Simulation parameters for the ROM at $T=0.19,0.21,0.23,0.24$ with $\tau_{\kappa}=20$.}\label{tableROM}
\end{table}

Numerical results for the ROM are obtained using the simulation scheme of Ref.~\cite{cherrier_role_2003}. The diagonal entries of the diagonal matrix $D$ are extracted according to the bimodal distribution
\begin{equation}
P(d) = \alpha \delta(d-1)+(1-\alpha) \delta(d+1),
\end{equation} 
where $\alpha$ specifies the percentage of positive eigenvalues of $D$. We here consider $\alpha = 0.3$, for which a RFOT phase with $T_\text{\tiny K} = 0.1803$ and $\TMCT = 0.2465$---with corresponding overlaps $q_\text{\tiny K} = 0.985$ and $q_\text{\tiny MCT} = 0.897$---is obtained.
An implementation of the planting method (which is based on the coincidence of the quenched and annealed averages above $T_\text{\tiny K}$) to produce equilibrated initial configurations is obtained by annealing both couplings and spins simultaneously. Table~\ref{tableROM} reports the simulation parameters. Recall that $\tau_\kappa$ is the time between sampled equilibrium configurations and $K$ the number of them.

\subsection{ROM Results}
Figure~\ref{FI1} shows the results for the time-dependent correlation function. The absence of aging, in particular, suggests that proper equilibrium is achieved. In addition, the sample-to-sample fluctuations of the overlap versus the energy are properly scattered along the average equilibrium curve (obtained analytically), parametrically in temperature.
 Interestingly, the sample-to-sample fluctuations of the susceptibilities 
(Fig.~\ref{scatRom}) here
follow a different pattern from those of the 3-spin (Fig.~\ref{3Susc}). For the ROM, they correspond to typical local susceptibilities at different temperatures, i.e., $\bm{\hat{\chi}}_{intra}^\text{\tiny FP}(\qEA)$ and $\bm{\hat{\chi}}_{intra}^\text{\tiny M}(\qEA)$, while for the 3-spin spherical model they follow atypical susceptibilities $\hat{\chi}_{intra}^\text{\tiny FP}(q)$ at the equilibrium temperature. The origin of this difference is not well understood, but might follow from the difference in equilibration scheme. (The 3-spin spherical model couplings are extracted around the planted configuration, while in the ROM they are relaxed together with the configuration.)
Figure~\ref{FI2} shows that measured local susceptibilities are consistent with the thermodynamic results.

\section{Fluctuations in the Random Lorentz Gas}\label{rlg}

We finally consider the random Lorentz gas (RLG), which is the simplest off-lattice model to exhibit a discontinuous MCT localization transition 
in the limit ${d\rightarrow\infty}$~\cite{biroli_unifying_2020,biroli_local_2021}. 
A possible construction of the system consists in planting a tracer at the origin due to the global translational invariance~\cite{biroli_unifying_2020},
and then dropping $N$ non-interacting obstacles
independently at random with probability
\begin{equation}
P(R_i) \propto e^{-\beta V(R_i)} \ ,
\end{equation}
with a radial potential $V(r)$ where $r$ is the distance from the origin.
The tracer particle is thus at equilibrium within a sea of other particles 
(obstacles) at a given inverse temperature $\beta$, see~\cite{biroli_unifying_2020,biroli_local_2021} for details.

\subsection{Free Energy}

The free energy of the tracer particle evolving in position $x$ within a cage defined by these $N$ obstacles is simply
\begin{equation}
F_\text{\tiny R} =  \ln\left[ \int dx e^{-\beta\sum_{i=1}^{N} V(x-R_i)} \right] \ ,
\end{equation} 
where the subscript $R$ underlines the dependence of the result on the distribution of the $N$ obstacles. 
The free energy is then averaged over all possible $N$ obstacle configurations,
\begin{equation}
F = \overline{F_\text{\tiny R}} \equiv \frac{\int dR e^{-\beta \sum_{i=1}^{N}V(R_i)} \ln[ \int dx e^{-\beta\sum_{i=1}^{N} V(x-R_i)} ]}{\int dR e^{-\beta \sum_{i=1}^{N}V(R_i)} } 
\ ,
\end{equation}
where the overline denotes an average over samples, hence over realizations of obstacle positions.
To evaluate this logarithm, we use the replica method, which requires obtaining the $n$th power of the partition function
\begin{equation}
Z^{n} = \int d^nx e^{-\beta\sum_{i=1}^{N} \sum_{a=1}^nV(x^{a}-R_i)}.
\end{equation}	
The average of this quantity 
 can be rewritten as 
\begin{equation}
\begin{split}
\overline{Z^n} &=\lim_{N\to\infty}\frac{\int dR e^{-\beta \sum_{i=1}^{N}V(R_i)} \int d^nx e^{-\beta\sum_{i=1}^{N} \sum_{a=1}^nV(x^{a}-R_i)} }{\int dR e^{-\beta \sum_{i=1}^{N}V(R_i)} } \\
&=  \lim_{N\to\infty} \int d^nx \Big [\frac{\int dR e^{-\beta V(R)}  e^{-\beta \sum_{a=1}^nV(x^{a}-R)} }{\int dR e^{-\beta V(R)} }\Big ]^N\ ,
\end{split}
\end{equation}
where
the second equality follows from the quenched particle positions being independent.
We now observe that because $V(R)$ is short-ranged,
$\exp[-\beta V(R)]$ is equal to one almost everywhere except around the origin, hence $\int dR \exp[-\beta V(R)]\approx \int dR= V$ is divergent.
One can then write
\begin{equation}\begin{split}
&\frac{\int dR e^{-\beta V(R)}  e^{-\beta \sum_{a=1}^nV(x^{a}-R)} }{\int dR e^{-\beta V(R)} } \\
&=\frac{\int dR [e^{-\beta V(R)}  e^{-\beta \sum_{a=1}^nV(x^{a}-R)}-1] + \int dR }{\int dR [ e^{-\beta V(R)} -1] + \int dR} \\
&\approx 1 - \frac{ \int dR [ e^{-\beta V(R)} -1] }{V} \\
&\quad + \frac{ \int dR [e^{-\beta V(R)}  e^{-\beta \sum_{a=1}^nV(x^{a}-R)}-1] }{V} + O(V^{-2}) \ ,
\end{split}
\end{equation}
which only holds in the glass phase, in which the $n$ copies of the
original tracer are all close to the origin, and the function $e^{-\beta \sum_{a=1}^nV(x^{a}-R)}$ also differs from one only
in the vicinity of the origin.
Taking the $N$th power for $N\to\infty$ at constant obstacle density $\rho = N/V$ then 
gives
\begin{equation}\label{Fre}
\begin{split}
&\overline{Z^n} = \exp\{-\rho \int dR [e^{-\beta V(R)} -1]\} \\ &\times \int d^nx \exp \Big \{\rho  \int dR [e^{-\beta V(R)}e^{-\beta \sum_{a=1}^nV(x^{a}-R)} - 1]\Big \}\ .
\end{split}
\end{equation}

\begin{figure}[t]
	\centering
	\includegraphics[width=1\columnwidth]{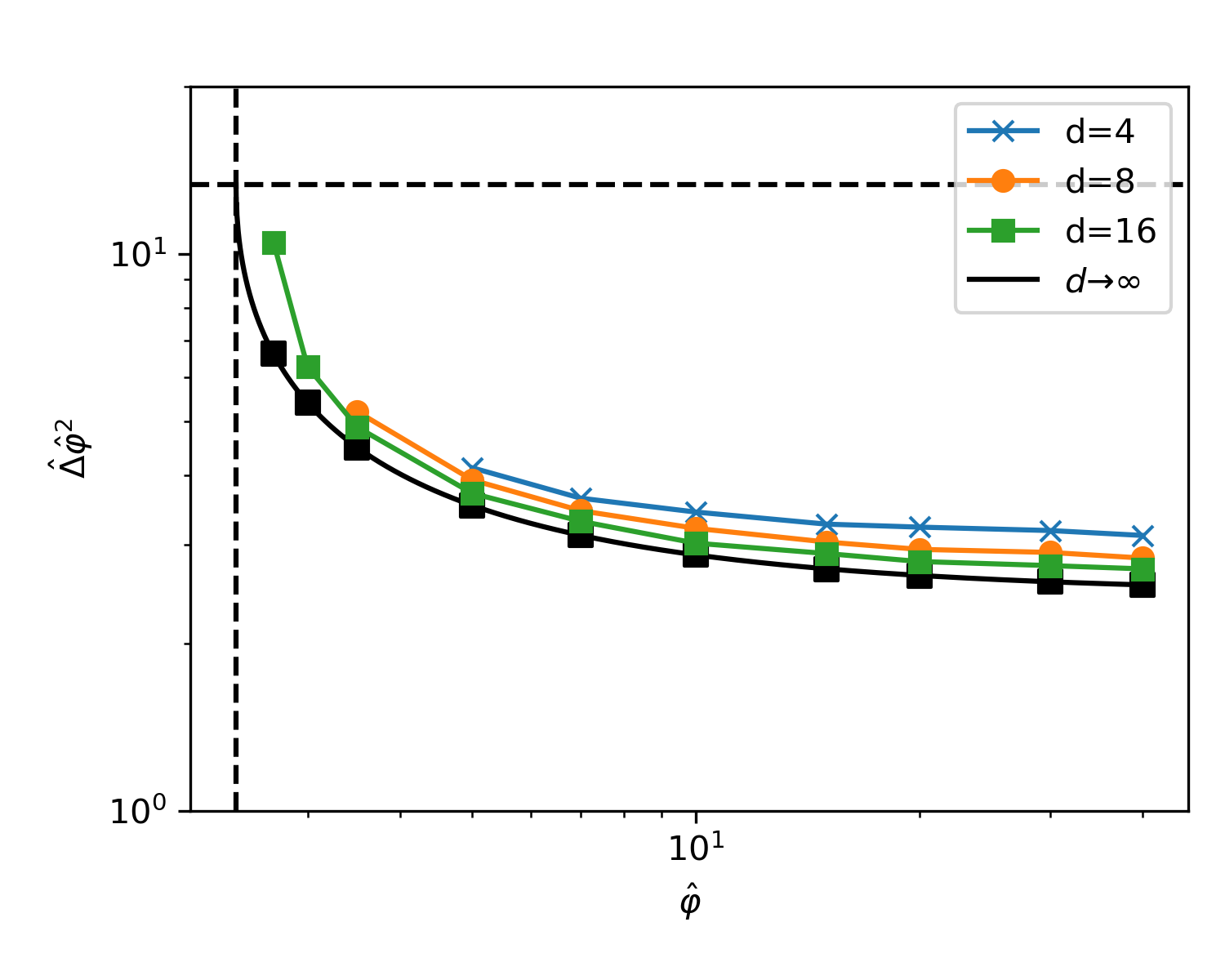}
	\caption{Rescaled equilibrium mean squared displacement (cage size) vs density for $d=4,8,16$. Each point on a line corresponds to a different scaled density $\hat{\varphi}=2.7,3,3.5,5,7,10,15,20,30,40$.  The analytical $d\to\infty$ result is shown in black. Dashed lines mark the MCT transition $\hat{\varphi}_{\MCT}$ and $\hat{\Delta}_{\MCT}$.}
	\label{fig:RLGmean}
\end{figure}

Because the pair potential $V(r)$ depends only on the distance between the planted particle and a given obstacle, it is symmetric under rotation. We can thus recast all integrals as purely radial expressions. In the limit of large dimension $d\to\infty$ all integrals further concentrate on a thin sphere of width $\frac{1}{d}$ around the optimal value, $\ell$, defined by the interaction radius of the potential~\cite{biroli_unifying_2020}. 
We thus change variables to $R=\ell(1+h/d)$ with rescaled variable $h$
and potential $\bar{v}(h) = V[\ell(1+\frac{h}{d})]$,  
\begin{equation}
\begin{aligned}
&\rho \int dR [e^{-\beta V(R)} -1] = \rho \Omega_d \int dr r^{d-1} [e^{-\beta V(r)} -1] \\
&\approx_{d\to\infty} d \hat\varphi \int dh e^{h} [e^{-\beta \bar{v}(h)} -1] \ ,
\end{aligned}
\end{equation}
$V_d=\Omega_d/d$ being the volume of
$d$-dimensional unit sphere
and $\hat\varphi=\rho V_d \ell^d/d$ being a scaled packing fraction that remains finite at the glass transition when $d\to\infty$~\cite{biroli_unifying_2020}. We see that $\overline{Z^n} \sim \exp(d)$, hence $d$ plays the role of the large parameter
in the saddle-point analysis.
In order to treat the integration over the $n$-times replicated tracer $d^n x$,
one can then follow the derivation of Ref.~\cite{parisi_theory_2020}, but with small adjustments. 
We do not reproduce the derivation here, but only give the list of adjustments and the final result.
First, \cite[Eq.~(4.46)]{parisi_theory_2020} is modified to
\begin{equation}
\langle y_a \rangle = \langle |x_a|^2 \rangle  = \alpha_{aa} \frac{\ell^2}{d} \ ,
\end{equation} 
and the fluctuations \cite[Eq.~(4.47)]{parisi_theory_2020} become
\begin{equation}
\langle y_a y_b \rangle - \langle y_a \rangle\langle y_b \rangle = \langle |x_a|^2 \rangle  = \alpha_{ab} \frac{\ell^2}{d^2} \ .
\end{equation} 
The replicated Mayer function, \cite[Eq.~(4.52)]{parisi_theory_2020}, 
also becomes
\begin{equation}\begin{split}
\overline{f}_{\text{eff}}(h) &= e^{-\beta \bar{v}(h)}e^{\sum_{a,b=1}^{n}\frac{\alpha_{ab}}{2}\partial_{h_a}\partial_{h_b}}\times \\ &\times[e^{-\beta\sum_{c=1}^n\bar{v}(h_c+\frac{\alpha_{cc}}{2})}]_{h_c=h}-1 \ .
\end{split}\end{equation}
The final result for the replicated energy, \cite[Eq.~(4.56)]{parisi_theory_2020}, takes the form 
\begin{equation}
\begin{aligned}\label{Free}
&F[\bbDelta]=\frac{d}{2}\ln[2 \det(-\bbDelta/2)(-\underline{1}^T\bbDelta^{-1}\underline{1})]\\
&+\frac{\hat{\varphi}}{2}\int dh e^h \{ e^{\sum_{a,b=1}^{n}-\frac{\Delta_{ab}}{4}\partial_{h_a}\partial_{h_b}}[e^{-\beta\sum_{c=1}^n\bar{v}(h_c)}-1] \}_{h_c=h} \ ,
\end{aligned}
\end{equation}
where $\bbDelta=(\Delta_{ab})\in\mathbbm{R}^{n\times n}$ is the matrix of squared displacements between replicas, which replaces the overlap matrix for spins.
The RLG free energy is then equivalent to that of an infinite-dimensional many-body (MB) system of particles~\cite{biroli_unifying_2020}, given the correspondence 
\begin{equation}
\frac{\hat{\varphi}_{\text{\tiny MB}}}{2}
\leftrightarrow \hat{\varphi}_{\text{\tiny RLG}} 
 \ , 
\qquad 
\frac{\Delta_{\text{\tiny MB}}}{2} \leftrightarrow \Delta_{\text{\tiny RLG}} \ .
\end{equation}
The equilibrium results for the mean squared displacement versus density, for a hard sphere potential $\bar{v}(h)=\infty$ for $h<0$ and zero otherwise,
are shown in Fig.~\ref{fig:RLGmean}.

\subsection{RSMM and Susceptibilities}

We now consider the general structure of the fluctuations around the RS solution,
\begin{equation}
\begin{aligned}
&\sum_{a\neq b,c\neq d}\mathbbm{M}^\text{\tiny RS}_{ab;cd}\delta \Delta_{ab}\delta\Delta_{cd}\\
&=m_{1}\sum_{a \neq b} \delta \Delta_{ab}^2 +  m_{2}\sum_{a  \neq  b  \neq  c}\delta \Delta_{ab}\delta \Delta_{bc} \\ & + m_{3}\sum_{a  \neq  b, c  \neq  d}\delta \Delta_{ab}\delta \Delta_{cd} \ .
\end{aligned}
\end{equation}
The RS saddle point corresponds to a matrix of squared displacements $\Delta_{ab} = \Delta(1-\delta_{ab})$~\cite{parisi_theory_2020}.
The derivative of the first term in Eq.~\eqref{Free}, i.e., the entropic contribution, then gives
%\begin{equation}
%\begin{aligned}
%&M^{RS_\text{\tiny ent}}_{ab;cd}\\
%&= \frac{1}{2}\Big \{ \frac{2}{\Delta^2} (\delta_{ac}\delta_{db} +\delta_{ad}\delta_{bc})-\frac{2}{n\Delta^2} (\delta_{ac}+\delta_{bd}+\delta_{ad}+\delta_{bc})+\frac{4}{n^2\Delta^2}\Big \}
%\end{aligned}
%\end{equation}
\begin{equation}\label{Ment}
\begin{aligned}
m_1^{_\text{\tiny ent}} &= \frac{1}{2}\frac{4}{\Delta^2} \ , \\
m_2^{_\text{\tiny ent}} &= -\frac{1}{2}\frac{8}{n\Delta^2} \ , \\
m_3^{_\text{\tiny ent}} &= \frac{1}{2}\frac{4}{n^2\Delta^2} \ ,
\end{aligned}
\end{equation}
while the interaction part has
\begin{equation}\label{eqM}
\begin{aligned}
\mathbbm{M}^\text{RS,{int}}_{ab;ab} &= \frac{1}{2}\hat{\varphi} \int dh e^h \Big [ e^{-\frac{\Delta}{4}(\sum_{c}\partial_{h_c})^2}
\times \\ &\times
(\partial^2_{h_a}\partial^2_{h_b})\prod_{c} g_\text{\tiny RS}(h_c)\Big ]\Big|_{h_c=h}\\
& = \frac{1}{2}\hat{\varphi} \int dh e^h e^{-\frac{\Delta}{4}\partial_{h}^2}g''_\text{\tiny RS}(h)^2 g_\text{\tiny RS}(h)^{n-2} \\
& = \frac{1}{2}\hat{\varphi} \int dh e^{h-\frac{\Delta}{4}}\big [\frac{g''_\text{\tiny RS}(h)}{g_\text{\tiny RS}(h)}\big ]^2 g_\text{\tiny RS}(h)^{n} \ , \\
\mathbbm{M}^\text{RS,{int}}_{ab;ac} &= \frac{1}{2}\hat{\varphi} \int dh e^{h-\frac{\Delta}{4}}\frac{g''_\text{\tiny RS}(h)}{g_\text{\tiny RS}(h)}\big [\frac{g'_\text{\tiny RS}(h)}{g_\text{\tiny RS}(h)}\big ]^2g_\text{\tiny RS}(h)^{n} \ , \\
\mathbbm{M}^\text{RS,{int}}_{ab;cd} &= \frac{1}{2}\hat{\varphi} \int dh e^{h-\frac{\Delta}{4}}\big [\frac{g'_\text{\tiny RS}(h)}{g_\text{\tiny RS}(h)}\big ]^4g_\text{\tiny RS}(h)^{n} \ ,
\end{aligned}
\end{equation}
where $a\neq b \neq c \neq d$ and 
\begin{equation}
\begin{aligned}
g_\text{\tiny RS}(h) &= e^{\frac{\Delta}{4}\partial^2_{h}}e^{-\beta \bar{v}(h)} \ , \\
g_\text{\tiny RS}'(h) &= \partial_{h}g_\text{\tiny RS}(h) \ , \\
g_\text{\tiny RS}''(h) &= \partial^2_{h}g_\text{\tiny RS}(h) \ .
\end{aligned}
\end{equation}
From the second to third line in the first Eq.~\eqref{eqM}, integration by parts is used so that the operator $e^{-\frac{\Delta}{2}\partial_{h}^2}$ acts on $e^{h}$ giving $e^{h-\frac{\Delta}{2}}$.
Given the change of parametrization in Eq.~\eqref{EqM}, Eq.~\eqref{eqM} can be recast as
\begin{equation}\label{Mint}
\begin{aligned}
&m_1^{_\text{\tiny int}}  =  
-\frac{1}{2} \hat{\varphi}\int dh e^{h-\frac{\Delta}{4}}\big \{\frac{g''_\text{\tiny RS}(h)}{g_\text{\tiny RS}(h)} - \big [\frac{g'_\text{\tiny RS}(h)}{g_\text{\tiny RS}(h)}\big ]^2\big \}^2 g_\text{\tiny RS}(h)^n \ ,\\
&m_2^{_\text{\tiny int}}  =  \\
&-\hat{\varphi}\int dh e^{h-\frac{\Delta}{4}}\big [\frac{g'_\text{\tiny RS}(h)}{g_\text{\tiny RS}(h)}\big ]^2\big \{\frac{g''_\text{\tiny RS}(h)}{g_\text{\tiny RS}(h)} - \big [\frac{g'_\text{\tiny RS}(h)}{g_\text{\tiny RS}(h)}\big ]^2\big \} g_\text{\tiny RS}(h)^n \ ,\\
&m_3^{_\text{\tiny int}} = 
-\frac{1}{4} \hat{\varphi} \int dh e^{h-\frac{\Delta}{4}}\big [\frac{g'_\text{\tiny RS}(h)}{g_\text{\tiny RS}(h)}\big ]^4 g_\text{\tiny RS}(h)^n\ .
\end{aligned}
\end{equation}
\begin{figure}[t]
	\centering
	\includegraphics[width=1.1\columnwidth]{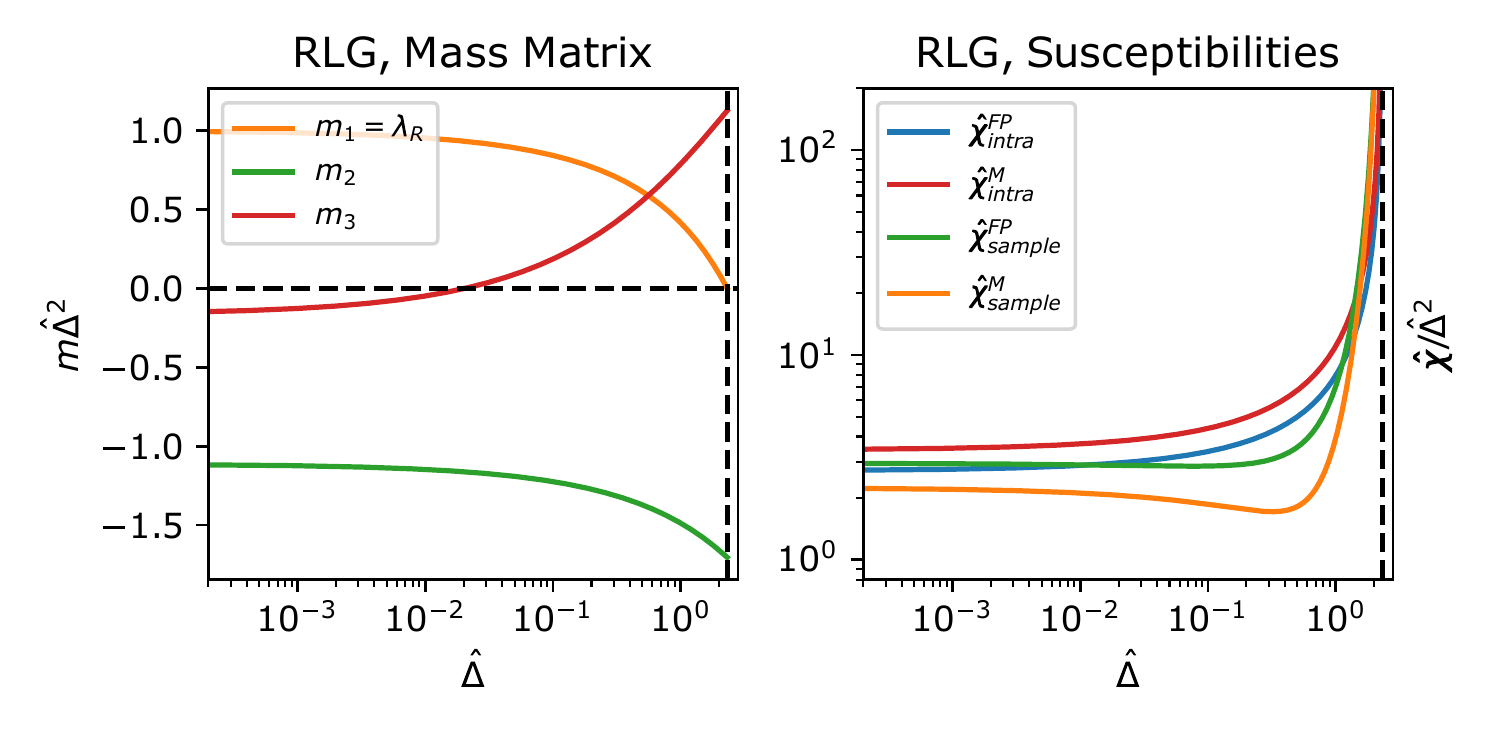}
	\caption{\textbf{(a)}: Equilibrium mass-matrix parameters $m_1,m_2,m_3$ for the RLG as a function of the equilibrium cage size $\hat{\Delta}$ (for $n=1$). \textbf{(b)}: Different typical susceptibilities (at the saddle point) as a function of  $\hat{\Delta}$.}\label{fig:Mass}
\end{figure}
The total mass matrix is then the sum of the entropic term in Eq.~\eqref{Ment} and the interaction term in Eq.~\eqref{Mint},
\begin{equation}\label{MassHS}
m_i =  m_i^{_\text{\tiny ent}} + m_i^{_\text{\tiny int}} \ ,\qquad
i=1,2,3 \ , 
\end{equation}
%\begin{equation}\label{MassHS}
%\begin{aligned}
%m_1 &=  \\
%&-\frac{1}{2} \Big \{ -\frac{4}{\Delta^2}+\hat{\varphi}\int dh e^{h-\frac{\Delta}{4}}\big (\frac{g''_\text{\tiny RS}(h)}{g_\text{\tiny RS}(h)} - \big (\frac{g'_\text{\tiny RS}(h)}{g_\text{\tiny RS}(h)}\big )^2\big )^2 g_\text{\tiny RS}(h)^n  \Big \} \\
%m_2 &=  \\
%&-\frac{1}{2} \Big \{ \frac{8}{n \Delta^2}+2 \hat{\varphi}\int dh e^{h-\frac{\Delta}{4}}\big (\frac{g'_\text{\tiny RS}(h)}{g_\text{\tiny RS}(h)}\big )^2\big (\frac{g''_\text{\tiny RS}(h)}{g_\text{\tiny RS}(h)} - \big (\frac{g'_\text{\tiny RS}(h)}{g_\text{\tiny RS}(h)}\big )^2\big ) g_\text{\tiny RS}(h)^n \Big \} \\
%m_3 &= \\
%&-\frac{1}{2} \Big \{-\frac{4}{n^2\Delta^2}+\frac{1}{2} \hat{\varphi} \int dh e^{h-\frac{\Delta}{4}}\big (\frac{g'_\text{\tiny RS}(h)}{g_\text{\tiny RS}(h)}\big )^4 g_\text{\tiny RS}(h)^n  \Big \} \\
%\end{aligned}
%\end{equation}
where $g_\text{\tiny RS}(h)=\Theta(h/\sqrt{\Delta})=\frac{1}{2}(1+\text{erf}[h/\sqrt{\Delta})]$ for a hard-sphere potential with interaction range $\ell$, the choice we make from now on.

The parameters $m_1,m_2,m_3$ and all the different susceptibilities derived in Sec.~\ref{SMM} are shown
in Fig.~\ref{fig:Mass}.
Appendix \ref{appendixE} reports the explicit limit of infinite scaled density $\hat{\varphi}$ (or equivalently $\Delta\to 0$).

%\subsection{Susceptibilities}

\begin{figure}[th]
	\centering
	\includegraphics[width=\columnwidth]{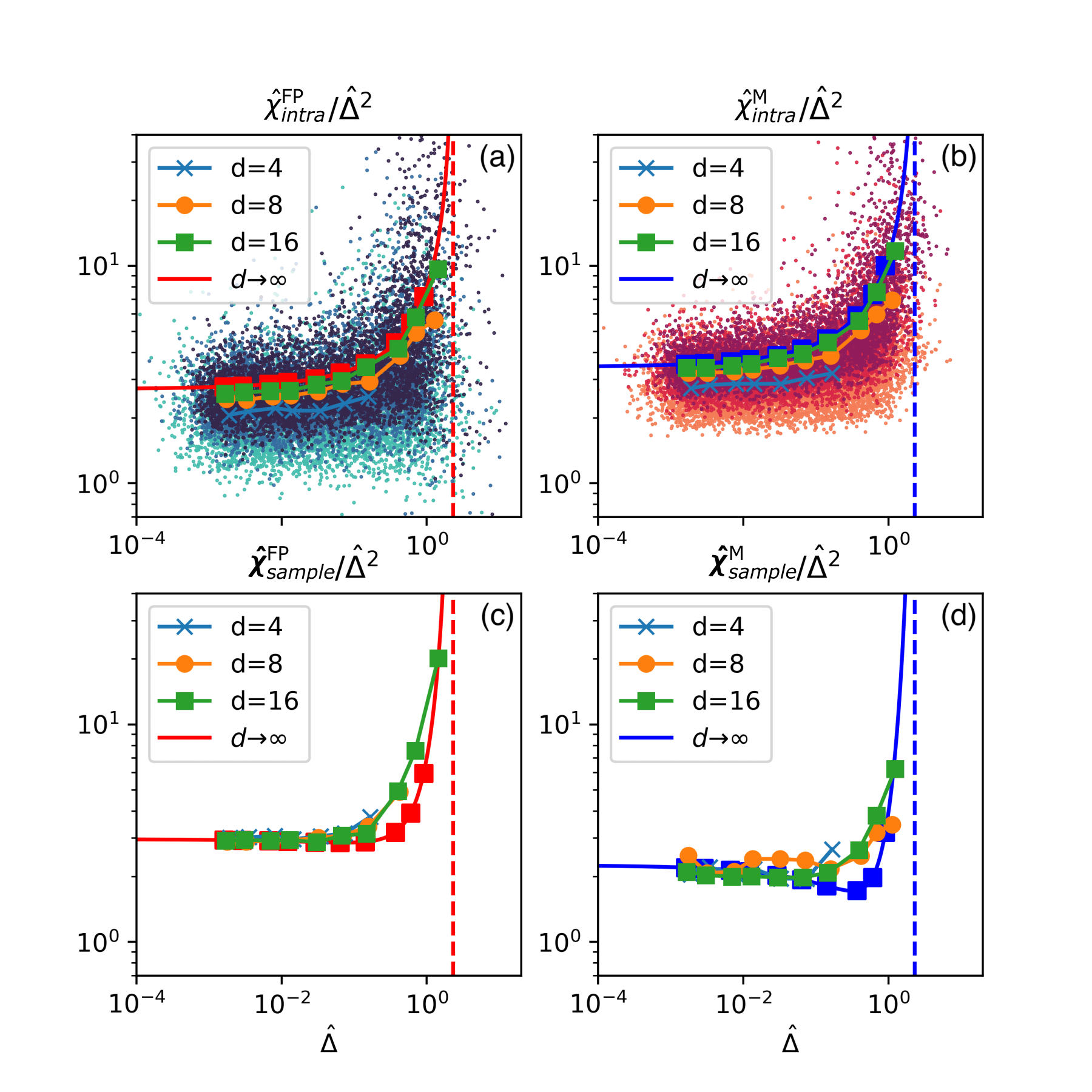}
		\includegraphics[width=0.9\columnwidth]{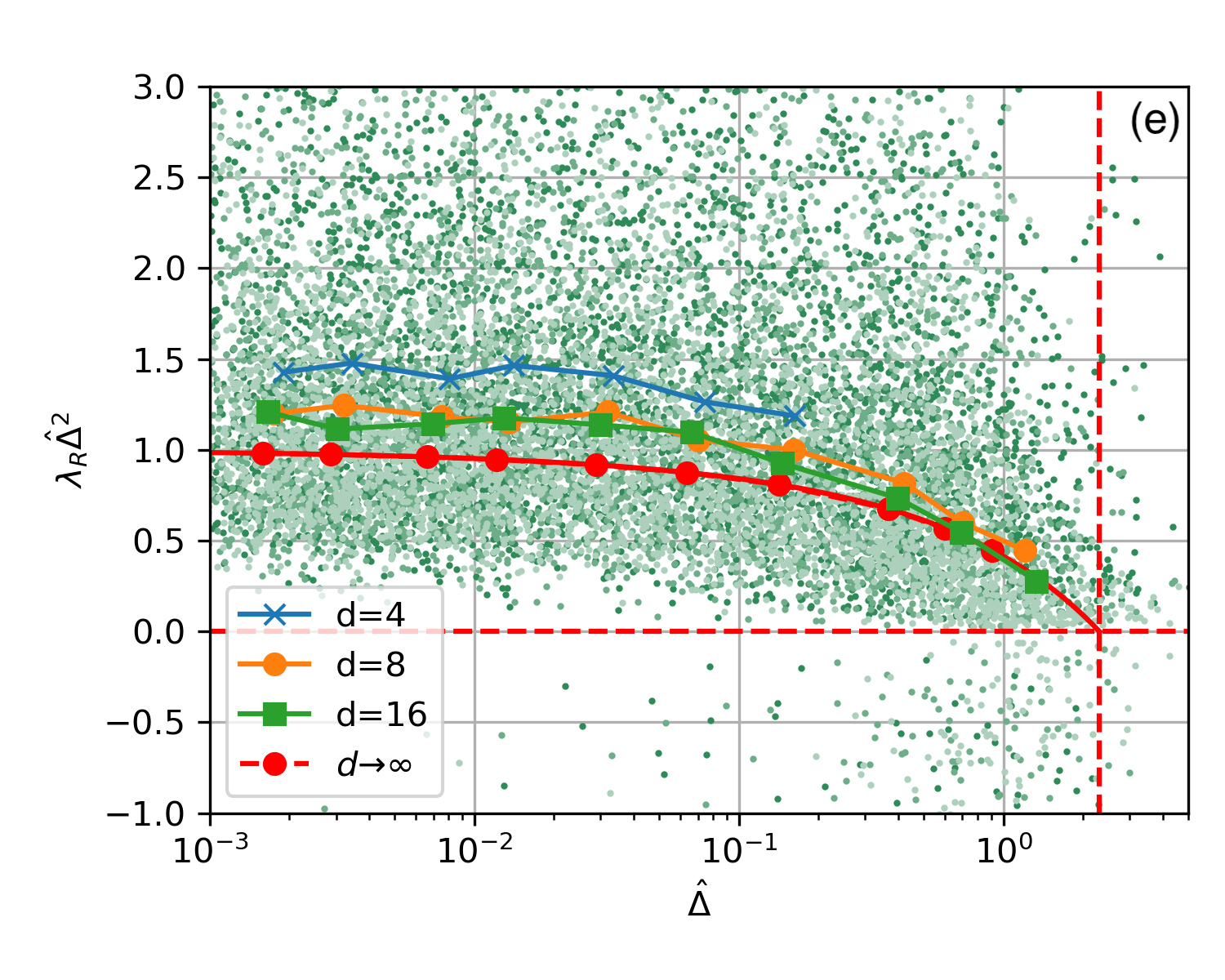}
	\caption{\textbf{(a),(b),(c),(d):} Four susceptibilities of the RLG as a function of $\hat{\Delta}$ for $d=4,8,16$, together with the ${d\to\infty}$ result. For intra-state susceptibilities single samples are shown as dots. The clouds of dots follow the typical line (red or blue) as in the ROM model. Further analysis of the $d$ scaling are reported in Ref.~\cite{biroli_local_2021}.
	\textbf{(e):} Rescaled RLG replicon fluctuation vs cage size. Results for $d=4,8,16$ are consistent with the $d\rightarrow\infty$ computation (red line). The dashed lines denote the MCT transition, at which the typical state opens up and the RS ansatz is no longer valid. Interestingly, even before that transition certain atypical states  have negative replicon fluctuations.
	}\label{fig:RLGchi1}
\end{figure}

\begin{figure}[th]
	\centering
	\includegraphics[width=1\columnwidth]{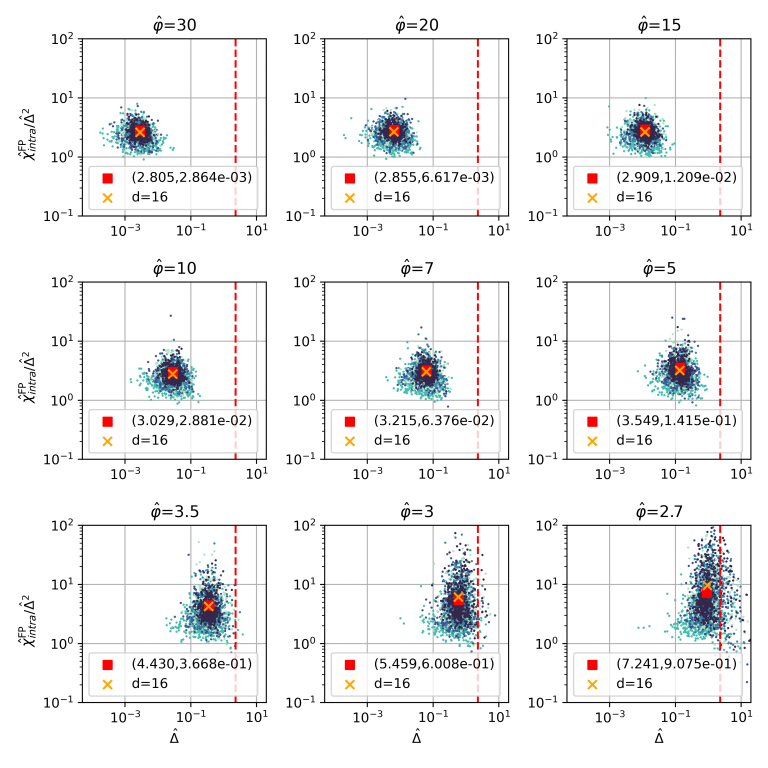}
	\caption{Scatter plot of the cage size $\hat{\Delta}$ vs thermal fluctuations $\hat{\chi}_{intra}^\text{\tiny FP}$ inside that cage. Each point represents a different cage, and each canvas corresponds to a different $\hat{\varphi}$. Different shades refer to different $d=4,8,12,16,20$. The red square denotes the $d\to\infty$ result, while the red dashed line denotes the MCT transition. The orange crosses denote the barycenter of the samples from simulations at $d=16$. }\label{fig:RLGchi2}
\end{figure}

\subsection{Simulations Details}

Numerical results for the RLG are obtained by simulating the tracer dynamics as in Ref.~\cite{biroli_unifying_2020}. For each instance, $N$ non-interacting obstacles are placed uniformly at random within a spherical shell centered at the origin. The shell has a unit inner radius and outer radius $r_\mathrm{max}$, hence $N$ is chosen at random from the Poisson distribution $p(N) = N_0^N \exp(-N_0)/N!$ and $N_0 = d\hat\varphi (r_\mathrm{max}^d - 1)$ is the expected number of obstacles. At $t=0$, the tracer is located at the origin and is assigned a unit velocity and a random orientation. Its position then evolves following a Newtonian dynamics. By this construction, the local environment for the tracer is identical to that of an infinite system as long as its displacement from the origin $r < r_\mathrm{max} - 1$.

For the data collection, the tracer mean squared displacement (MSD, $\langle r^2_{01} \rangle$) and mean fourth-power displacement $\langle r^4_{01} \rangle$ ---for a given obstacle distribution---are averaged over tracer initial position at $t=0$ and different final positions at times $\delta t, 2 \delta t, 3 \delta t, ..., 2^{10} \delta t$, respectively. The tracer reference position used to compute these moments can also be averaged along the trajectory, which then gives $\langle r^2_{12} \rangle$ and $\langle r^4_{12} \rangle$, respectively. For each choice of density, $N_\mathrm{sample}$ independent runs with different obstacle configurations are conducted. By definition (see Sec.~\ref{secth} and~\ref{secdyn}), we then have
\begin{equation}
\begin{aligned}
    \Delta &= \overline{\langle r^2_{01} \rangle} = \overline{\langle r^2_{12} \rangle} \ , \\
    \bm{\chi}_{tot} &= \overline{\langle r^4_{01} \rangle} - \Delta^2 = \overline{\langle r^4_{12} \rangle} - \Delta^2 \ , \\
    \bm{\chi}^\text{\tiny FP}_{intra} &= \overline{\langle r^4_{01} \rangle - \langle r^2_{01} \rangle^2} \ , \\
    \bm{\chi}^\text{\tiny FP}_{sample} &= \overline{\langle r^2_{01} \rangle^2} - \Delta^2 \ , \\
    \bm{\chi}^\text{\tiny M}_{intra} &= \overline{\langle r^4_{12} \rangle - \langle r^2_{12} \rangle^2} \ , \\
    \bm{\chi}^\text{\tiny M}_{sample} &= \overline{\langle r^2_{12} \rangle^2} - \Delta^2 \ .
\end{aligned}
\end{equation}
Properly scaled quantities are used to compare the finite-$d$ simulation with the infinite-$d$ theoretical predictions~\cite{biroli_local_2021},
\begin{equation}
\begin{split}
\hat\Delta &= d \Delta,
\end{split}
\quad \text{and} \quad
\begin{split}
\hat\chi &= d^3 \chi.
\end{split}
\end{equation}

In practice, $r_\mathrm{max}$ is taken such that less than $1\%$ of the tracers escape during individual runs. The choice of elementary time interval $\delta t$ is chosen such that $\delta t > t_\mathrm{plateau}$, the time needed for the MSD over different sample runs to reach a long-time plateau (see Ref.~\cite[Fig.~2(a)]{biroli_unifying_2020}). For each condition, $N_\mathrm{sample}$ is of the order of $10^4$. Note, however, that $10^3$ samples are evaluated for $d=20$ with small $\hat\varphi \le 4$, because larger $r_\mathrm{max}$ (thus larger $N$) and longer $\delta t$ are then necessary to prevent the tracer from escaping, and hence observe the plateau.
Results in Figs.~\ref{fig:RLGchi1} and \ref{fig:RLGchi2} show a clear consistency between theory and
numerical simulations.

\section{Conclusion} 
\label{conclusion}

In this article, we have studied equilibrium fluctuations by means of the replica method in three different mean-field models that present a random first order transition (RFOT). Following the approach developed in Ref.~\cite{franz_field_2011} we have derived explicit formulas connecting the replica symmetric mass matrix and two different kinds of susceptibilities, sample-to-sample and intra-state fluctuations. The results are found to perfectly describe $1/N$ (or $1/d$) \textit{small} fluctuations of the overlaps around their thermodynamic value in the case of disordered model belonging to the standard RS class and to the RFOT class (both close and far from criticality). 
The problem of \textit{large} fluctuations remains open. Simulations suggest that for some models (e.g. the ROM) large fluctuations coincide with typical fluctuations at different temperatures, while in some other models (e.g. the spherical 3-spin) this is not the case. 

Reverse-engineering the process, one could imagine having a disordered system at equilibrium for which the Hamiltonian is not known but can be described by building a local RS free energy potential that capture its small fluctuations in the thermodynamic limit. Following the numerical approach presented here, it is possible to derive from the two intra-state susceptibilities, $\bm{\chi}_{intra}^\text{\tiny FP}$ and $\bm{\chi}_{intra}^\text{\tiny M}$, the relative $m_1$ and $m_2$ of each state and from the sample-to-sample susceptibility $\bm{\chi}_{sample}$ the third parameter $m_3$. One can therefore build the mass matrix, which is then the Gaussian overlap action (effective potential) that describes the equilibrium fluctuations of this system. One could also vary temperature---or tune other parameters---and explore how the mass matrix parameters vary. Moreover, as evinced in the analysis of the spherical 3-spin simulations, it is possible to define the replicon fluctuations of a single state, which are fluctuations that may foresee the possible breaking of the state (Gardner transition) upon lowering temperature. In general, it would be interesting to see if the study of single states by the replica action can provide insight into the free energy landscape structure. 

Continuing in the path of this work, two interesting extensions are also possible. First, one could try to evaluate fluctuations in models that present a 1RSB saddle structure. In this case, the hierarchy of fluctuations has an added level induced by the presence of \textit{clusters} of states. Is it possible to evaluate these susceptibilities and compare the results with actual simulations? Second, we have explored small fluctuations and have argued that occasionally these seem to be compatible with large fluctuations. Is it possible to make these statements more quantitative and, in general, explore large deviations from the typical values, around the thermodynamic limit?
Both these directions have been already somehow explored in the $p$-spin model. In Ref.~\cite{crisanti_spherical_1992} the mass matrix of the 1RSB saddle point in the $p$-spin model has been evaluated. The missing step is to relate these results to actual susceptibilities. Moreover, Ref.~\cite{franz_large_2019} has attempted to systematically explore large deviations in the $p$-spin model. It would be interesting to compare these predictions with numerical simulations.

As mentioned in the introduction, another direction that is in principle achievable by combining mean-field dynamical analysis and the replica method~\cite{rizzo_path_2021} is to study instantonic paths between different states. This analysis could lead to the evaluation of the escape-time from a given state, thus relating Gaussian fluctuations to relative instantonic escape rates. An effort of this type would be in the same spirit of what has been qualitatively done in numerical simulations of structural glasses \cite{berthier_self-induced_2021}. 

%Finally, we note that the replica method has been shown to be effective in a variety of different calculations: in the evaluations of the typical spectrum of Hessian matrices~\cite{edwards_eigenvalue_1976}, in the calculation of large deviations of the free energy (not just Gaussian fluctuations)~\cite{rizzo_replica_2014}, in the counting of the number of saddles of different index in a complex landscape~\cite{cavagna_stationary_1998}, and many others. Despite the wide use of the method, the underlying physical intuition remains elusive. We hope that this work brings a new light on the matter. \fz{I would remove this paragraph.}

Data relevant to this work have been archived and can be accessed at the Duke Digital Repository~\cite{mfdata}.

\medskip
\acknowledgments

We thank Ludovic Berthier, Silvio Franz, Pierre Le Doussal, Federico Ricci-Tersenghi and Tommaso Rizzo for helpful discussions. This project has received funding from the European Research Council (ERC) under the European Union's Horizon 2020 research and innovation programme (grant agreement n. 723955 - GlassUniversality) and from the Simons Foundation (Grant No. 454935 to G.B.; Grant No. 454937 to P.C.;  Grant No. 454955 to F.Z.). The simulations were carried out on the Duke Compute Cluster and Open Science Grid~\cite{osg07,osg09}, supported by National Science Foundation award 1148698, and the U.S. Department of Energy's Office of Science. 
%\end{acknowledgments}

%\clearpage

%\section{References}
%{\small
%	[1] S. Franz, G. Parisi, F. Ricci-Tersenghi, and T. Rizzo, "Field Theory of Fluctuations in Glasses”
%	[2] S. Franz, H. Jacquin, G. Parisi, P. Urbani, and F. Zamponi, "Static replica approach to critical correlations in glassy systems”
%	[3] Yi Hu, "Cage distributions in random Lorentz gas" (note September 2, 2020)
%	[4] A. Crisanti and H.-J. Sommers, "The spherical p-spin interaction spin glass model: the statics,” Z. Physik B 
%	[5] G. Parisi, U. Pierfrancesco, and F. Zamponi, "Theory of Simple Glasses: Exact Solutions in Infinite Dimensions"
%	[6] J. Baik, E. Collins-Woodfin, P. L. Doussal, and H. Wu, Spherical Spin Glass Model with External Field
%	[7] R. Cherrier, D. S. Dean, and A. Lefèvre, "Role of the interaction matrix in mean-field spin glass models", 2003
%	[8] E. Marinari, G. Parisi, and F. Ritort, "Replica field theory for deterministic models. II”, 1994
%	[9] T. Sarlat, A. Billoire, G. Biroli, and J.-P. Bouchaud, "Predictive power of MCT: numerical testing and finite size scaling for a mean field spin glass”, 2009
%	[10]G. Folena, S. Franz, and F. Ricci-Tersenghi, Gradient Descent Dynamics in the Mixed P-Spin Spherical Model: Finite-Size Simulations and Comparison with Mean-Field Integration, 2021
%}

\bibliography{bibliography}

%\clearpage

\onecolumngrid

\appendix

\section{Diagonalization of the Replica Symmetric Mass Matrix}\label{AppendixA}

Given the quadratic fluctuations of the overlap around the replica symmetric solution
\begin{equation}
\delta^2 \mQ{}{RS}
\equiv \frac{1}{2}\sum_{a\neq b;c\neq d}\mathbbm{M}^\text{\tiny RS}_{a b;c d}\delta \hat{q}_{ab}\delta \hat{q}_{cd}
= \frac{1}{2} \big ( m_1 \sum_{a\neq b}\delta \hat{q}_{ab}^2 + m_2 \sum_{a\neq b \neq c}\delta \hat{q}_{ab}\delta \hat{q}_{ac} + m_3 \sum_{a\neq b;c\neq d}\delta \hat{q}_{ab}\delta \hat{q}_{cd}\big) \ ,
\end{equation}
we obtain the following eigenvalue equations
\[m_1\delta \hat{q}_{ab}+m_2\frac{1}{2}\sum_{c}(\delta \hat{q}_{ac}+\delta \hat{q}_{bc}) + m_3\sum_{c\neq d}\delta \hat{q}_{cd} = \lambda \delta \hat{q}_{ab} \ .\]
In order to diagonalize $\mathbbm{M}^\text{\tiny RS}$ we then need to decompose the space of fluctuations based on the kernels (null spaces) of the operators $\frac{1}{2}\sum_{c}(\delta \hat{q}_{ac}+\delta \hat{q}_{bc})$ and $\sum_{c\neq d}\delta \hat{q}_{cd}$. 
%\subsubsection{Replicon eigenspace}
The \textit{replicon} eigenspace corresponds to vectors that are kernels of both operators. 
An example of such a vector is:

\[|v_\text{\tiny R} \rangle = 
\begin{cases}
\delta \hat{q}_{13} = \delta \hat{q}_{31} = 1 \\
\delta \hat{q}_{14} = \delta \hat{q}_{41} = -1 \\
\delta \hat{q}_{23} = \delta \hat{q}_{32} = -1 \\
\delta \hat{q}_{24} = \delta \hat{q}_{42} = 1 \\
\end{cases}\]
with all other element being $0$. This eigenvector corresponds to the eigenvalue $\lambda_\text{\tiny R} = m_1$.
%\subsubsection{Anomalous eigenspace}
The \textit{anomalous} eigenspace is orthogonal to the first and belongs to the kernel of the second operator, i.e., $\sum_{a\neq b}\delta \hat{q}_{ab} = 0$. One such vector is

\[|v_\text{\tiny A} \rangle = 
\begin{cases}
\delta \hat{q}_{1a} = \delta \hat{q}_{a1} = 1 \qquad \forall a\neq 1 \ , \\
\text{all other elements} = \frac{2}{2-n} \ ,
\end{cases}\]
which corresponds to a symmetric matrix in which the first row differs from all the others. This eigenvector corresponds to the eigenvalue $\lambda_\text{\tiny A} = m_1+m_2\frac{1}{2}(n-1+1+(n-2)\frac{2}{2-n})= m_1+m_2\frac{n-2}{2}$.
%\subsubsection{Longitudinal eigenspace}
Finally, the \textit{longitudinal} eigenspace, which is orthogonal to the first two, corresponds to the homogeneous vector
\[|v^L \rangle = 
\begin{cases}
\delta \hat{q}_{ab} = 1 \qquad \forall a\neq b
\end{cases},\]
and is associated with the eigenvalue 
$\lambda_\text{\tiny L} = m_1+m_2\frac{1}{2}(2(n-1))+m_3 n(n-1)= m_1+m_2(n-1)+m_3 n(n-1)$.

%\subsubsection{Dimension of each eigenspace}

The dimension of each eigenspace is built in inverse order: $\mu_\text{\tiny L} = 1$ because there is only one vector;
$\mu_\text{\tiny A} = n-1$ because in this space each eigenvector is built from a row ($n$ rows) and must be orthogonal to the $L$ eigenspace (-1); $\mu_\text{\tiny R} = \frac{n(n-1)}{2}-n$ because it must fill all the dimensions of symmetric matrices ($\frac{n(n-1)}{2}$) while being orthogonal to the first two eigenspaces ($-n$).

\section{$p$-spin with homogeneous external field}\label{AppendixB0}

In the main text (section \ref{2spin}), we discuss the 2-spin spherical model with both a homogeneous and a Gaussian external field.  While the saddle point is the same in the two cases, fluctuations differs. In this Appendix we derive the free energy of the $p$-spin in the homogeneous case, and show the analogy with the Franz-Parisi potential. 

Given the Hamiltonian $H = -\sum^N_{ij} J_{ij}s_is_j-h\sum_is_i$, we want to evaluate the replicated free energy.
We use the definition 
\begin{equation}
\overline{H[s]H[s']}-\overline{H[s]}\ \overline{H[s']}=N f\big (\frac{s\cdot s'}{N}\big )\ ,
\end{equation}
to describe the fluctuations of the Hamiltonian. $f(q)=1/2q^2$ in the 2-spin model.
The replicated partition function then reads
\begin{equation}
 \overline{\exp (-\beta n f)}= \overline{\text{Tr}_{s^n} e^{-\beta \sum_{a=1}^{n} H[s^a]} } =\text{Tr}_{s^n} e^{-N \Big [-\beta h \big (\frac{\bm{1}\cdot s}{N}\big )+ \frac{1}{2}\beta^2 \sum^{n}_{a,b} f\big (\frac{s\cdot s'}{N}\big )\Big]}\ .
\end{equation}
In order to change variables from spins to magnetization $m=\big (\frac{\bm{1}\cdot s}{N}\big )$ and overlaps $q=\big (\frac{s\cdot s'}{N}\big )$ we use two different Lagrange multipliers $\gamma_a$, $\lambda_{ab}$, the first conjugated to the magnetization and the second to the overlaps.
Therefore, we have
\begin{equation}\label{B3}
e^{-N \Big [-\beta h \sum_a^n m_a+ \frac{1}{2}\beta^2 \sum^{n}_{a,b} f(q_{ab})\Big ]} e^{-i N\Big[  \sum_{a}^n \gamma_a m_{a}+\frac{1}{2}\sum_{a,b}^n \lambda_{ab} q_{ab}\Big ]}
\text{Tr}_{s^n} e^{iN \Big [\sum_{a}^n \gamma_a s_{a}+ \frac{1}{2}\sum_{a,b}^n \lambda_{ab}  s_{a}s_{b}\Big ]} \ .
\end{equation}
where the logarithm of the last term gives 
$-\frac{N}{2} (\ln \det (-i\lambda) -i\sum_{a,b}^n \gamma_a [\lambda^{-1}]_{ab}\gamma_b  )$.
Extremizing Eq.~\eqref{B3} with respect to $\gamma_a$ and then $\lambda_{ab}$ we obtain the free energy as a function of the overlaps and the magnetizations
% \begin{equation}
% \gamma_{a}=-\sum_{b} \lambda_{ab}m_b.
% \end{equation}
% The free energy therefore reads
% \begin{equation}\label{Eqmlq}
% \beta h \sum_a m_a- \frac{1}{2}\beta^2 \sum^{n}_{a,b} f(q_{ab})+i\frac{1}{2}\sum_{a,b} m_a \lambda_{ab}m_b -i\frac{1}{2}\sum_{a,b} \lambda_{ab} q_{ab}-\frac{1}{2}\ln \det (-i\lambda)
% \end{equation}
% Again, extremizing with respect to $\lambda_{ab}$ gives
% \begin{equation}
% i\lambda_{ab}=\big [ q_{ab}-m_am_b\big ]^{-1}
% \end{equation}
% Substituting back gives
\begin{equation}\label{Eqmq}
\beta h \sum_a m_a- \frac{1}{2}\beta^2 \sum^{n}_{a,b} f(q_{ab})-\frac{1}{2}\ln \det (\mQ{}{}-mm^T),
\end{equation}
where for convenience we write $q_{ab}-m_am_b$ in matrix form as $\mQ{}{}-mm^T$.
This is \emph{equivalent} to a Franz-Parisi potential that has the external field $h$ as a confining potential ($-\beta' =h$ and $ f(p_{a})=p_a = m_a$),
\begin{equation}
-\beta \beta' \sum_a f(p_{a})- \frac{1}{2}\beta^2 \sum^{n}_{a,b} f(q_{ab})-\frac{1}{2}\ln \det (\mQ{}{}-pp^T)
\end{equation}
where $\beta'$ is the inverse temperature of the reference configuration (see \cite{folena_mixed_2020} for more details). In short, fixing an external field $h$ is analogous to fixing the overlap with the reference configuration $p$. Averaging over a Gaussian distribution of external fields is analogous to averaging over different reference configurations.
% Considering a RS ansatz  $q_{ab}=(1-q)\delta_{ab}+q$ and $m_a=m$, we obtain the free energy (considering the derivative $\partial_n$ at $n=0$)
% \begin{equation}
% \beta h  m- \frac{1}{2}\beta^2 \big ( f(1)-f(q)\big )-\frac{1}{2}\big [  \ln(1-q) + \frac{q-m^2}{1-q}\big ].
% \end{equation}
% This expression is analogous to the Franz-Parisi RS potential (see Ref.~\cite[Eq.~(31)]{franz_recipes_1995}).
% We then want to marginalize the distribution with respect to $m$, and obtain
% \begin{equation}
% m = -\beta h (1-q)
% \end{equation}
% which plugged back gives the free energy (equivalent to \cite[Eq.~(4.4)]{crisanti_spherical_1992})
% \begin{equation}
% - \frac{1}{2}\beta^2 h^2(1-q) - \frac{1}{2}\beta^2 \big [ f(1)-f(q)\big ]-\frac{1}{2}\big [  \ln(1-q) + \frac{q}{1-q}\big ]
% \end{equation}
% If we identify $h^2=\overline{h}^2$, this RS free energy is the same as that obtained in Eq.~\eqref{overRS} ($n=0$) for the Gaussian case.

To study the fluctuations of the overlap we need the free energy to be written only as a function of the overlap, thus we extremize Eq.~\eqref{Eqmq} with respect to $m_a$. To do that we expand the last term in powers of $m_a$ around $n=0$, as it is done in Ref.~\cite[Eq.~(3.12)]{crisanti_spherical_1992} (despite a typo). 
\begin{equation}
\ln \det (\mQ{}{}-mm^T) = \ln \det (\mQ{}{}) - \sum^n_{a,b}\big [ \mQ{}{}\big ]^{-1}_{ab}m_am_b - \frac{1}{2}\big (\sum^n_{a,b}\big [ \mQ{}{}\big ]^{-1}_{ab}m_am_b \big )^2+O(n^3),
\end{equation}
and evaluate the saddle point in $m$ at the second order in $n$,
\begin{equation}
m_a = -\beta \sum_b q_{ab} h_b+O(n^2) \ .
\end{equation}
Plugging back into  Eq.~\eqref{Eqmq} and considering that the field is homogeneous in the replica index, i.e., $h_a=h\ \forall a$, we obtain (Ref.~\cite[Eq.~(3.17)]{crisanti_spherical_1992})
\begin{equation}
F(\mQ{n}{})=\frac{1}{2}\beta^2 h^2 \sum^n_{a,b} q_{ab} + \frac{1}{2}\beta^2 \sum^{n}_{a,b} f(q_{ab})+\frac{1}{2}\ln \det (\mQ{}{})-\frac{1}{4}\beta^4 h^4 (\sum^n_{a,b} q_{ab})^2.
\end{equation}
This expression provides the free energy of a generic $p$-spin with homogeneous external field for generic $\mQ{n}{}$ ansatz. We notice that the only difference with the free energy of the $p$-spin with external Gaussian field (Eq.~\eqref{overlapAc} together with Eq.~\eqref{eq:fq}) is the last term, which contributes to the mass matrix $m_3$, and thus to disorder fluctuations  ($\bm{\chi}_{dis}$). As expected, taking a homogeneous field suppresses disorder fluctuations. 

\section{Cumulants Method for Franz-Parisi Fluctuations}\label{AppendixB}

In this Appendix we show how to evaluate $\bm{\hat{\chi}}_{intra}^\text{\tiny FP}$,$\bm{\hat{\chi}}_{sample}^\text{\tiny FP}$ using a cumulant expansion (over disorder) of the Franz-Parisi potential $V_{\text{\tiny FP}}(p)$. 
Let's consider a random potential $V$ whose fluctuations are implicitly encoded in some quenched couplings $J$ (which can include the reference configuration).
This potential has a given distribution $P(V)$ which should be inferred given the distribution of the couplings $P(J)$.
In order to do that, we can expand $P(V)$ in cumulants, using the replica method, Eq.~\eqref{rep_met}.

The replica method is typically used to evaluate the first-order cumulant
\begin{equation}
\overline{V} = \lim_{n\to 0} \partial_n \overline{Z^n} = \lim_{n\to 0}  \partial_n \overline{\exp(n V)} \ ,
\end{equation}
where $V=\ln(Z)$ and $\overline{\bullet}=\int J P(J) (\bullet)$.
Therefore, $\overline{\exp(n V)}=W(n)$ can be seen as the cumulant generating function of the disorder distribution $P(V)$. Further derivatives evaluated at $n=0$ then give further cumulants,
\begin{equation}
\begin{aligned}
\overline{V} &= \partial_n W(n)|_{n=0}\\
\overline{V^2}-\overline{V}^2 &= \partial^2_{n} W(n)|_{n=0}\\
\ldots
\end{aligned}
\end{equation}
The cumulant generating function can also be evaluated for the Franz-Parisi potential, for which the average over disorder include now both the reference configuration and quenched couplings. We then have
\begin{equation}
\begin{aligned}
\overline{V_{\text{\tiny FP}}(p)} &= \partial_n W(p;n)|_{n=0}\\
\overline{V_{\text{\tiny FP}}(p_1)V_{\text{\tiny FP}}(p_2)}-\overline{V_{\text{\tiny FP}}(p_1)}\overline{V_{\text{\tiny FP}}(p_2)} &= \partial_{n_1}\partial_{n_2} W(p_1,p_2;n_1,n_2)|_{n_1,n_2=0}\\
\ldots
\end{aligned}
\end{equation} 
where $p_1,p_2$ are the overlap with the reference configuration. (Note that the disorder average is evaluated having fixed the overlap with the reference configuration, but considering the ensemble of all possible  reference configurations at equilibrium.) 

Therefore the intra-state FP susceptibility is given by fluctuations of the averaged potential
\begin{equation}\label{chithapp}
\bm{\hat{\chi}}_{intra}^\text{\tiny FP}  = \frac{1}{\overline{\beta V''_{\text{\tiny FP}}(p)}}\bigg|_{p=\qEA},
\end{equation}
where $V'(p)\equiv d_pV(p)$ and $V''(p)\equiv d^2_pV(p)$.
In order to evaluate $\bm{\chi_\text{het}}$ we are interested in fluctuations of the minimum of the potential. Following Sec.~\ref{oneD}, we have
\begin{equation}
d_p V_{\text{\tiny FP}}(p) = 0 \quad \Longrightarrow \quad  d_p^2 \overline{V_{\text{\tiny FP}}(p)} \delta p+d_p \delta V_{\text{\tiny FP}}(p) = 0 \quad \Longrightarrow \quad \delta p = \frac{\delta V'_{\text{\tiny FP}}(p)}{ \overline{V''_{\text{\tiny FP}}(p)} },
\end{equation}
where $\delta V'_{\text{\tiny FP}}(p) = V'_{\text{\tiny FP}}(p)-\overline{V'_{\text{\tiny FP}}(p)}$ is the fluctuation of the derivative of the potential and $\delta p= p-p^*$ are the small fluctuations around the minimum of the potential $p^* $, such that  $d_p \overline{V(p^*)}=0$.
The variance of small fluctuations of the overlap with the reference configuration is therefore
\begin{equation}\label{chiH}
\bm{\hat{\chi}}_{sample}^\text{\tiny FP} =  N\overline{\delta p^2}\big |_{p=\qEA}=\frac{\overline{ \big(\delta  V'_{\text{\tiny FP}}(p)\big)^2}}{ {\overline{V_{\text{\tiny FP}}''(p)}}^2 } \bigg |_{p = \qEA} =  \frac{\partial_{n_1}\partial_{n_2} d_{p_1}  d_{p_2} W(p_1,p_2;n_1,n_2)|_{n_1,n_2=0}^{p_1=p_2=\qEA}}{ [{\bm{\hat{\chi}}_{intra}^\text{\tiny FP}}]^{-2}}.
\end{equation}

In the following we will threat in details thermal fluctuations and heterogeneous fluctuations with the cumulant method applied to the FP potential, showing concrete calculations with the $p$-spin spherical model. We then show the equivalence between the mass matrix method employed in the main text and the cumulant method.
To do so, we consider the mass matrix of Gaussian fluctuations around the RFOT saddle point. This saddle point --- if we consider a Franz-Parisi potential at equilibrium --- is equivalent to the saddle point of the Monasson potential with $n+1$ replicas. In the FP formulation, a special replica $s_0$ breaks the permutation symmetry $S^{n+1}$ in the $0$ direction. However, the mass matrix around the RS solution, if we include this special direction $0$, is equal in both cases.
%given a reference configuration $s_{0}$ at equilibrium at temperature $T$ we evaluate the free energy of a second one at a given overlap $p$ (or distance) from the first one. 
%Normally in the FP approach $s_{0}$ is a typical configuration, here we also consider the influence of small fluctuations of $s_{0}$ around the typical case.

\subsection{Thermal Fluctuations}
We assume that the intra-state FP susceptibility is given by Eq.~\eqref{chithapp}, reported here in the averaged form
\begin{equation}
	\bm{\hat{\chi}}_{intra}^\text{\tiny FP}  = \frac{1}{\beta\bm{V''_{\text{\tiny FP}}}(p)}\bigg|_{p\quad\text{s.t}\quad\bm{V'}_{\text{\tiny FP}}(p)=0} =  N \Big ( \overline{\langle  p^2\rangle_{\text{\tiny FP}}-\langle p\rangle^2_{\text{\tiny FP}}} \Big )\ ,
\end{equation}
where $\bm{V_{\text{\tiny FP}}}(p) = \overline{V_{\text{\tiny FP}}(p)}$ is the averaged potential and the overline denotes averaging over the quenched disorder and over the reference configuration.
We want to evaluate the second-order total derivative $\bm{V''_{\text{\tiny FP}}}$, in particular around the saddle minimum of the potential $\bm{V'}_{\text{\tiny FP}}=0$. At this point, we assume an RFOT ansatz (RS+1) for the overlap matrix
\begin{equation}\label{RSn1}
\mQ{n+1}{RS} = \begin{pmatrix}  
1&p\\
p&(1-q)\mathbbm{1}^{n}+q \mathbbm{J}^{n,n} \\
\end{pmatrix}\ ,
\end{equation}
where $\mathbbm{1}^{n}$ stands for identity matrix in $n$ dimensions, $\mathbbm{J}^{n,n}$ stands for a $n\times n$ matrix of ones, and the $q$ overlap must be extremized, i.e., $-\beta \bm{V_{\text{\tiny FP}}}(p) = \text{Ext}_{q} \big [\partial_n F(\mQ{n+1}{RS})|_{n=0}\big ]= -\text{Ext}_{q} \big [V(p;q)]$.
Therefore, the second-order total derivative can be written as
\begin{equation}\label{thFP}
    \beta d^2_p\bm{V_{\text{\tiny FP}}}(p) = \partial^2_p  V(p;q) - \frac{\left(\partial_p\partial_q V(p;q)\right)^2}{\partial^2_q V(p;q)}\bigg|_{q\quad\text{s.t}\quad\partial_q V(p;q)=0} \ .
\end{equation}
where we have used the identity $d_p[\partial_q V(p;q)]= (\partial_p+\frac{dq}{dp}\partial_q )[\partial_q V(p;q)] = 0$, which implies $\frac{dq}{dp}= - \frac{\partial_p\partial_q V(p;q)}{\partial^2_q V(p;q)}$.
At the minimum of the FP potential, i.e., $\bm{V'}_{\text{\tiny FP}}=0=\partial_p V(p;q)$, which corresponds to the condition $p=q=\qEA$, as argued in the main text, the intra-state FP susceptibility evaluated with the Franz-Parisi potential is equal to that evaluated with the Monasson potential (see Fig.~\ref{fig:3spin}). In the following we check this equivalence in the $p$-spin spherical model and then show that it holds for any RFOT mean-field model.

\subsubsection{Example: equivalence for the $p$-spin spherical model}
For a generic overlap matrix $\mQ{n}{}$, the Franz-Parisi action of the $p$-spin spherical model reads
\begin{equation}
\begin{aligned}
\exp(N F_{\text{\tiny FP}}(p;\mQ{n}{})) &= \overline{\Tr_{s_0} \frac{e^{-\beta H[s_0]}}{Z} \ln \big [ \Tr_{s^n} e^{-\beta \sum_{a=1}^{n} H[s_a]} \delta(p-s_a\cdot s_0/N)\big ]} \\
=&\Tr_{s^n} e^{\frac{N}{2}\beta^2 \sum^{n}_{a,b=0} (\overline{H[s_a]H[s_b]}-\overline{H[s_a]}\,\overline{H[s_b]} \big )}\\
=& \Tr_{\mQ{n}{}}\det(\mQ{n+1}{})^\frac{N}{2} \exp \Big (\frac{N}{2}\sum^{n}_{a,b=0} \beta^2f(q_{ab})\Big ) \ ,
\end{aligned}
\end{equation}
where  $F_{\text{\tiny FP}}(p;\mQ{n}{}) = F(\mQ{n+1}{})$ is defined in Eq.~\eqref{overlapAc} and $f(q_{ab})=\overline{H[s_a]H[s_b]}-\overline{H[s_a]}\,\overline{H[s_b]}$ is the Hamiltonian covariance which defines the $p$-spin model and have discarded subdominant term in$ $N. Further details of this calculation can be found in \cite{folena_mixed_2020}. Inserting the RS ansatz \eqref{RSn1}, we get the action
\begin{equation}
\begin{aligned}\label{Vact}
V(p;q) &= -\lim_{n \to 0} \partial_n F_{\text{\tiny FP}}(p;\mQ{n}{})\\
&= -\frac{1}{2}\Big (\ln (1-q)+\frac{q-p^2}{1-q}+\beta^2(f(1)+2f(p)-f(q))\Big )+\textit{const} \ . \\
\end{aligned}
\end{equation}
The FP potential is obtained by extremizing this action with respect to $q$, which gives
\begin{equation}\label{spFP}
\beta^2 f'(q)=\frac{q-p^2}{(1-q)^2}
\end{equation}
thus implicitly defining the dependence of $q^*(p)$.
Given this action and using Eq.~\eqref{thFP}, the intra-state FP susceptibility as a function of overlap with the reference configuration is
\begin{equation}
\bm{\hat{\chi}}_{intra}^\text{\tiny FP}(p) = \frac{1}{\bm{\beta V''_{\text{\tiny FP}}}(p)} = \Big ( \frac{1}{1-q}-\beta^2 f''(p)+\frac{2 p^2}{(1-q) \left(1+q-2 p^2-(1-q)^3 \beta^2f''(q)\right)}\Big )^{-1} \ ,
\end{equation}
where $q=q^*(p)$. At $p=q=\qEA$ this expression is equal to the susceptibility evaluated with the mass matrix method applied to the Monasson potential (compare with $\bm{\hat{\chi}}^\text{\tiny RFOT,FP}_{intra}(q_{EA})$ in  Eq.~\ref{Chth}).

\subsubsection{Proof of the equivalence with the mass matrix method}
Here we show the equivalence of the cumulants method and the mass matrix method for the intra-state FP susceptibility at the typical (equilibrium) overlap $\qEA$. We want to evaluate the second-order total derivative $d^2_{p}V(p)$
\begin{equation}\label{total_der}
\begin{aligned}
d^2_{p}\overline{V_{\text{\tiny FP}}(p)} =&\partial_n (\partial_p + \sum_{ab}^{n}\frac{\partial q_{ab}}{\partial p}{\partial_q}_{ab} ) (\partial_p + \sum_{cd}^n\frac{\partial q_{cd}}{\partial p}{\partial_q}_{cd} ) F(p;\mQ{n}{})\Big|_{n=0}\\
&\partial_n \Big [\partial^2_p F(p;\mQ{n}{})-\sum_{ab}^n \sum_{cd}^n (\partial_{q_{ab}}\partial_p F(p;\mQ{n}{}))(\partial_{q_{ab}}\partial_{q_{cd}} F(p;\mQ{n}{}))^{-1}(\partial_{q_{cd}}\partial_p F(p;\mQ{n}{})) \Big ]\Big|_{n=0},
\end{aligned}
\end{equation}
where we have used the equation $\partial_{q_{ab}} F(p;\mQ{n}{}) =0$ because fluctuations are evaluated at the saddle point.  Also,  $d_p \partial_{q_{ab}} F(p;\mQ{n}{}) = (\partial_p F(p;\mQ{n}{})+ \sum_{ab}\frac{\partial q_{ab}}{\partial p}{\partial_q}_{ab} F(p;\mQ{n}{})) =0$. 
At this point we restrict our calculation to RS overlap matrices $\mQ{}{RS}$, for which the mass matrix
\begin{equation}\label{massM}
\mathbbm{M}^\text{\tiny RS}_{a\neq b;c\neq d} \equiv \frac{m_1}{2} (\delta_{a c}\delta_{ b d}+\delta_{a d}\delta_{ b c}) +  \frac{m_2}{4} (\delta_{a c}+\delta_{a d}+\delta_{ b c}+\delta_{ b d}) + m_3 
\end{equation}
is given in terms of three parameters $m_1,m_2,m_3$. We notice that $\partial_p = \sum_a \partial_{q_{a0}}$, and therefore the previous equation can be rewritten as
\begin{equation}\label{Tot}
d^2_{p}F(p;\mQ{n}{}) = \sum_{ab}\mathbbm{M}_{0a;0b}-\sum_{ab}\sum_{c\neq d}\sum_{e \neq f}M_{0a;cd}\mathbbm{M}^{-1}_{cd;ef}\mathbbm{M}_{ef;b0}.
\end{equation}

Given two different equilibrium trajectories from an equilibrium configuration $s_0$, their mutual overlap has ---typically for different disorders--- the same Gaussian fluctuations that the overlap between each trajectory and the reference configuration $s_0$.
\begin{equation}
\overline{\langle \delta \hat{q}_{01}^2 \rangle} = 	\overline{\langle \delta \hat{q}_{12}^2 \rangle}
\end{equation}    
This statement is a direct consequence of the fact that FP at equilibrium ($\beta'=\beta$) and M at equilibrium ($x=1$) present the same mass matrix, and therefore
\begin{equation}
\mathbbm{M}_{01;01}  = \mathbbm{M}_{12;12}\qquad \mathbbm{M}_{01;02}  = \mathbbm{M}_{12;13}\qquad \mathbbm{M}_{01;12}  = \mathbbm{M}_{12;23}\qquad \mathbbm{M}_{01;23}  = \mathbbm{M}_{12;34}
\end{equation}
We can proceed in the evaluation of each term of Eq.~\eqref{Tot} given Eq.~\eqref{massM}:
\begin{equation}\label{Meq}
\begin{aligned}
\mathbbm{M}_{0a;0b} &= \frac{m_1}{2}\delta_{ab} +\frac{m_2}{4}(\delta_{ab} +1)+m_3\\
\mathbbm{M}_{0a;cd} &= \frac{m_2}{2}(\delta_{ac} +\delta_{ad})+m_3\\	
\mathbbm{M}^{-1}_{cd;ef} &\equiv \mathbbm{G}_{cd;ef}  =  \frac{g_1}{2} (\delta_{ac}\delta_{bd}+\delta_{ad}\delta_{bc}) +  \frac{g_2}{4} (\delta_{ac}+\delta_{ad}+\delta_{bc}+\delta_{bd}) + g_3  ,
\end{aligned}
\end{equation}
where $\mathbbm{G}$ is the inverse of $\mathbbm{M}$ and its coefficients are:
\begin{equation}
\begin{aligned}\label{inv}
g_1&=\frac{1}{m_1}\\
g_2&=-\frac{2 m_2}{[m_1 (2 m_1+m_2 (n-2)]}\\
g_3&=\frac{-2 m_1 m_3+m_2^2+m_2 m_3 n}{m_1 [2 m_1+m_2 (n-2)] [m_1+(n-1) (m_2+m_3 n)]}.\\
\end{aligned}
\end{equation} 
We next proceed with evaluating Eq.~\eqref{Tot}. Because the average over disorder comes from the derivative over $n$ evaluated in $n=0$, only terms proportional to $n$ must then be considered
\begin{equation}
\begin{aligned}
d^2_{p}F(p;\mQ{n}{})  &= \sum_{ab} [\frac{m_1}{2}\delta_{ab} +\frac{m_2}{4}(\delta_{ab} +1)+m_3]-\sum_{ab}\sum_{c\neq d}\sum_{e \neq f}[\frac{m_2}{2}(\delta_{ac} +\delta_{ad})+m_3]\Big ( \mathbbm{G}_{cd;ef} \Big )[\frac{m_2}{2}(\delta_{be} +\delta_{bf})+m_3]\\
&= n(\frac{m_1}{2} +\frac{m_2}{4})+O(n^2) -	[\frac{m_2}{2}+O(n)]\Big ( \sum_{c\neq d}\sum_{e \neq f} \mathbbm{G}_{cd;ef} \Big )[\frac{m_2}{2}+O(n)]\\
&= n(\frac{m_1}{2} +\frac{m_2}{4}) + n\frac{m^2_2}{4}\Big ( g_1-g_2 \Big )+O(n^2).
\end{aligned}
\end{equation}
Recalling the intra-state FP susceptibility definition in Eq.~\eqref{chithapp},  and given the inversion formula in Eq.~\eqref{inv} evaluated at $n=0$, we have
\begin{equation}\label{chiDYN}
\bm{\hat{\chi}}_{intra}^\text{\tiny FP} =\frac{1}{\beta \overline{V''_{\text{\tiny FP}}(\qEA)}} = 	\frac{1}{\partial_n d^2_{p}F(p;\mQ{n}{})\big |_{n=0,p=\qEA}} = \frac{4}{m_1}+\frac{4}{m_2-2m_1}\bigg |_{q=\qEA},
\end{equation}
which is equal to that obtained by dynamical considerations (see Eq.~\eqref{atypicalth}), which gives $\bm{\hat{\chi}}_{intra}^\text{\tiny FP} = 4(\mathbbm{G}_{12:12}-\mathbbm{G}_{12;23}) =4(\frac{g_1}{2}+\frac{g_2}{4}) = \frac{4}{m_1}+\frac{4}{m_2-2m_1}$, where now $g_1,g_2$ are evaluated at $n=1$.

\subsection{Sample-to-sample Susceptibility}

The sample-to-sample susceptibility can be evaluated using Eq.~\eqref{chiH}. We thus need to evaluate the derivatives of the second-order cumulant $\partial_{n_1}\partial_{n_2} W(p_1,p_2;n_1,n_2)|_{n_1,n_2=0}$ of the FP potential. Following the same kind of prescription used for the first-moment calculation we introduce the two-block RS matrix
\begin{equation}\label{RSn1n2}
\mQ{n_1+n_2+1}{RS} = \begin{pmatrix}  
1&p_1&p_2\\
p_1&(1-q_1)\mathbbm{1}^{n_1}+q_1 \mathbbm{J}^{n_1,n_1} &q_{12}\mathbbm{J}^{n_1,n_2}\\
p_2&q_{12}\mathbbm{J}^{n_2,n_1} &(1-q_2)\mathbbm{1}^{n_2}+q_2\mathbbm{J}^{n_2,n_2} ,
\end{pmatrix},
\end{equation}
which depends on two constraint $p_1,p_2$ and three parameters $q_1,q_2,q_{12}$. Recall that $\mathbbm{1}^{n}$ is the identity matrix in $n$ dimensions and  $\mathbbm{J}^{n_1,n_2}$ is a $n_1\times n_2$ matrix of ones. The parameters $q_1$ and $q_2$ are derived from the first cumulant extremization, i.e., $\partial_q V(p;q)=0$, which gives the optimal value dependence $q^*(p)$ in Eq.~\eqref{spFP}. Finally, we need to find $q_{12}$ and therefore (following the entropy maximization principle) we extremize the action
\begin{equation}
    \bm{W}\text{\tiny FP}(p_1,p_2) =\text{Ext}_{q_{12}} \big [\partial_{n_1}\partial_{n_2} F_{\text{\tiny FP}}(\mQ{n_1+n_2+1}{RS}) |^{q_1=q^*(p_1),q_2=q^*(p_2)}_{n_1=n_2=0}\big ]= \text{Ext}_{q_{12}} \big [ W(p_1,p_2;q_1^*(p_1),q_2^*(p_2),q_{12}) \big ] \ . 
\end{equation}
Next, we wish to evaluate the total second derivative
\begin{equation}
d_{p_1}d_{p_2} \bm{W}\text{\tiny FP}(p_1,p_2) =  \begin{pmatrix}
 1 &\frac{\partial q^*_1}{\partial p_1} & \frac{\partial q^*_{12}}{\partial p_1}
 \end{pmatrix}
 \cdot
 \begin{pmatrix}  
	\partial_{p_1}\partial_{p_2} &\partial_{p_1}\partial_{q_2}&\partial_{p_1}\partial_{q_{12}}\\
	\partial_{q_1}\partial_{p_2} &\partial_{q_1}\partial_{q_2}&\partial_{q_1}\partial_{q_{12}}\\
	\partial_{q_{12}}\partial_{p_2} &\partial_{q_{12}}\partial_{q_2}&\partial_{q_{12}}\partial_{q_{12}}\\
\end{pmatrix}\bm{W}\text{\tiny FP}
\cdot 
\begin{pmatrix}
1\\ \frac{\partial q^*_2}{\partial p_2} \\ \frac{\partial q^*_{12}}{\partial p_2}
\end{pmatrix}.
\end{equation}
In order to evaluate $\frac{\partial q^*_1}{\partial p_1} (\frac{\partial q^*_2}{\partial p_2})$, we use the same trick as in Eq.~\eqref{thFP},
\begin{equation}\label{derF}
d_p[\partial_q V(p;q)]= (\partial_p+\frac{dq}{dp}\partial_q )[\partial_q V(p;q)] = 0 \quad \Rightarrow \quad \frac{d q^*}{d p} = -\partial_{p}\partial_{q} V/\partial^2_{q} V \ .
\end{equation}
The same procedure for $\frac{\partial q_{12}}{\partial p_1} (\frac{\partial q_{12}}{\partial p_2})$ gives
\begin{equation}
\frac{d}{d p_1} (\partial_{q_{12}} W) = \partial_{p_1}\partial_{q_{12}} W + \frac{\partial q_1^*}{\partial p_1}\partial_{q_1}\partial_{q_{12}} W+ \frac{\partial q_{12}^*}{\partial p_1}\partial^2_{q_{12}} W = 0 \quad \Rightarrow \quad \frac{\partial q_{12}^*}{\partial p_1} = -( \partial_{p_1}\partial_{q_{12}} W+ \frac{\partial q_1^*}{\partial p_1}\partial_{q_1}\partial_{q_{12}}W)/\partial^2_{q_{12}} W.
\end{equation}
Here and in the following $W$ stands for $W(p_1,p_2;q_1,q_2,q_{12})$. Using these two formulas we obtain
\begin{equation}\label{Wder}
\begin{aligned}
d_{p_1}d_{p_2} \bm{W}_\text{\tiny FP}(p_1,p_2)  = &\big(\partial_{p_1}\partial_{p_2}W-\frac{\partial_{p_1}\partial_{q_{12}}W\partial_{q_{12}}\partial_{p_2}W}{\partial^2_{q_{12}}W}\big)\\
+&\big(\partial_{q_1}\partial_{p_2}W-\frac{\partial_{q_1}\partial_{q_{12}}W\partial_{q_{12}}\partial_{p_2}W}{\partial^2_{q_{12}}W}\big) (-\frac{\partial_{p_1}\partial_{q_1} V}{\partial^2_{q_1} V})
+\big(\partial_{p_1}\partial_{q_2}W-\frac{\partial_{p_1}\partial_{q_{12}}W\partial_{q_{12}}\partial_{q_2}W}{\partial^2_{q_{12}}W}\big) (-\frac{\partial_{p_2}\partial_{q_2} V}{\partial^2_{q_2} V})\\
+&\big(\partial_{q_1}\partial_{q_2}W-\frac{\partial_{q_1}\partial_{q_{12}}W\partial_{q_{12}}\partial_{q_2}W}{\partial^2_{q_{12}}W}\big) (-\frac{\partial_{p_1}\partial_{q_1} V}{\partial^2_{q_1} V})(-\frac{\partial_{p_2}\partial_{q_2} V}{\partial^2_{q_2} V}).
\end{aligned}
\end{equation}
Because at the end we wish to evaluate the limit $p_1=p_2=p$, it follows that $q_1=q_2=q$, which are related by the $q^*(p)$ given by the extremization of $V$. In this case also $q_{12}=q$, as is evident by noticing that the matrix given in Eq.~\eqref{RSn1n2} becomes equal to the first-cumulant matrix in Eq.~\eqref{RSn1}. Moreover, if we are interested in the equilibrium case (reference replica $s_{0}$ at the same conditions of all others), we have $p=q=\qEA$.

\subsubsection{Example: equivalence for the $p$-spin spherical model}
In the $p$-spin spherical model the second cumulant action reads
\begin{equation}
\begin{aligned}
\exp\big(N F_{\text{\tiny FP}}(p_1,p_2;\mQ{n_1+n_2}{})\big) 
&= \overline{\Tr_{s_0} \frac{e^{-\beta H[s_0]}}{Z} \big [ \Tr_{s^{n_1+n_2}} e^{-\beta \sum_{a=1}^{n_1} H[s_a]} \delta(p_1-s_a\cdot s_0/N)e^{-\beta \sum_{a=n_1+1}^{n_2} H[s_a]} \delta(p_2-s_a\cdot s_0/N)\big ]} \\
=& \Tr_{\mQ{n_1+n_2}{}}\det(\mQ{n_1+n_2+1}{})^{\frac{N}{2}} \exp \Big (\frac{N}{2}\sum^{n_1+n_2}_{a,b=0} \beta^2f(q_{ab})  \Big ) \ ,
\end{aligned}
\end{equation}
where $F_{\text{\tiny FP}}(p_1,p_2;\mQ{n_1+n_2}{})=F(\mQ{n_1+n_2+1}{})$. Inserting the RS ansatz from Eq.~\eqref{RSn1n2}, after some long manipulation we get the action
\begin{equation}\label{WW}
\begin{aligned}
W(p_1,p_2;q_1,q_2,q_{12}) &= \lim_{n_1,n_2 \to 0} \partial_{n_1}\partial_{n_2}F_{\text{\tiny FP}}(p_1,p_2;\mQ{n_1+n_2}{})\\
&=\beta^2f(q_{12})-\frac{1}{2}\frac{(q_{12}-p_1p_2)^2}{(1-q_1)(1-q_2)} \ .
\end{aligned}
\end{equation}
While $q_1$ and $q_2$ are given by extremizing the first cumulant $V(p;q)$, we need to fix $q_{12}$ by extremizing $W(p_1,p_2;q_1(p_1),q_2(p_2),q_{12})$, and get
\begin{equation}
\beta^2f'(q_{12})=\frac{q_{12}-p_1p_2}{(1-q_1)(1-q_2)} .
\end{equation}
Because we want to study the case $p_1=p_2=p$ which implies $q_1=q_2=q$, extremizing Eq.~\eqref{WW} we get $f'(q_{12})=\frac{1}{2}\frac{q_{12}-p^2}{(1-q)^2}$, which is the same as Eq.~\eqref{spFP} satisfied by $q$. Therefore, $q_1=q_2=q_{12}$.
Finally, using Eq.~\eqref{Wder} we evaluate the total derivative
\begin{equation}
\begin{aligned}
\overline{ \big( \delta V'_\text{\tiny FP}(p) \big ) ^2} = d_{p_1}d_{p_2} \bm{W}_\text{\tiny FP}(p_1,p_2)|_{p_1=p_2=p} &= 
\frac{q-2 p^2}{(q-1)^2} -\frac{p^2}{(q-1)^2 \left(\beta ^2 (q-1)^2 f''(q)-1\right)}\\
&-\frac{4 \beta ^2 p^2 \left(p^2-q\right) f''(q)}{\left(\beta ^2 (q-1)^2 f''(q)-1\right) \left(\beta ^2 (q-1)^3 f''(q)-2 p^2+q+1\right)}\\
&-\frac{2 \left(p^3-p q\right)^2 \left(\beta ^2 (q-1)^2 f''(q)+1\right)}{(q-1)^2 \left(\beta ^2 (q-1)^2 f''(q)-1\right) \left(\beta ^2 (q-1)^3 f''(q)-2 p^2+q+1\right)^2}\\
\end{aligned}
\end{equation}
where $q=q^*(p)$. This expression corresponds to small fluctuations of the derivative of the FP potential at fixed overlap $p$. 
The sample-to-sample susceptibility is obtained using Eq.~\eqref{chiH}, i.e., $\bm{\hat{\chi}}^\text{\tiny FP}_{sample} = \overline{ \big( [\delta V']^\text{\tiny FP}_{intra}(p) \big)^2}[\bm{\hat{\chi}}_{intra}^\text{\tiny FP}]^2$.
For the typical case, $q=p=\qEA$ and we recover the result obtained in the main text using the mass matrix on the M potential (compare with $\bm{\hat{\chi}}^\text{\tiny RFOT, FP}_{inter}$ in Eq.~\eqref{Chihet}).

\subsubsection{Proof of the equivalence with the mass matrix method}
To conclude we prove that the cumulant method and the mass method give the same sample-to-sample (or inter-state) susceptibility for a general RFOT mean-field model (without external field) at the equilibrium overlap $\qEA$.

Consider that $d_{p_1}$ and $d_{p_2}$ are total derivatives, thus following the same computation as in Eqs.~\eqref{total_der} and \eqref{Tot} we have
\begin{equation}\label{Tot2}
d_{p_1} d_{p_2}W(p_1,p_2; n_1,n_2) = \sum_{a}^{n_1}\sum_{b}^{n_2}\mathbbm{M}_{0a;0b}-\sum_{a}^{n_1}\sum_{b}^{n_2}\sum_{c\neq d}\sum_{e \neq f}\mathbbm{M}_{0a;cd}\mathbbm{M}^{-1}_{cd;ef}\mathbbm{M}_{ef;b0},
\end{equation}
where the index $a$ runs from 1 to $n_1$ and the index  $b$ from $n_1+1$ to $n_2$. 
Using Eq.~\eqref{Meq}, the first term of Eq.~\eqref{Tot2} reads
\begin{equation}\label{FRST}
\sum_{a}^{n_1}\sum_{b}^{n_2}\mathbbm{M}_{0a;0b} = n_1n_2(\frac{m_2}{4}+m_3),
\end{equation}
while the second term of Eq.~\eqref{Tot2} can be rewritten as
\begin{equation}\label{eqSUM}
\begin{aligned}
\sum_{a}^{n_1}\sum_{b}^{n_2}\sum_{c\neq d}\sum_{e \neq f}&\big [\frac{m_2}{2}(\delta_{ac}+\delta_{ad})+ m_3\big ]\mathbbm{G}_{cd;ef}\big [\frac{m_2}{2}(\delta_{be}+\delta_{bf})+ m_3\big ] = \Gamma_1+2\Gamma_2+\Gamma_3\\
\Gamma_1 \equiv & n_1 n_2 m_3^2 \sum_{c\neq d}\sum_{e \neq f}\mathbbm{G}_{cd;ef}+\\
\Gamma_2 \equiv & n_1 m_3 \frac{m_2}{2}  \sum_{b}^{n_2} \sum_{c\neq d}\sum_{e \neq f} (\delta_{ac}+\delta_{ad})\mathbbm{G}_{cd;ef} \\
\Gamma_3 \equiv & \Big (\frac{m_2}{2}\Big )^2 \sum_{a}^{n_1} \sum_{b}^{n_2} \sum_{c\neq d}\sum_{e \neq f} (\delta_{ac}+\delta_{ad})(\delta_{be}+\delta_{bf})\mathbbm{G}_{cd;ef}\\
\end{aligned}
\end{equation}
where $\mathbbm{G}_{cd;ef}=\mathbbm{M}^{-1}_{cd;ef}$. 

%\noindent\makebox[\linewidth]{\rule{\paperwidth}{0.4pt}}
At this point we notice that the sum $\sum_{c\neq d}$ (or $\sum_{e \neq f}$) can be rewritten as
\begin{equation}
\sum_{c\neq d} = \sum_{c\neq d}^{n_1}+\sum_{c\neq d}^{n_2} + \sum_{c}^{n_1} \sum_{d}^{n_2}+ \sum_{c}^{n_2} \sum_{d}^{n_1}.
\end{equation}
There are then four possible combinations of contraction of the tensor $\mathbbm{G}_{cd;ef}$:
\begin{equation}
\begin{aligned}
G^\text{\tiny I} &= \sum_{c\neq d}^{n_1}\sum_{e\neq f}^{n_1} \mathbbm{G}_{cd;ef} = n_1(n_1-1)g_1+n_1(n_1-1)^2g_2+{n_1}^2(n_1-1)^2g_3\\
G^\text{\tiny II} &= \sum_{c\neq d}^{n_1}\sum_{e\neq f}^{n_2} \mathbbm{G}_{cd;ef} = n_1(n_1-1)n_2(n_2-1)g^3\\
G^\text{\tiny III} &= \sum_{c\neq d}^{n_1}\sum_{e}^{n_1} \sum_{f}^{n_2} \mathbbm{G}_{cd;ef} = n_1(n_1-1)n_2\frac{g_2}{2}+n_1(n_1-1) n_1 n_2 g_3\\
G^\text{\tiny IV} &= \sum_{c}^{n_1} \sum_{d}^{n_2}\sum_{e}^{n_1} \sum_{f}^{n_2} \mathbbm{G}_{cd;ef} = n_1n_2\frac{g_1}{2}+({n_1}^2n_2+n_1{n_2}^2)\frac{g_2}{4}+{n_1}^2{n_2}^2g_3\\
\end{aligned}
\end{equation}

%\noindent\makebox[\linewidth]{\rule{\paperwidth}{0.4pt}}
Therefore, the three terms of Eq.~\eqref{eqSUM} can be rewritten as:
\begin{equation}
\begin{aligned}
\Gamma_1 = n_1 n_2 m_3^2 \sum_{c\neq d}\sum_{e \neq f}\mathbbm{G}_{cd;ef} &= n_1 n_2 m_3^2 \Big [  n(n-1)g_1+n(n-1)^2g_2+{n}^2(n-1)^2g_3\Big ]\\
\Gamma_2 = 2 n_1 m_3 \frac{m_2}{2} \sum_{c\neq d}\Big ( \sum_{e \neq f}^{n_2} \sum_{e}^{n_1} \sum_{f}^{n_2}\Big )\mathbbm{G}_{cd;ef} &= n_1 m_3 \frac{m_2}{2} \Big (G^\text{\tiny I} + G^\text{\tiny II}+4G^\text{\tiny III}+2G^\text{\tiny IV}\Big )\\
\Gamma_3 = \Big  (\frac{m_2}{2}\Big )^2\Big ( \sum_{c \neq d}^{n_1} \sum_{c}^{n_1} \sum_{d}^{n_2}\Big )\Big ( \sum_{e \neq f}^{n_2} \sum_{e}^{n_1} \sum_{f}^{n_2}\Big )\mathbbm{G}_{cd;ef} &=  \Big (\frac{m_2}{2}\Big )^2 \Big (G^\text{\tiny II}+2G^\text{\tiny III}+G^\text{\tiny IV}\Big )\\
\end{aligned}
\end{equation}
where in the first line $n=n_1+n_2$. Taking the derivative with respect to $n_1$ and $n_2$ and substituting  $n_1 = n_2 = 0$ we finally obtain the three terms:
\begin{equation}\label{OTH}
\begin{aligned}
\Gamma_1 &= 0\\
\Gamma_2 &= m_3 \frac{m_2}{2} (-g_1+g_2)\\
\Gamma_3 &= \Big  (\frac{m_2}{2}\Big )^2  (g_3-g_2+\frac{g_1}{2})\\
\end{aligned}
\end{equation}

Finally, using Eqs.~\eqref{FRST} and \eqref{OTH}  we obtain that (see Eq.~\eqref{chiH})
\begin{equation}
\begin{aligned}
\bm{\hat{\chi}}_{sample}^\text{\tiny FP} =  &\Big [  (\frac{m_2}{4}+m_3) -  2 m_3 \frac{m_2}{2} (-g_1+g_2) -\Big ( \frac{m_2}{2}\Big )^2  (g_3-g_2+\frac{g_1}{2})\Big ] [\bm{\hat{\chi}}_{intra}^\text{\tiny FP}]^{2}\\
= & \frac{4 m_2 (m_2+m_3)-2 m_1 (m_2+4 m_3)}{m_1^2 (2 m_1-m_2)},
\end{aligned}
\end{equation}
where $g_1,g_2,g_3$ are given by Eq.~\eqref{inv} with $n=0$.
This expression is exactly that  obtained by dynamical considerations (see Eq.~\eqref{eqhet}), which gives $\bm{\hat{\chi}}^{\text{\tiny FP}}_{sample} = 4\mathbbm{G}_{12;23} =g_2+4 g_3$, where $g_2,g_3$ are now evaluated at $n=1$.

%\section{$p$-spin Additionals}

\section{Analytical Formulas for the Mixed $p$-spin Spherical Model}
\label{AppendixC}
Recall that the function $f(q)$ characterize the specific model considered, e.g., $f_2(q) = \frac{1}{2}(q^2+h^2q)$ and $f_3(q) = \frac{1}{2}q^3$.
Given the mass parameters for the $p$-spin model (Eq.~\eqref{MassM}):
\begin{equation}
\begin{aligned}
m_1(q) &= \frac{2}{(1-q)^2} - 2\beta^2f''(q)\\
m_2(q)  &= -\frac{4q}{(1-q)^2[1+(n-1)q]} \\
m_3(q)  &= \frac{2q^2}{(1-q)^2[1+(n-1)q]^2} \qquad 
 \Big \{ +2\beta^4 h^4 \quad \text{if homogeneous }h \Big \}\\ 
\end{aligned}
\end{equation}
the various susceptibilities defined in Sec.~\ref{SMM} can be computed. The factor $2\beta^4 h^4$ is added in $m_3(q)$ only for a non-random homogeneous external field $h$, as discussed in Appendix \ref{AppendixB0}.
\begin{equation}\label{Chth}
\begin{aligned}
&\bm{\hat{\chi}}^\text{\tiny RS,FP}_{intra}(q) = \frac{1}{m_2(q)-m_1(q)}+\frac{3}{m_1(q)}
= -\frac{(q-1)^2 \left[\beta ^2 (q-1)^3 f''(q)+2 q+1\right]}{\left[\beta ^2 (q-1)^2 f''(q)-1\right] \left[\beta ^2 (q-1)^3 f''(q)+q+1\right]}\\
&\bm{\hat{\chi}}^\text{\tiny RFOT,FP}_{intra}(q) = 4 \left[\frac{1}{m_2(q)-2 m_1(q)}+\frac{1}{m_1(q)}\right]
= \frac{(q-1)^2 \left[ -\beta^2(q-1)^2 f''(q)+2 q+1\right]}{-\beta ^2 \left[q^3-3 q+2\right] f''(q)+\beta ^4 (q-1)^4 f''(q)^2+q+1}.
\end{aligned}
\end{equation}
(The RS superscript corresponds to the case $n=0$, and RFOT to $n=1$.)
\begin{equation}\label{Chdyn}
\begin{aligned}
&\bm{\hat{\chi}}^\text{\tiny RS,M}_{intra}(q) = \frac{2}{m_2(q)-m_1(q)}+\frac{4}{m_1(q)}
=-\frac{(q-1)^2 \left[\beta ^2 (q-1)^3 f''(q)+3 q+1\right]}{\left[\beta ^2 (q-1)^2 f''(q)-1\right] \left[\beta ^2 (q-1)^3 f''(q)+q+1\right]}\\
&\bm{\hat{\chi}}^\text{\tiny RFOT,M}_{intra}(q) = \frac{8}{m_2(q)-2 m_1(q)}+\frac{6}{m_1(q)} 
=\frac{(q-1)^2 \left[ -\beta^2(q-1)^2 f''(q)+3 q+1\right]}{-\beta ^2 \left(q^3-3 q+2\right) f''(q)+\beta ^4 (q-1)^4 f''(q)^2+q+1.}
\end{aligned}
\end{equation}
%We inserting $n=0$ for a global RS solution and $n=1$ for a RFOT system.
These are the two local susceptibilities and for any finite $N$ the overlap value can fluctuate. For big enough systems, the fluctuations of $q$ around $\qEA$ is captured by Gaussian fluctuations, i.e., by the sample-to-sample susceptibilities:
\begin{equation}\label{Chihet}
\begin{aligned}
\bm{\hat{\chi}}^\text{\tiny RS,FP}_{dis} &= \frac{3 m_2(\qEA){}^2-m_1(\qEA) \left(m_2(\qEA)+4 m_3(\qEA)\right)}{m_1(\qEA) \left[m_1(\qEA)-m_2(\qEA)\right]{}^2}\\
\bm{\hat{\chi}}^\text{\tiny RFOT,FP}_{inter} &= \frac{4 m_2(\qEA) \left(m_2(\qEA)+m_3(\qEA)\right)-2 m_1(\qEA) \left[m_2(\qEA)+4 m_3(\qEA)\right]}{m_1(\qEA){}^2 \left[2 m_1(\qEA)-m_2(\qEA)\right]}
\end{aligned}
\end{equation}
\begin{equation}\label{Chivar}
\begin{aligned}
\bm{\hat{\chi}}^\text{\tiny RS,M}_{dis} &= \frac{2 \left[m_2(\qEA){}^2-2 m_1(\qEA) m_3(\qEA)\right]}{m_1(\qEA) \left[m_1(\qEA)-m_2(\qEA)\right]{}^2}\\
\bm{\hat{\chi}}^\text{\tiny RFOT,M}_{inter} &= \frac{4 m_2(\qEA) \left[m_2(\qEA)+m_3(\qEA)\right]-8 m_1(\qEA) m_3(\qEA)}{m_1(\qEA){}^2 \left[2 m_1(\qEA)-m_2(\qEA)\right]}\\
\end{aligned}
\end{equation}
where we have used the fact that $\bm{\hat{\chi}}_{inter}=0$ in RS models and $\bm{\hat{\chi}}_{dis}=0$ in RFOT without external field, as commented in section \ref{Discussion}.
In RFOT models without external field the total susceptibility is
\begin{equation}
\bm{\hat{\chi}}^\text{\tiny RFOT}_{tot} = \bm{\hat{\chi}}^\text{\tiny FP}_{inter} +\bm{\hat{\chi}}_{intra}^\text{\tiny FP}(\qEA)= \bm{\hat{\chi}}^\text{\tiny M}_{inter}+\bm{\hat{\chi}}^\text{\tiny M}_{intra}(\qEA) \ ,
\end{equation}
while in RS models is
\begin{equation}
\bm{\hat{\chi}}^\text{\tiny RS}_{tot} = \bm{\hat{\chi}}^\text{\tiny FP}_{dis} +\bm{\hat{\chi}}_{intra}^\text{\tiny FP}(\qEA)= \bm{\hat{\chi}}^\text{\tiny M}_{dis}+\bm{\hat{\chi}}^\text{\tiny M}_{intra}(\qEA) \ .
\end{equation}

\section{Numerical Results for the 3-spin Model at $\mathbf{T=0.6}$}\label{AppendixC2}

In this appendix we report the results of the numerical simulation of the 3-spin spherical model at the temperature $T=0.6$. The same plots presented in Section \ref{3spinsec} for $T=0.59$ are here reproduced for $T=0.6$ (Figs.~\ref{EqCorr2},\ref{3Susc2},\ref{fig21}). These plots confirms the agreements of RFOT fluctuations and the analytical predictions. In Fig.~\ref{3Susc2} the clouds of susceptibilities follows for $q>\qEA$ the atypical branch at the same temperature (dotted and dashed-dotted), as it is the case for $T=0.59$ (see Fig.\ref{3Susc}a). We observe that at $T=0.6$ (less stable state), the gap between sample-to-sample (green/yellow) fluctuations and intra-state fluctuations (blue/red) is increased (Fig.~\ref{fig21}:left), in comparison with the $T=0.59$ case (Fig.~\ref{fig:3spin-3}). This is expected since at the transition ($T=\TMCT\approx0.61237)$ we have $\bm{\hat{\chi}}_\text{sample} \propto \bm{\hat{\chi}}_{sample}^2\sim N^{1/4}$.

{
\centering
\begin{table}[!ht]
	\begin{tabular}{|c|c|c|c|c|c|c|}
	%\multicolumn{7}{c}{3-spin; $T=0.6$; } \\
	\hline 
	$N$ & 400 & 800 & 1600 & 3200 & 6400 & 12800 \\ 
	\hline 
	$N_\mathrm{sample}$ & 322 & 70 & 80 & 98 & 46 & 10\\ 
	\hline 
	$K$ & 100 & 50 & 20 & 20 & 100 & 20 \\ 
	\hline 
\end{tabular}
\caption{Simulation parameters for the (RFOT) 3-spin at $T=0.6$ with $\tau_{\kappa}=21$}\label{table3spin2}
\end{table}

\begin{figure}[!h]
\includegraphics[width=0.45\columnwidth]{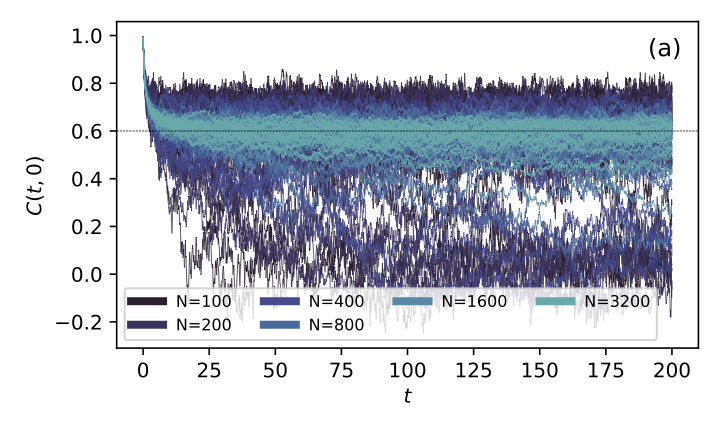}
\includegraphics[width=0.38\columnwidth]{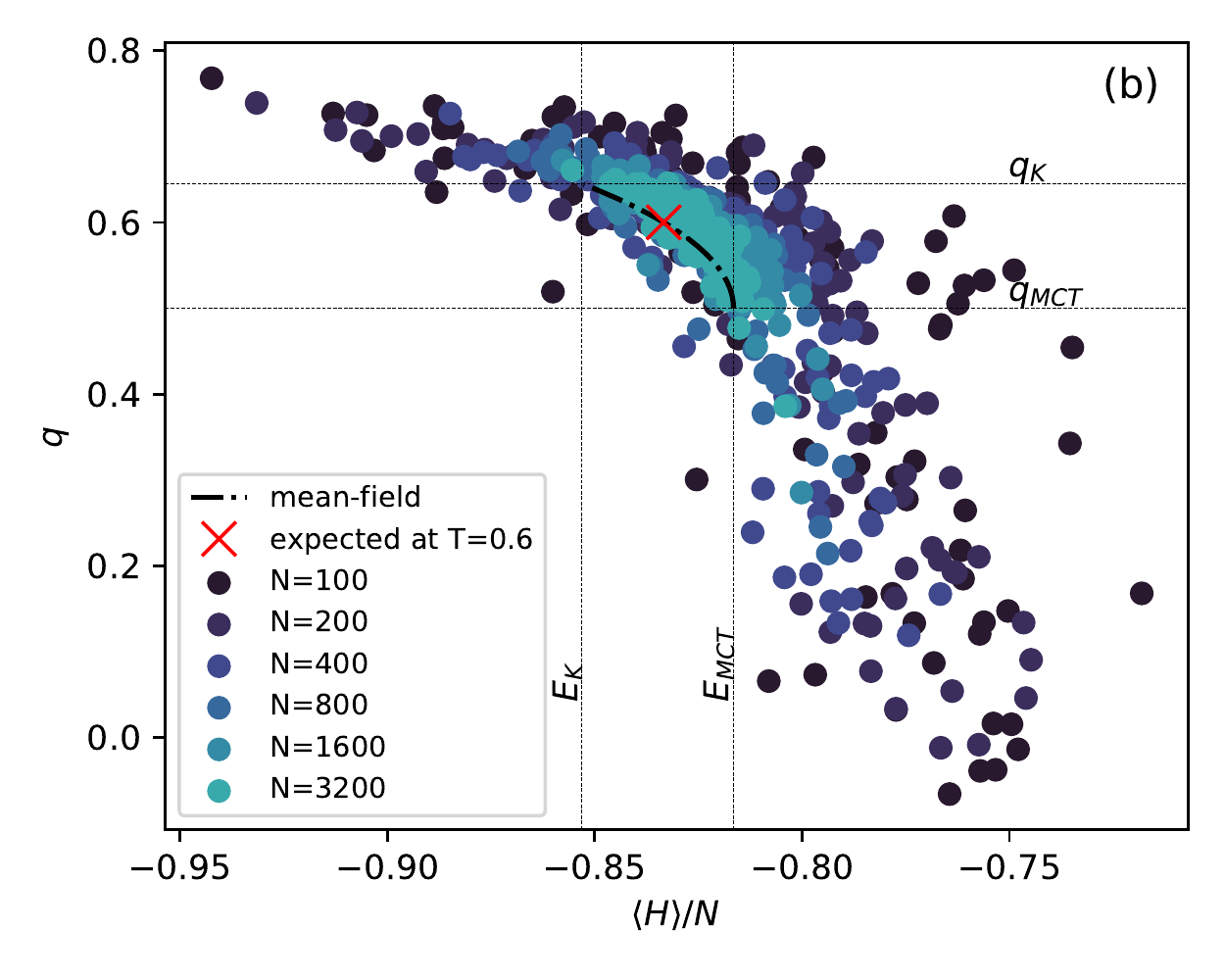}
\caption{Equilibrium dynamics in the 3-spin spherical model at $T=0.6 \lesssim \TMCT$, for different $N=100,200,400,800,1600,3200$ (to be compared with Fig.~\ref{EqCorr2} at $T=0.59$).
		\textbf{a:} Time evolution of $C(t,0)=\sum_i^Ns_i(t)s_i(0)/N$.
		\textbf{b:} Scatter plot of equilibrium overlap vs equilibrium energy in the 3-spin below $\TMCT$. Each point represents a different sample (and state) for a single equilibrium trajectory. The red cross indicates the thermodynamic expectation at that temperature.}\label{EqCorr2}
\includegraphics[width=0.45\columnwidth]{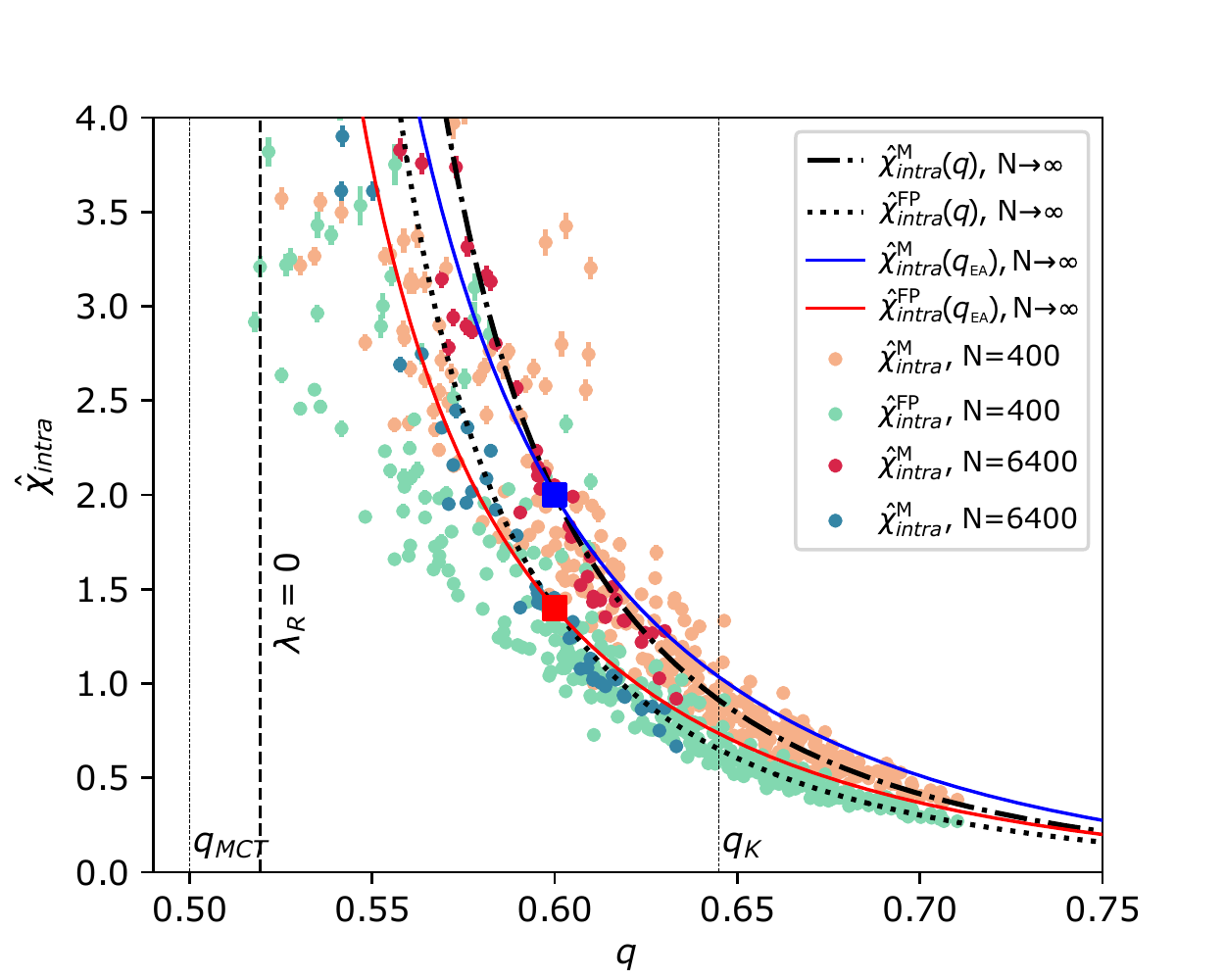}
\caption{Scatter plot of the overlap versus intra-state susceptibilities in the 3-spin at $T=0.6$ (to be compared with Fig.~\ref{EqCorr2} at $T=0.59$). On each sample the bar indicates the estimate of the error in evaluating the intra-state susceptibility. The dotted line represents the expected $q$ vs $\bm{\hat{\chi}}_{intra}^\text{\tiny FP}(q)$ at fixed temperature (Eq.~\eqref{Chth}), and the dashed-dotted for $\bm{\hat{\chi}}_{intra}^\text{\tiny M}(q)$ (Eq.~\eqref{Chdyn}). The squared red and blue dots mark the typical value in the thermodynamic limit at that temperature, i.e., $\bm{\hat{\chi}}_{intra}^\text{\tiny FP}(\qEA)$ and $\bm{\hat{\chi}}_{intra}^\text{\tiny M}(\qEA)$.}\label{3Susc2}
\includegraphics[width=0.4\columnwidth]{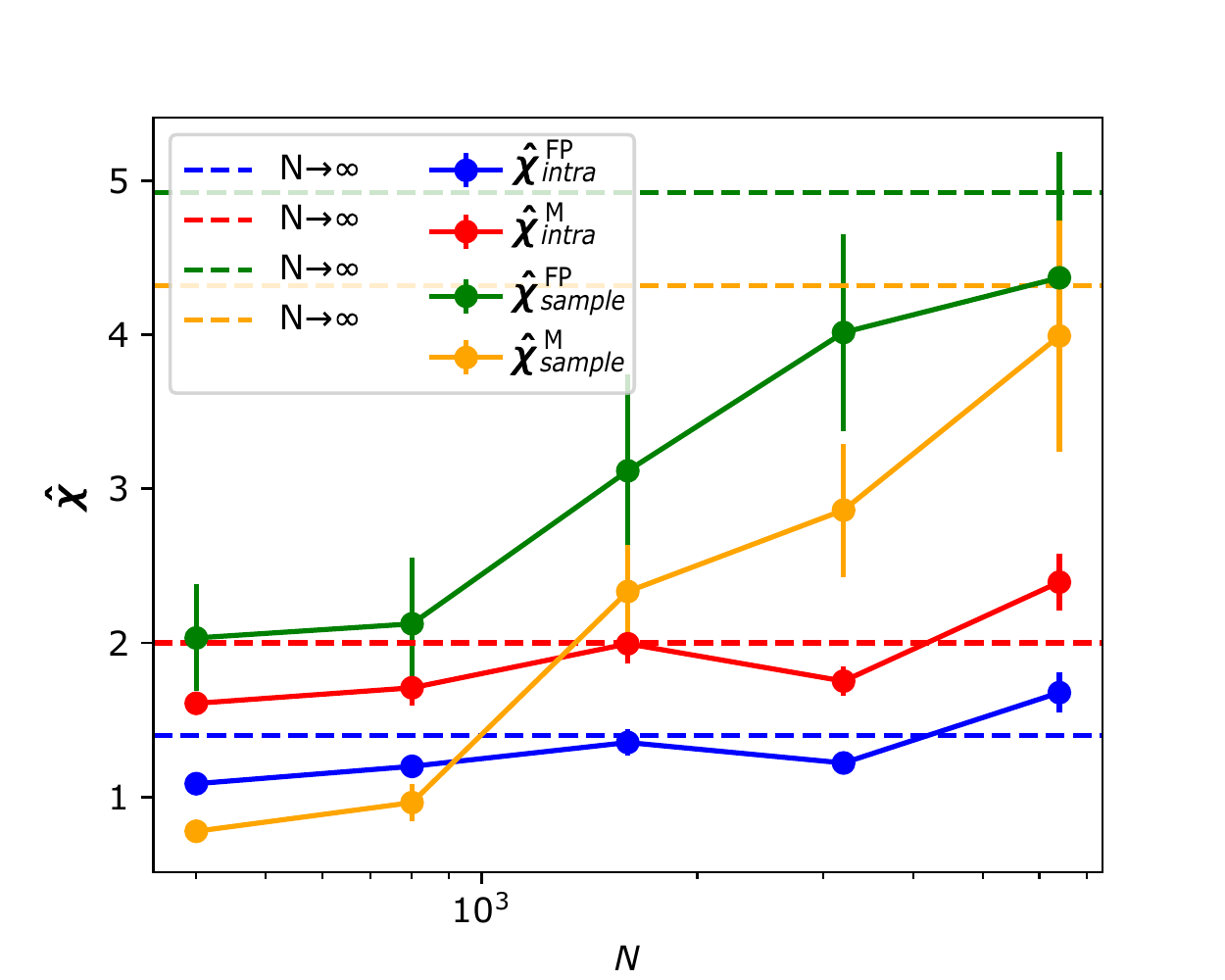}
\includegraphics[width=0.4\columnwidth]{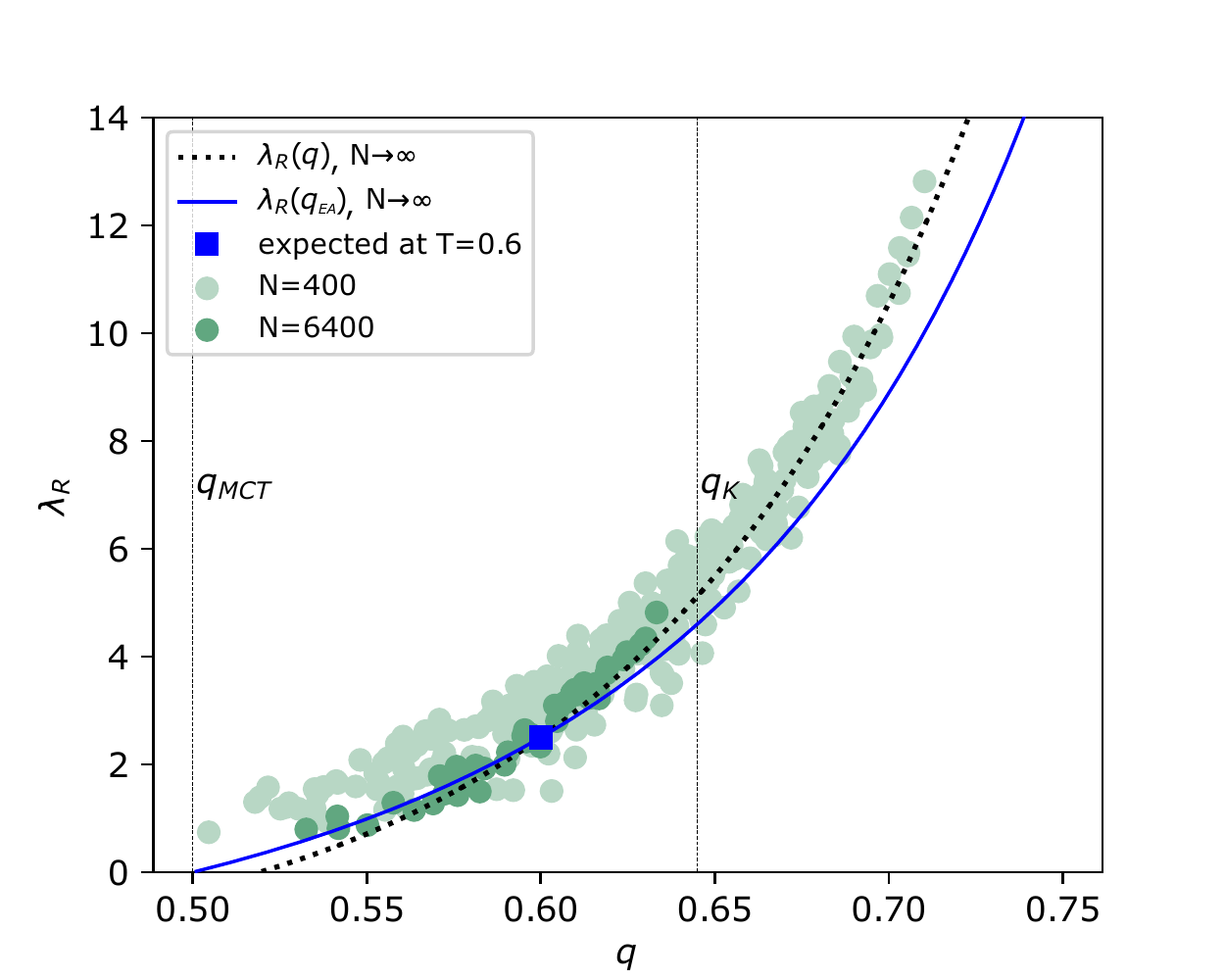}
\caption{\textbf{Left:} Size scaling of the intra-state susceptibilities $\bm{\hat{\chi}}_{intra}^\text{\tiny FP}=\overline{\hat{\chi}_{intra}^\text{\tiny FP}}$, $\bm{\hat{\chi}}_{intra}^\text{\tiny M}=\overline{\hat{\chi}_{intra}^\text{\tiny M}}$ and of the sample-to-sample susceptibilities $\bm{\hat{\chi}}_{sample}^\text{\tiny FP}$, $\bm{\hat{\chi}}_{sample}^\text{\tiny M}$ in the 3-spin at $T=0.6$. The large error in the sample-to-sample susceptibility follows from the relatively small number of samples considered (see Table~\ref{table3spin2}).
\textbf{Right:} Overlap of the state vs replicon eigenvalue $\lambda\text{\tiny R}(q)=2/(2\bm{\hat{\chi}}_{intra}^\text{\tiny FP}(q)-\bm{\hat{\chi}}_{intra}^\text{\tiny M}(q))$, as defined in Eq.~\eqref{rep}.  The dotted line is $\lambda\text{\tiny R}(q)$ in the thermodynamic limit at the respective temperature. The line $\lambda\text{\tiny R}(\qEA)$ for different temperatures (blue) intersects the point $(0,q_\text{\tiny MCT}=0.5)$. 
}\label{fig21}
\end{figure}
}

\section{Derivative of a Function of $\mQ{}{RS}$ in the ROM}\label{appendixD}
In this Appendix, we provide the additional details of the derivation for analyzing the ROM. We specifically wish to compute
\begin{equation}\label{Qk}
\partial_{q_{cd}}[\mQ{}{}^k]_{ab} = \sum_{p=0}^{k-1} ([\mQ{}{}^p]_{ac}[\mQ{}{}^{k-1-p}]_{db}+[\mQ{}{}^p]_{ad}[\mQ{}{}^{k-1-p}]_{bc}).
\end{equation}
For the RS solution $\mQ{}{RS}= d\delta_{ab}+e$, using Eq.~\eqref{RS} we have
\begin{equation}
[\mQ{}{}^p] = d^p\mathbb{I}+\frac{1}{n}(u^p-d^p)\mathbb{J},
\end{equation}
where $u = d+e$.
This matrix has non-diagonal elements $[\mQ{}{}^p]_{ab} = -\frac{1}{n}d^p[1-(\frac{u}{d})^p]$ and diagonal elements $[\mQ{}{}^p]_{aa} = q_0^p\{1 -\frac{1}{n}[1-(\frac{u}{d})^p]\}$. Therefore, there exist three different values for Eq.~\eqref{Qk}, depending on the choice of index. Using the geometric sum relation $\sum_{i=0}^{k-1}x^i = (1-x^k)/(1-x)$ we obtain
\begin{equation}
\begin{aligned}
\partial_{q_{12}}[\mQ{}{}^k]_{34} &= \sum_{p=0}^{k-1} \frac{2}{n^2} d^{k-1}\big \{ [1-(\frac{u}{d})^p][1-(\frac{u}{d})^{k-1-p}]\big \}\\
&= \frac{2}{n^2} \big ( k u^{k-1} + k q_0^{k-1}-2\frac{u^k-d^k}{u-d} \big )\equiv A_1\\
\partial_{q_{12}}[\mQ{}{}^k]_{13} &= %\sum_{p=0}^{k-1} \frac{2}{n^2} q_0^{k-1} \big ( (1-(\frac{u}{q_0})^p)(1-(\frac{u}{q_0})^{k-1-p})\big ) - \frac{1}{n}q_0^{k-1}(1-(\frac{u}{q_0})^p)\big )\\
A_1 - \frac{1}{n}kq_0^{k-1} +\frac{1}{n} \frac{d^k-u^k}{d-u}\\
\partial_{q_{12}}[\mQ{}{}^k]_{12} &= %\sum_{p=0}^{k-1} \frac{2}{n^2} q_0^{k-1} \big ( (1-(\frac{u}{q_0})^p)(1-(\frac{u}{q_0})^{k-1-p})\big ) - \frac{2}{n}q_0^{k-1}(1-(\frac{u}{q_0})^p)+q_0^{k-1}\big )\\
A_1 - \frac{2}{n}kd^{k-1} +\frac{2}{n} \frac{d^k-u^k}{d-u}+kd^{k-1}\\
\end{aligned}
\end{equation}
Given a generic function of the overlap matrix $f(\mQ{}{})$, we obtain that in the RS case
\begin{equation}\label{fQ}
\begin{aligned}
\partial_{q_{12}}[f(\mQ{}{})]_{34} 
&=  \frac{2}{n^2} f'(u) + \frac{2}{n^2} f'(d)-\frac{4}{n^2}\frac{f(u)-f(d)}{u-d} \\
\partial_{q_{12}}[f(\mQ{}{})]_{13} &= \frac{2}{n^2} f'(u) + (\frac{2}{n^2}-\frac{1}{n}) f'(d)-(\frac{4}{n^2}-\frac{1}{n})\frac{f(u)-f(d)}{u-d} \\
\partial_{q_{12}}[f(\mQ{}{})]_{12} &= \frac{2}{n^2} f'(u) + (\frac{2}{n^2}-\frac{2}{n}+1) f'(d)-(\frac{4}{n^2}-\frac{2}{n})\frac{f(u)-f(d)}{u-d} .
\end{aligned}
\end{equation}

\section{High Density Behavior of the RLG}\label{appendixE}

%\subsection{High Density Limit $\hat{\varphi}\to\infty$}	
In this Appendix, we consider the high density limit for the RLG at equilibrium ($n=1$). We start with the equilibrium density equation. Changing variable $h=\sqrt{\Delta}x$ gives
\begin{equation}
\hat{\varphi}^{-1} = \frac{\Delta}{2} \int_{-\infty}^{\infty}dh e^{h-\frac{\Delta}{4}} \frac{g'_\text{\tiny RS}(h)^2}{g_\text{\tiny RS}(h)} = \frac{\sqrt{\Delta}}{2} e^{-\frac{\Delta}{4}}\int_{-\infty}^{\infty}dx e^{\sqrt{\Delta}x} \frac{\Theta'(x)^2}{\Theta(x)}.
\end{equation}
The high density (h.d.) limit corresponds to $\Delta\to0$, hence  we obtain
\begin{equation}
\hat{\varphi}^{-1}_\mathrm{h.d.} \sim \frac{\sqrt{\Delta}}{2}\int_{-\infty}^{\infty}dx \frac{\Theta'(x)^2}{\Theta(x)} = \frac{\sqrt{\Delta}}{\pi}\int_{-\infty}^{\infty}dx\frac{e^{-2x^2}}{[1+\text{erf}(x)]} = K_\mathrm{h.d.}\sqrt{\Delta} \qquad K_\mathrm{h.d.} = 0.6387...
\end{equation}
We then evaluate $m_1,m_2,m_3$ in the high density limit, again changing variable $h=\sqrt{\Delta}x$:
\begin{equation}
\begin{aligned}
{m_1} &=  \frac{2}{\Delta^2}-\frac{1}{\Delta}\frac{\int_{-\infty}^{\infty} dh e^{h-\frac{\Delta}{4}}\big \{\frac{g''_\text{\tiny RS}(h)}{g_\text{\tiny RS}(h)} - \big [\frac{g'_\text{\tiny RS}(h)}{g_\text{\tiny RS}(h)}\big ]^2\big \}^2 g_\text{\tiny RS}(h)}{ \int_{-\infty}^{\infty}dh e^{h-\frac{\Delta}{4}} \frac{g'_\text{\tiny RS}(h)^2}{g_\text{\tiny RS}(h)}} = \frac{1}{\Delta^2} \Big \{ 2 -  \frac{\int_{-\infty}^{\infty}dx e^{\sqrt{\Delta}x} \big [\frac{\Theta''(x)}{\Theta(x)}-(\frac{\Theta'(x)}{\Theta(x)})^2\big ]^2\Theta(x)}{\int_{-\infty}^{\infty}dx e^{\sqrt{\Delta}x} \frac{\Theta'(x)^2}{\Theta(x)}}\Big \}\\
{m_2} &= -\frac{1}{\Delta^2} \Big \{ 4 + 2\frac{\int_{-\infty}^{\infty}dx e^{\sqrt{\Delta}x} (\frac{\Theta'(x)}{\Theta(x)})^2 \big [\frac{\Theta''(x)}{\Theta(x)}-(\frac{\Theta'(x)}{\Theta(x)})^2\big ]\Theta(x)}{\int_{-\infty}^{\infty}dx e^{\sqrt{\Delta}x} \frac{\Theta'(x)^2}{\Theta(x)}}\Big \}\\
{m_3} &= \frac{1}{\Delta^2} \Big \{ 2 - \frac{\int_{-\infty}^{\infty}dx e^{\sqrt{\Delta}x} (\frac{\Theta'(x)}{\Theta(x)})^4\Theta(x)}{\int_{-\infty}^{\infty}dx e^{\sqrt{\Delta}x} \frac{\Theta'(x)^2}{\Theta(x)}}\Big \}\\
\end{aligned}
\end{equation}
Taking the high-$d$ high density limit $\Delta\to0$ gives:
\begin{equation}
\begin{aligned}
{m_1}_\mathrm{h.d.} &\sim \frac{1}{\Delta^2} \Big \{ 2 - \frac{\int_{-\infty}^{\infty}dx \frac{8 e^{-4 x^2} \left\{\sqrt{\pi } e^{x^2} x [\text{erf}(x)+1]+1\right\}^2}{\pi ^2 [\text{erf}(x)+1]^3}}{2K_\mathrm{h.d.}} \Big \} = \frac{2-1}{\Delta^2}=\frac{1}{\Delta^2}\\
{m_2}_\mathrm{h.d.} &\sim  -\frac{1}{\Delta^2} \Big \{ 4 + 2 \frac{\int_{-\infty}^{\infty}dx\frac{e^{-4 x^2} \left\{-8 \sqrt{\pi } e^{x^2} x [\text{erf}(x)+1]-8\right\}}{\pi ^2 [\text{erf}(x)+1]^3}}{2K_\mathrm{h.d.}} \Big \} = -\frac{4-\frac{1.8438...}{0.6387...}}{\Delta^2}=-\frac{1.1131...}{\Delta^2}\\
{m_3}_\mathrm{h.d.} &\sim \frac{1}{\Delta^2} \Big \{ 2 - \frac{1}{2}\frac{\int_{-\infty}^{\infty}dx\frac{8 e^{-4 x^2}}{\pi ^2 [\text{erf}(x)+1]^3}}{2K_\mathrm{h.d.}} \Big \}= \frac{2-\frac{5.5313...}{4 \times 0.6387...}}{\Delta^2} = \frac{0.1777...}{\Delta^2} .
\end{aligned}
\end{equation}

\section{Nomenclature for the Susceptibilities in Different References}\label{appendixH}

\begin{table}[ht]
    {\large
	\centering
	\begin{tabular}{|c|c|c|c|c|c|c|c|}
		%\multicolumn{7}{c}{3-spin; } \\
		\hline 
		This work & $\bm{\hat{\chi}}_{intra}^\text{\tiny FP}$ &  $\bm{\hat{\chi}}_{sample}^\text{\tiny FP}$ & $\bm{\hat{\chi}}_{dis}^\text{\tiny FP}$ &
		$\bm{\hat{\chi}}_{tot}$ &
		$\bm{\hat{\chi}}_{intra}^\text{\tiny M}$ & $\bm{\hat{\chi}}_{sample}^\text{\tiny M}$ & $\bm{\hat{\chi}}_{dis}^\text{\tiny M}$ \\
		\hline 
		Ref.\cite{berthier_structure_2007} & 
		$\chi_{4,C}^{iso}$ &  
		$\delta_{4,C}$ & 
		& 
		$\chi_{4,C}$ &
		& 
		&
		\\
		\hline 
		Ref.\cite{franz_field_2011,franz_static_2013} & 
		$\chi_{th}$ &  
		$\chi_{het}$ & 
		$\chi_{dis}$ & 
		$\chi_{tot}$ & 
		 & 
		 & 
		 \\
		\hline 
		Ref.\cite{baik_spherical_2021} & 
		& 
		&  
		& 
		& $\sigma^2_{\mathfrak{R}}$
		& 
		& $\propto \mathcal{S}_{N}$
		\\
		\hline 
	\end{tabular}
	}
\end{table}

\end{document}